\documentclass[%
preprint,
 amsmath,amssymb,
 aps,
]{revtex4-2}

\usepackage{graphicx}
\usepackage{dcolumn}
\usepackage{bm}
\usepackage{braket}
\usepackage{graphicx}
\usepackage{epstopdf}
\usepackage{mwe}    
\usepackage{subfig}
\usepackage[T1]{fontenc}
\usepackage{caption} 
\usepackage{amssymb}
\usepackage{amsmath}
\usepackage{booktabs}
\usepackage{tabularx}
\usepackage{multirow}

\usepackage[T1]{fontenc}
\usepackage{caption}
\usepackage{floatrow}
\usepackage[font=normalsize]{caption}

\usepackage{xcolor}

\newcommand{\tensr}[1]{\bm{\mathsf{#1}}} 

\begin{document}

\preprint{PREPRINT}

\title{On Boundary Conditions for Lattice Kinetic Schemes for Magnetohydrodynamics Bounded by Walls with Finite Electrical Conductivities}

\author{Eman Yahia}
\email{Eman.Yahia@ucdenver.edu}
\affiliation{Department of Mechanical Engineering, University of Colorado Denver, 1200 Larimer street, Denver, CO  80204, U.S.A.\\}

\author{Kannan N. Premnath}
\email{Kannan.Premnath@ucdenver.edu}
\affiliation{Department of Mechanical Engineering, University of Colorado Denver, 1200 Larimer street, Denver, CO  80204, U.S.A.\\}

\date{\today}

\begin{abstract}
Magnetohydrodynamic (MHD) flow of liquid metals through conduits play an important role in the proposed configurations for harnessing fusion energy, and various other engineering and scientific problems. The interplay between the magnetic fields and the fluid motion gives rise to complex flow physics. Moreover, the electrical conductivity of the bounding solid walls relative to that of the fluid, in addition to the Hartmann number, has a major influence on the structure of the flow field and the induced magnetic fields in such cases. An effective approach to represent such effects is via the Shercliff boundary condition for thin conducting walls relating the induced magnetic field and its wall normal gradient at the boundary via a parameter referred to as the wall conductance ratio~\cite{shercliff1956flow}. Within the framework of the lattice Boltzmann (LB) method for MHD, a lattice kinetic scheme (LKS) involving a vector distribution function for the magnetic fields was proposed by Dellar~\cite{dellar2002lattice}. However, the prior LB studies only considered boundary implementations for the limiting special cases involving either insulating or perfectly conducting walls without accounting for the finite conductivity effects of the container walls. In this paper, we present two new boundary schemes that enforce the Shercliff boundary condition in the LB schemes for MHD. It allows for the specification of the wall conductance ratio to any desired value based on the actual conductivities and length scales of the container and the wall thicknesses and can be made to vary freely between the special cases of zero (for insulating walls) and infinity (for perfectly conducting walls). One approach is constructed using a link-based formulation involving a weighted combination of the bounce back and anti-bounce back of the distribution function for the magnetic field and the other approach involves an on-node moment-based implementation. Both the boundary schemes are parameterized by the relaxation time of the distribution function of the magnetic field and the wall conductance ratio, and prescribe the closure relations for the incoming distribution functions from the walls. Moreover, their extensions to representing moving walls are presented. The boundary schemes are validated for canonical body force driven or shear driven MHD flows for a wide range of the values of the wall conductance ratio. The accuracy and second order grid convergence of these new and more general boundary schemes for LKS for MHD are demonstrated.
\end{abstract}

\maketitle



\section{Introduction}
The modern study of magnetohydrodynamics (MHD) can be traced back to the pioneering works of Hartmann~\cite{hartmann1937theory,hartmann1937hg} and Alfv\'{e}n~\cite{alfven1942existence}. The former performed an analytical and experimental investigation of liquid metals in channels, while the latter predicted the existence of the MHD waves, which was then confirmed experimentally~\cite{lundquist1949experimental}. Much of the early investigations centered on MHD problems related to astrophysics and the geodynamo~\cite{elsasser1955hydromagnetism,molokov2007magnetohydrodynamics}. On the other hand, there have been increasing number of technological problems involving MHD in power generation and materials processing applications~\cite{davidson1999magnetohydrodynamics}, and in systems such as tokamaks for the demonstration of the nuclear fusion process~\cite{muller2013magnetofluiddynamics}.

Since a successful harnessing of fusion energy would represent a major breakthrough advance in meeting the future energy needs with essentially no carbon footprint, there have many ongoing research endeavors in this direction, including the experimental ITER project, aimed to overcome the associated challenges. From a fluid mechanics perspective, fusion energy systems are of great current interest as they involve a variety of configurations involving the flow of liquid metals under high magnetic fields and subjected to different conditions thereby representing rich avenues for studying and understanding the multiphysics associated with wall bounded MHD flows (see e.g.,~\cite{morley2000liquid,smolentsev2010mhd,rhodes2018magnetohydrodynamic,smolentsev2021physical,mistrangelo2021mhd}). For example, the transport of the lead-lithium alloy and confined by solid wall structures is being considered both as a breeder and coolant as part of the liquid blanket module in the proposed fusion reactor configurations.

The fluid motions of the liquid metals under imposed magnetic fields are represented by the Navier-Stokes (NS) equations augmented with a locally varying Lorentz force that arises from the interaction between the electrically conducting fluid in motion and the magnetic field. The evolution of the components of the magnetic field is based on a combined representation of the different laws given by the Maxwell's equations of electromagnetism and referred to as the magnetic induction equation, which is, in turn, influenced by the fluid motions~\cite{shercliff1965,moreau1990magnetohydrodynamics,davidson2001}. Taken together, these macroscopic equations for MHD, which are briefly summarized in the next section (see Sec.~\ref{sec:MacroEqns}), are, in general, strongly coupled system for the spatio-temporal variations of the velocity and magnetic fields.

An important element associated with solving the MHD equations is the representation of the boundary conditions. While the velocity field is subjected to the well-known no-slip conditions at the boundary, the boundary conditions for the magnetic field requires some further consideration. Since the metallic fluid in motion is confined by a solid wall, whose material, in general, can have a finite electrically conductivity, the effect of the latter needs to be represented on the former consistently. Indeed, the electrical properties of the solid wall can significantly affect the structure of both the magnetic and velocity fields for the coupled MHD system (see e.g.,~\cite{muller2013magnetofluiddynamics,blishchik2021extensive,blishchik2021observation}). One approach to accommodate this would be to solve the magnetic induction equations within both the liquid and the solid and then impose the appropriate conjugate continuity conditions at their boundary locations. However, due to the possibly large disparities between the electrical conductivities of the liquid and solid, such an approach becomes stiff and computationally challenging to solve.

An alternative and effective strategy would be to impose the effect of the solid with its finite conductivity only as a boundary condition under a thin conducting wall assumption first proposed by Shercliff~\cite{shercliff1956flow}, which avoids the need to directly solve for the magnetic field variations within the solid. Essentially, this approach, whose derivation is given for completeness in Sec.~\ref{sec:ShercliffBC}, involves relating the induced magnetic field and its wall normal gradient at the boundary via a parameter referred to as the wall conductance ratio $c_w$~\cite{muller2013magnetofluiddynamics}. Some early analytical studies that demonstrate the influence of this parameter on certain wall-bounded canonical MHD problems are presented in Refs.~\cite{chang1961duct,hunt1965magnetohydrodynamic,yu1970convective}. Moreover, over the years, the Shercliff boundary condition accommodating the electrical properties of the wall have been widely used in various numerical investigations that simulate MHD within confining geometries, including those related to the fusion engineering applications -- for example, see Refs.~\cite{singh1984finite,sterl1990numerical,mistrangelo2006three,vantieghem2009velocity,mistrangelo2017magnetohydrodynamic,arlt2017effect,mistrangelo2021mhd}. Indeed, the main characteristic electro-magnetic parameters in such MHD problems are the Hartmann number, which is related to the relative effect of the imposed magnetic field strength and defined later, and the wall conductance ratio. Thus, a key feature of any effective numerical scheme for realistic liquid metal MHD in wall bounded conduits requires a strategy that accommodates the finite conductivity of walls via the Shercliff boundary condition.

The lattice Boltzmann (LB) method~\cite{mcnamara1988use,higuera1989boltzmann,he1997theory,dellar2013interpretation} is a mesoscopic computational technique for fluid dynamics and other transport problems, and has seen remarkable success during the last three decades~\cite{benzi1992lattice,lallemand2021lattice}. Emerging from the lattice gas automata with further contributions from certain key elements of the kinetic theory, it has evolved into a highly parallelizable numerical algorithm that computes the distribution functions varying as a result of a local collision step, which is followed by a lock-step advection or streaming along certain discrete particle velocity directions. Such discrete velocities, which are referred to as the lattice, respect certain symmetric properties in order to recover the required invariance properties associated with the macroscopic behavior of the fluid motions, such as the isotropy of the viscous stress tensor. It may be noted that the standard convention to represent a lattice set is $DdQq$, where $d$ represents the number of spatial dimensions and $q$ denotes the number of discrete particle directions. The effect of collision is often modeled via a relaxation process that represents the approach of the distribution functions or their moments to their equilibria at certain characteristic relaxation times. The boundary conditions on the velocity field in simulating the NS equations, for example, need to be specified in terms of the incoming distribution functions from the boundary nodes. Since the boundary conditions have a major influence on the numerical accuracy, stability and computational efficiency, the development of the required closure relations has attracted significant focus and much effort (see e.g., Ref.~\cite{kruger2017lattice}). For representing the no-slip velocity boundary condition, one of the most popular approaches is the link-based bounce-back type boundary scheme and its variants (e.g., Refs.~\cite{ladd1994numerical,bouzidi2001momentum}). As an alternative in this regard, one could enforce a subset of the hydrodynamic moments of the distribution functions directly on the lattice nodes~ (e.g., Refs.~\cite{noble1995consistent,bennett2010lattice,bennett2012lattice,reis2012lattice,dellar2013moment}).

The LB method has the following advantages compared to the more conventional numerical techniques. First, its streaming step is linear, while its collision step models all nonlinearity locally and are thus characterized by its simpler implementations; on the other hand, given that the convective term of the Navier-Stokes equations is nonlinear and nonlocal, their direct solution techniques based on classical approaches require more cumbersome implementation procedures. Moreover, the pressure field is computed locally via an equation of state in the LB method, which avoids the need to solve the time-consuming Poisson-type elliptic equation required by the conventional methods. Second, the local algorithmic procedure of the LB approach lends itself for a natural implementation on modern parallel computing systems for efficient simulations of large size flow problems. Third, the boundary conditions associated with complex geometries can be accommodated with ease in the LB method via relatively simple rules on the particle distribution functions on Cartesian grids, which circumvents the often time-consuming grid generation process required with traditional approaches. And, fourth, the coupling of the complex multiphysics phenomena with fluid motions can be represented readily in the LB formulations via utilizing additional distribution functions and specialized collision models, with their implementations following the same general framework of the collide-and-stream algorithmic steps.

The LB method has also been applied for simulating MHD. Some of the earlier approaches~\cite{chen1991lattice,martinez1994lattice} in such cases were inspired from the developments in the lattice gas automata~\cite{chen1992magnetohydrodynamics}. These, along with other formulations~\cite{succi1991lattice,mendoza2008three}, led to complicated algorithms for simultaneously computing the velocity and magnetic fields, which compromised their implementation and efficiency. Recognizing the features unique to the vectorial magnetic induction equation, Dellar~\cite{dellar2002lattice} proposed a separate evolution equation for the vector distribution function, referred to as the lattice kinetic scheme (LKS) for representing the dynamics of the magnetic field. On the other hand, the usual scalar distribution function is solved for the velocity field. This strategy resulted a significant simplification when compared to the earlier approaches LB approaches for MHD. It was further extended by Pattison \emph{et al}~\cite{pattison2008progress}, in particular, in using grid refinement approaches for resolving the Hartmann layers at high Hartmann number MHD flows relevant to fusion MHD problems and a more general collision model for the velocity field solver for enhanced stability. Furthermore, a moment system representation for the evolution of the vector distribution function was proposed and analyzed by Dellar~\cite{dellar2009moment}, who also extended it for handling a class of MHD problems with current-dependent resistivity~\cite{dellar2013lattice}. Also, recently, De Rosis \emph{et al}~\cite{de2021one} presented a one-stage, simplified LBM for efficient simulation of MHD. Here, we mention that the prior investigations involving LB schemes for MHD have provided relatively limited focus on the boundary conditions for the vector distribution functions for the magnetic field. Most of the studies treat the wall to be insulated -- a limiting special case situation, which was implemented using an anti-bounce back scheme in Ref.~\cite{dellar2002lattice}, an extrapolation method in Ref.~\cite{pattison2008progress} and a moment method in Ref.~\cite{dellar2013moment}. As such, none of the prior investigations seem to have explicitly considered the finite electrical properties of the wall, which, as mentioned above, can have significant influence on both the magnetic and velocity field distributions within the fluid in motion, and hence are essential in faithfully representing realistic wall-bounded MHD flows, especially involving liquid metals.

This paper addresses the above mentioned limitation and constructs and analyzes two new boundary schemes for the vector distribution function for MHD by incorporating the actual electrical properties of the wall via the Shercliff condition for thin conducting walls~\cite{shercliff1956flow}. The first approach is constructed using a link-based formulation, which, as we shall see, would be based on a weighted combination of the bounce back and anti-bounce back of the distribution function for the induced magnetic field. The second boundary scheme developed for the conducing walls will be an on-node moment based approach. The resulting closure relations for the incoming distribution functions from the walls for both the boundary schemes will be shown to be parameterized by the relaxation time of the distribution function of the magnetic field and the wall conductance ratio. Both the schemes will be further generalized for representing MHD involving moving walls. The resulting boundary condition specification approaches can use the wall conductance ratio based on the true electrical conductivity of the wall relative to that of the fluid and also involving the actual length scale of the container relative to the wall thickness. We will then test and validate our boundary formulations for well-defined class of body force driven as well as shear driven MHD flows for a wide range of the wall conductance ratio, and also investigate the order of accuracy for grid convergence.

The outline of this paper is as follows. Section~\ref{sec:MacroEqns} presents the governing macroscopic equations for MHD, which is then followed by a derivation of the Shercliff thin wall boundary condition in Sec.~\ref{sec:ShercliffBC}. The LB algorithms for evolving a vector distribution function for the magnetic field and a scalar distribution function for the velocity field are discussed briefly in Secs.~\ref{sec:LKSmagneticfield} and~\ref{sec:LBMvelocityfield}, respectively. Then, Secs.~\ref{sec:linkbasedShercliffBC} and~\ref{sec:momentbasedShercliffBC} present the derivations of the new link-based and on-node moment-based boundary schemes, respectively, for the magnetic induction equation in wall-bounded geometries. The results for validation and grid convergence of the two boundary schemes for the magnetic distribution function for the canonical body-force driven and shear-driven MHD flows are discussed in Sec.~\ref{sec:resultsanddiscussion}. Finally, a summary of the developments and conclusions arising from this work as well as its future outlook are provided in Sec.~\ref{sec:summaryandconclusions}. Moreover, various additional details supporting the developments of this work are presented in the appendices. For completeness,~\ref{sec:momentbasedBCvelocity} discusses the moment-based formulation for imposing the no-slip velocity condition, and the two boundary schemes constructed in the main sections are extended for moving walls in~\ref{sec:bcsmovingwalls}. Finally, a detailed derivation of the analytical solution of the shear-driven MHD flow between two parallel plates with contrasts in the wall conductance ratios, which is utilized for validation of the proposed boundary schemes, is presented in~\ref{sec:analyticalsolutionMHDCouetteflow}.

\section{Macroscopic Equations of MHD}\label{sec:MacroEqns}
The governing equations for MHD is a coupled system of equations involving the evolution of the density and momentum fields given by the standard Navier-Stokes equations including the Lorentz force, and a magnetic induction equation representing the variation of the magnetic fields. It may be noted that the magnetic induction equation is obtained from the Maxwell's equations of electromagnetism by combining the Faraday's law, Amp\`{e}re-Maxwell's law and Ohm's law under the constraints of the solenoidal magnetic field, i.e., no magnetic monopoles~\cite{shercliff1965,moreau1990magnetohydrodynamics,davidson2001,muller2013magnetofluiddynamics}. Thus, the MHD equations can be written as
\begin{subequations}\label{eq:NSE-MHD}
\begin{eqnarray}
\partial_t \rho + {\bm{\nabla}}\cdot (\rho \bm{u})   &=& 0,\label{eq:NSE-MHD1} \\
\partial_t (\rho \bm{u})+ {\bm{\nabla}} \cdot (\rho \bm{u}\bm{u}) &=& -{\bm{\nabla}} p+{\bm{\nabla}} \cdot \tensr{T}+\bm{F}_{Lorentz}+\bm{F}_{ext},\label{eq:NSE-MHD2}\\
\partial_t \bm{B} + {\bm{\nabla}}\cdot(\bm{u}\bm{B}-\bm{B}\bm{u})   &=& \eta \nabla^2\bm{B},\quad\quad \eta = \frac{1}{\sigma\mu_m},\label{eq:NSE-MHD3}
\end{eqnarray}
\end{subequations}
where $\rho$, $\bm{u}$ and $\bm{B}$ are the density field, velocity field, and magnetic field, respectively, $p$ is the pressure, $\bm{F}_{Lorentz}$ is the Lorentz force, $\bm{F}_{ext}$ is an external body force. Here, $\tensr{T}$ is the viscous stress tensor given by $\tensr{T}= \rho\nu\left[\bm{\nabla}\bm{u}+(\bm{\nabla}\bm{u})^\dag\right]$, where $\nu$ is the kinematic viscosity of the fluid, and $\eta$ is the magnetic resistivity, which is related to the electrical conductivity of the fluid $\sigma$ and the magnetic permeability $\mu_m$. The Lorentz force $\bm{F}_{Lorentz}$ appearing in Eq.~(\ref{eq:NSE-MHD2}) reads as
\begin{equation}
\bm{F}_{Lorentz} = \bm{J}\times \bm{B},
\end{equation}
where
\begin{equation}
\bm{J} = \frac{1}{\mu_m}\bm{\nabla}\times \bm{B}
\end{equation}
is the local current density in the fluid. The above system of equations are to be supplemented with appropriate boundary conditions at the interface between the fluid and the solid walls. The velocity field satisfies the standard no-slip condition. A treatment of the boundary condition for the magnetic field that accounts for the finite wall conductivity is discussed next.

\section{Shercliff Boundary Condition on the Induced Magnetic Field for Thin Conducting Walls}\label{sec:ShercliffBC}
\begin{figure}[ht]
\centering
\includegraphics[width=0.5\linewidth]{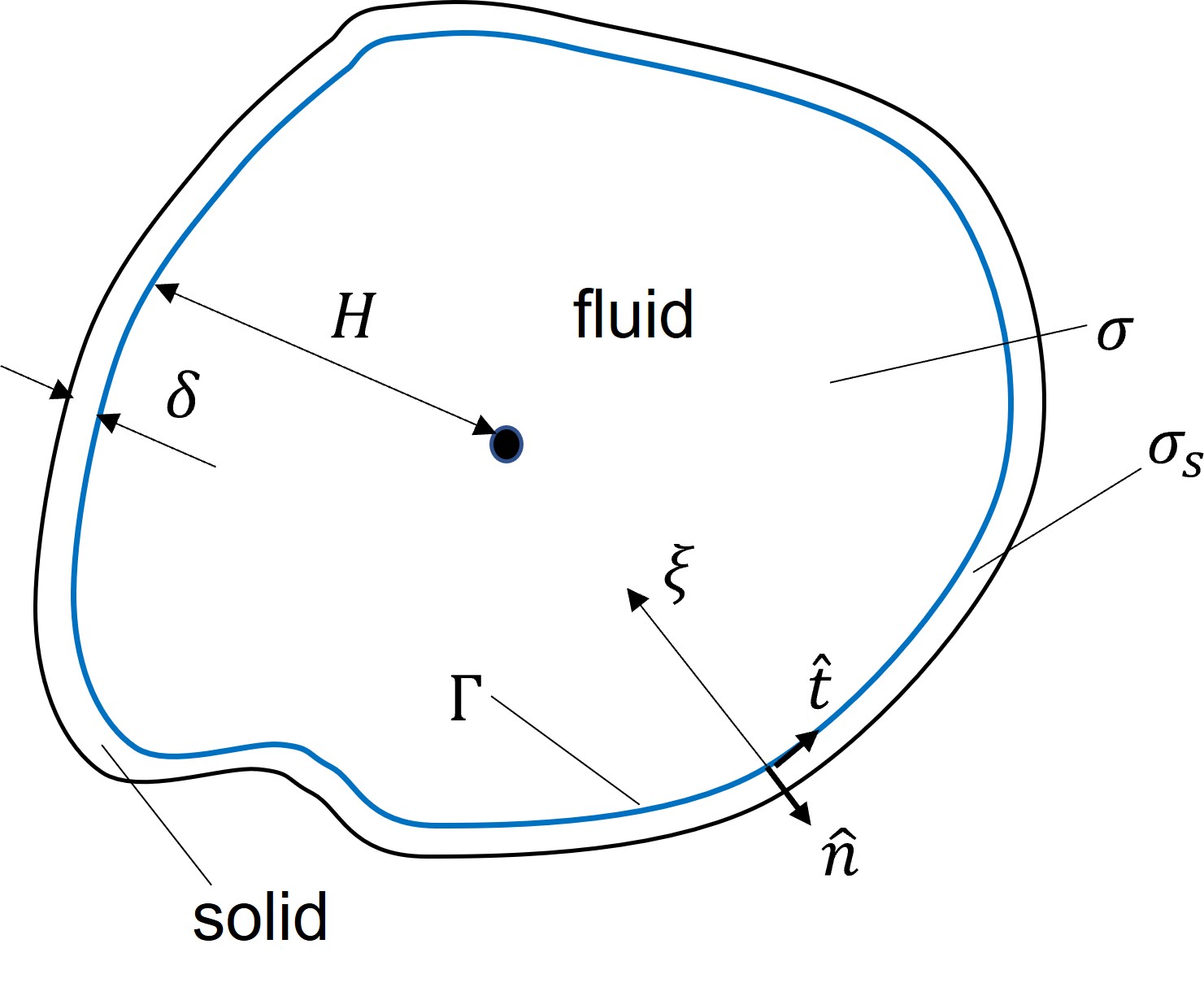}
\caption{Illustration of a cross section of an electrically conducting fluid bounded by a solid wall with a finite electrical conductivity for the derivation of the Shercliff thin wall boundary condition.}
\label{fig:fig_schematic_Shercliff_BC} 
\end{figure}

We will now present a derivation of the of the boundary condition on the induced magnetic field for thin conducting walls proposed by Shercliff~\cite{shercliff1956flow}. Figure~\ref{fig:fig_schematic_Shercliff_BC} illustrates a cross section of a fluid with an electrical conductivity $\sigma$ bounded by a solid wall whose electrical conductivity is $\sigma_s$, where $H$ is the characteristic length scale of the confinement of the fluid and $\delta$ is the wall thickness. The boundary between the fluid and solid is represented by $\Gamma$ with an outward unit normal $\hat{n}$ and a unit tangent $\hat{t}$.

Let $B$ denotes the cross-section normal component of the induced magnetic field in the fluid, while $B_s$ represents a similar component within the solid wall.
On the interface $\Gamma$, the no-slip velocity condition, and the continuity of the tangential component of the electrical field $E$, and of the cross-section normal component of the induced magnetic field $B$ are satisfied, which read as
\begin{equation}
\bm{u}=\bm{u}_s=0,\quad\quad E_t = E_{t,s}, \quad\quad B = B_s, \quad\quad \mbox{on}\;\; \Gamma.\label{eq:interfaceconditionsMHD}
\end{equation}
Using the Ohm's law for the current density $\bm{J}=\sigma(\bm{E}+\bm{u}\times\bm{B})$ in the tangential direction along $\Gamma$, and by exploiting the conditions appearing in the no-slip condition, i.e., $J_t|_\Gamma=\sigma E_t|_\Gamma$ or $E_t|_\Gamma=J_t|_\Gamma/\sigma$, the continuity condition of the electric field in the above (Eq.~(\ref{eq:interfaceconditionsMHD})) becomes
\begin{equation}
\left(\frac{J_t}{\sigma}\right)_\Gamma=\left(\frac{J_{t,s}}{\sigma_s}\right)_\Gamma. \label{eq:interfaceconditionelectricfield}
\end{equation}
Then, writing the tangential current density $J_t$ as a conjugation of the normal derivative of magnetic field component $B$, i.e., $J_t=(1/\mu_m)\partial B/\partial \hat{n}$, Eq.~(\ref{eq:interfaceconditionelectricfield}) can be rewritten as
\begin{equation}
\left(\frac{1}{\sigma}\frac{\partial B}{\partial \hat{n}}\right)_\Gamma=\left(\frac{1}{\sigma_s}\frac{\partial B_s}{\partial \hat{n}}\right)_\Gamma, \label{eq:interfaceconditionelectricfield2}
\end{equation}
It may be noted that the induced magnetic field $B$ in the fluid satisfies a component of the magnetic induction equation given in Eq.~(\ref{eq:NSE-MHD3}), while the component $B_s$ in the solid obeys
\begin{equation}
\bm{\nabla}\cdot(\eta_s \bm{\nabla}B_s)= 0,\quad\quad \eta_s = 1/(\sigma_s\mu_m). \label{eq:magneticfieldequationinsolid}
\end{equation}
To proceed further, following Shercliff~\cite{shercliff1956flow}, we now assume that the solid conducting walls are relatively thin, i.e., $\delta \ll H$. This means that the solution of the Laplace's equation (Eq.~(\ref{eq:magneticfieldequationinsolid})) for $B_s$ can be well approximated by a function \emph{linear} in the spatial coordinates, analogous to the solution of the Laplace's equation within a membrane. Let $\xi$ be the local coordinate normal to the wall (see Fig.~\ref{fig:fig_schematic_Shercliff_BC}). Since $\hat{n}$ is the outward unit normal by definition, it follows that $\xi=-\hat{n}H$. Then, we can prescribe an ansatz $B_s=c_1\xi+c_2$ and subject to $B_s(\xi=0)=B_s|_\Gamma$ and $B_s(\xi=-\delta)=0$, whose solution immediately yields $B_s(\xi)=B_s|_\Gamma(1+\xi/\delta)$ or $B_s(\hat{n})=B_s|_\Gamma(1-\hat{n}H/\delta)=B|_\Gamma(1-\hat{n}H/\delta)$, where the last expression arises from the continuity of the magnetic field at the interface via Eq.~(\ref{eq:interfaceconditionsMHD}). Then, the wall normal gradient of the induced magnetic field within the solid wall evaluated at the interface becomes the following:
\begin{equation}
\left(\frac{\partial B_s}{\partial \hat{n}}\right)_\Gamma=-B|_\Gamma\frac{H}{\delta}. \label{eq:magneticfieldgradientinsolid}
\end{equation}
Using the above expression (Eq.~(\ref{eq:magneticfieldgradientinsolid})), which essentially is only related to the magnetic field of interest from the fluid side evaluated at the interface and the inherent length scales of the problem, in Eq.~(\ref{eq:interfaceconditionelectricfield2}), we obtain
\begin{equation}
\left(\frac{1}{\sigma}\frac{\partial B}{\partial \hat{n}}\right)_\Gamma=-\frac{1}{\sigma_s}\frac{H}{\delta}B|_\Gamma. \label{eq:interfaceconditionelectricfield3}
\end{equation}
Finally, defining
\begin{equation}
c_w=\frac{\sigma_s\delta}{\sigma H} \label{eq:wallconductanceratiodefinition}
\end{equation}
as the \emph{wall conductance ratio} parameter, which is a dimensionless quantity, and using it in Eq.~(\ref{eq:interfaceconditionelectricfield3}), we obtain the Shercliff boundary condition for the induced magnetic field involving a thin conducting wall as~\cite{muller2013magnetofluiddynamics}
\begin{equation}
\frac{\partial B}{\partial \hat{n}}+\frac{B}{c_w}=0\quad\quad \mbox{on}\;\; \Gamma \label{eq:ShercliffBCunitnormal}
\end{equation}
given in terms of the unit normal $\hat{n}$ gradient. This can also be rewritten in terms of the local spatial coordinate, via $\xi=-\hat{n}H$, as
\begin{equation}
-\frac{\partial B}{\partial \xi}+\frac{B}{c_w H}=0\quad\quad \mbox{on}\;\; \Gamma \label{eq:ShercliffBClocalcoordinate}
\end{equation}
It may be noted that this general conducting wall boundary condition degenerates to the following special cases:
\begin{eqnarray}
&&c_w=0:\quad B|_\Gamma=0\quad\quad\quad\quad \mbox{(Perfectly insulating wall)}.\nonumber\\
&&c_w\rightarrow\infty: \left(\frac{\partial B}{\partial \xi}\right)_\Gamma = 0\quad\quad \mbox{(Perfectly conducting wall)}.\nonumber
\end{eqnarray}
The boundary condition given in Eq.~(\ref{eq:ShercliffBCunitnormal}) or (\ref{eq:ShercliffBClocalcoordinate}) is of mixed or Robin type relating the induced magnetic field with its wall-normal gradient at the boundary via the wall conductance ratio. Here, we emphasize that the Shercliff boundary condition avoids the need for computing conjugate MHD solutions requiring the calculation of the magnetic field within the solid walls $B_s$ explicitly. While this can be achieved in principle by computing the magnetic fields in both the fluid and solid media with the necessary matching of the conditions at the interface, it often leads to numerically stiff situations when there is large disparities in the conductivities leading to computational challenges and inefficiencies with time consuming calculations. Indeed, the thin wall boundary condition is widely used in studying wall-bounded MHD of liquid metals in various applications (see e.g.,~\cite{chang1961duct,hunt1965magnetohydrodynamic,yu1970convective,singh1984finite,sterl1990numerical,mistrangelo2006three,vantieghem2009velocity,mistrangelo2017magnetohydrodynamic,arlt2017effect,mistrangelo2021mhd}).
Hence, in this work, we develop new LB boundary schemes that implements this boundary condition that effectively accounts for the finite electrical conductivity of the wall. Also, it would be interesting to develop LB formulations for solving conjugate MHD problems which may be useful in certain cases and will form a subject for a future investigation.

\section{Lattice Kinetic Scheme (LKS) for Magnetic Induction Equation}\label{sec:LKSmagneticfield}
\begin{figure}[ht]
\centering
\includegraphics[width=0.7\linewidth]{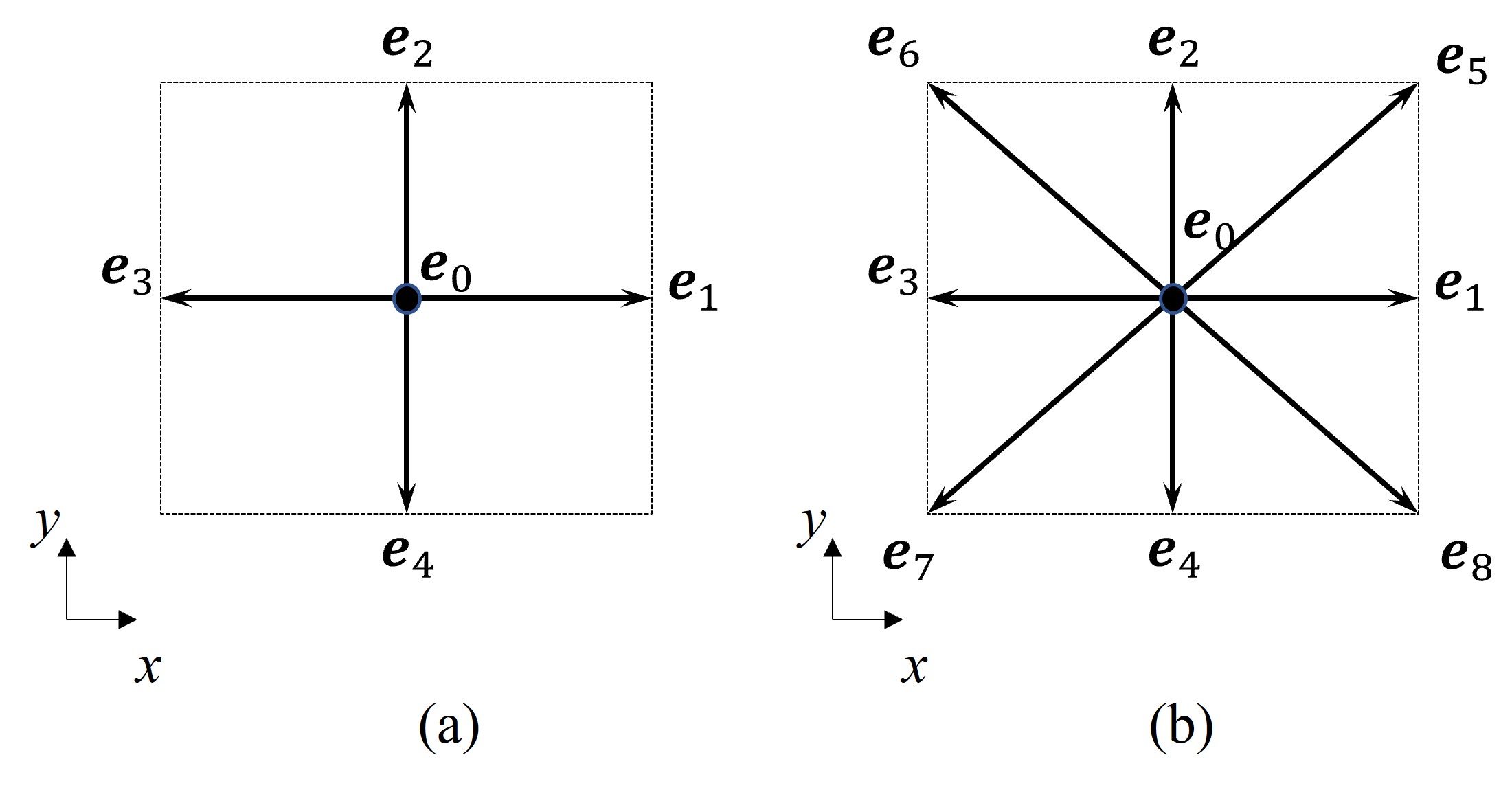}
\caption{(a) D2Q5 lattice for solution of the magnetic induction equation using a vector distribution function and (b) D2Q9 lattice for the solution of the Navier-Stokes equations using a scalar distribution function.}
\label{fig:fig_D2Q9_D2Q5_lattice} 
\end{figure}

Before constructing the boundary schemes, we will first briefly discuss the solution procedure for the MHD equations given in Eq.~(\ref{eq:NSE-MHD}) using the LB methods. In this regard, we will now look at the LKS introduced by Dellar~\cite{dellar2002lattice} for solving the magnetic induction equation (Eq.~(\ref{eq:NSE-MHD3})) to obtain the magnetic field. For maintaining simplicity of our presentation without losing generality, we will only consider the LB formulations in two-dimensions (2D) and involving the simplest collision model in the following. Recognizing the inability of scalar distribution functions to represent the antisymmetric tensor $\bm{u}\bm{B}-\bm{B}\bm{u}$ appearing as a flux term in the left side of the magnetic induction equation given in Eq.~(\ref{eq:NSE-MHD3}), Dellar~\cite{dellar2002lattice} proposed the use of a vector distribution function, whose LB evolution will now be discussed. The D2Q5 lattice used for this purpose is shown in Fig.~\ref{fig:fig_D2Q9_D2Q5_lattice}(a).

The 2D LKS in the index notation reads as
\begin{equation}\label{eq:LKSformagneticfield}
g_{\alpha i}(\bm{x}+\bm{e}_\alpha\delta_t,t+\delta_t)-g_{\alpha i}(\bm{x},t)=-\frac{1}{\tau_m}(g_{\alpha i}-g_{\alpha i}^{eq}),
\end{equation}
where $i\in \{x,y\}$ represent the Cartesian spatial directions and $\alpha =\{0,1,2,3,4\}$ denote the discrete particle directions. The right side of Eq.~(\ref{eq:LKSformagneticfield}) represents the relaxation of the $i$-th component of the vector distribution function to its equilibrium $g_{\alpha i}^{eq}$ with a relaxation time $\tau_m$ during a time step $\delta_t$. The left side of Eq.~(\ref{eq:LKSformagneticfield}) shows the perfect-shift streaming of $g_{\alpha i}$ along the characteristic directions. Here, the equilibrium distribution function $g_{\alpha i}^{eq}$ can be written as
\begin{equation}\label{eq:geqmagneticfield}
g_{\alpha i}^{eq}=W_\alpha\left\{B_i+\frac{e_{\alpha j}}{c_{sm}^2}(u_jB_i-B_ju_i)\right\},
\end{equation}
which is related to the magnetic field vector component $B_i$ and the rank-2 antisymmetric tensor $u_jB_i-B_ju_i$, and the associated weights $W_\alpha$ are given by
\begin{equation}
W_\alpha=
 \begin{cases}
     1/3   & \quad \alpha = 0\\
     1/6   & \quad \alpha = 1,2,3,4
  \end{cases}
\end{equation}
Here, typically $c_{sm}^2=1/3$. Then, defining the first velocity moment of the vector distribution function $g_{\alpha i}$ as the following rank-2 tensor $\Lambda_{ji}$ in the component form as
\begin{equation}
\Lambda_{ji}=\sum_{\alpha=0}^4 e_{\alpha j}g_{\alpha i},
\end{equation}
whose equilibrium parts are represented by
\begin{equation}\label{eq:rank2antisymmtensoreq}
\Lambda_{ji}^{(0)}=\sum_{\alpha=0}^4 e_{\alpha j}g_{\alpha i}^{eq}=u_jB_i-B_ju_i,
\end{equation}
and the (leading) non-equilibrium parts based on the Chapman-Enskog analysis~\cite{chapman1990mathematical} (discussed in detail later) are given by
\begin{equation}\label{eq:rank2antisymmtensornoneq}
\Lambda_{ji}^{(1)}=\sum_{\alpha=0}^4 e_{\alpha j}g_{\alpha i}^{(1)}=\sum_{\alpha=0}^4 e_{\alpha j}(g_{\alpha i}-g_{\alpha i}^{eq})=-\tau_mc_{sm}^2\partial_jB_i.
\end{equation}
The solution of the LKS (Eq.~(\ref{eq:LKSformagneticfield})), by rewriting it in the form of collide-and-stream steps, yields the distribution function $g_{\alpha i}$, whose zeroth and first velocity moments yield the magnetic field $B_i$ and the current density $J_i$, respectively, which can be written as
\begin{equation}\label{eq:zerothmomentofg}
B_i=\sum_{\alpha =0}^4 g_{\alpha i},
\end{equation}
and
\begin{equation}
J_i=\frac{1}{\mu_m}\epsilon_{ijk}\partial_jB_k=-\frac{1}{\mu_m}\frac{1}{\tau_mc_{sm}^2}\epsilon_{ijk}\Lambda_{jk}^{(1)}=-\frac{1}{\mu_m}\frac{1}{\tau_mc_{sm}^2}\epsilon_{ijk}(\Lambda_{jk}-\Lambda_{jk}^{(0)}),
\end{equation}
where $\epsilon_{ijk}$ is the rank-3 permutation (Levi-Civita) tensor. The above lattice kinetic algorithm simulates the magnetic induction equation (Eq.~(\ref{eq:NSE-MHD3})), where the magnetic resistivity $\eta$ is related to the relaxation time $\tau_m$ via
\begin{equation}
\eta = \frac{1}{\sigma\mu_m}=c_{sm}^2\left(\tau_m-\frac{1}{2}\right)\delta_t.
\end{equation}

\section{Lattice Boltzmann Method (LBM) for Fluid Flow}\label{sec:LBMvelocityfield}
We will now present a LB scheme for solving the flow field part of the MHD equations (Eq.~(\ref{eq:NSE-MHD1}) and (\ref{eq:NSE-MHD2})). Again, for maintaining brevity, we now present the simplest LBM for simulating the fluid motion based on the single relaxation time (SRT) model~\cite{qian1992lattice}, which can readily replaced with any of the more general or sophisticated collision model when required. The 2D SRT-LBM for computing the hydrodynamics is based on the scalar distribution function $f_{\alpha}$ that evolves on the D2Q9 lattice (see Fig.~\ref{fig:fig_D2Q9_D2Q5_lattice}(b)) having a greater degree of isotropy than that used in the previous section to recover the viscous stress tensor. It can be written as
\begin{equation}\label{eq:LBMforflowfield}
f_{\alpha}(\bm{x}+\bm{e}_\alpha\delta_t,t+\delta_t)-f_{\alpha}(\bm{x},t)=-\frac{1}{\tau}(f_{\alpha}-f_{\alpha}^{eq})+\left(1-\frac{1}{2\tau}\right)S_\alpha\delta_t, \end{equation}
where the equilibrium distribution function $f_\alpha^{eq}$ and the source term $S_\alpha$, which incorporates the various body forces including the Lorentz force, can be written, respectively, as
\begin{equation}
f_{\alpha}^{eq}=w_\alpha\rho\left\{1+\frac{\bm{e}_\alpha\cdot\bm{u}}{c_s^2}+\frac{(\bm{e}_\alpha\cdot\bm{u})^2}{2c_s^4}-\frac{\bm{u}\cdot\bm{u}}{2c_s^2}\right\},
\end{equation}
and
\begin{equation}
S_{\alpha}=\frac{(\bm{e}_\alpha-\bm{u})}{\rho c_s^2}\cdot(\bm{F}_{Lorentz}+\bm{F}_{ext})f_{\alpha}^{eq}.
\end{equation}
Here, the weights $w_\alpha$ are given by
\begin{equation}
w_\alpha=
 \begin{cases}
     4/9   & \quad \alpha = 0\\
     1/9   & \quad \alpha = 1,2,3,4\\
     1/36  & \quad \alpha = 5,6,7,8
  \end{cases}
\end{equation}
In the above, typically, $c_s^2=1/3$. The solution of the LBM (Eq.~(\ref{eq:LBMforflowfield})), by rewriting it in the form of collide-and-stream steps (see e.g.,~\cite{kruger2017lattice}), results in the distribution function $f_\alpha$, whose various velocity moments give the fluid density and the velocity fields, which can be represented as
\begin{eqnarray}
\rho&=&\sum_{\alpha=0}^8 f_\alpha,\\
\rho\bm{u}&=&\sum_{\alpha=0}^8 f_\alpha\bm{e}_\alpha+\frac{1}{2}(\bm{F}_{Lorentz}+\bm{F}_{ext})\delta_t,
\end{eqnarray}
and the pressure is given by $p=\rho c_s^2$.
These hydrodynamic fields satisfy the NS equations (Eq.~(\ref{eq:NSE-MHD1}) and (\ref{eq:NSE-MHD2})), where the kinematic viscosity of the fluid $\nu$ is related to the relaxation time $\tau$ as
\begin{equation}
\nu=c_s^2\left(\tau-\frac{1}{2}\right)\delta_t.
\end{equation}

\section{Link-based Boundary Scheme for Implementation of Shercliff Boundary Condition for the Magnetic Field}\label{sec:linkbasedShercliffBC}
\begin{figure}[ht]
\centering
\includegraphics[width=0.7\linewidth]{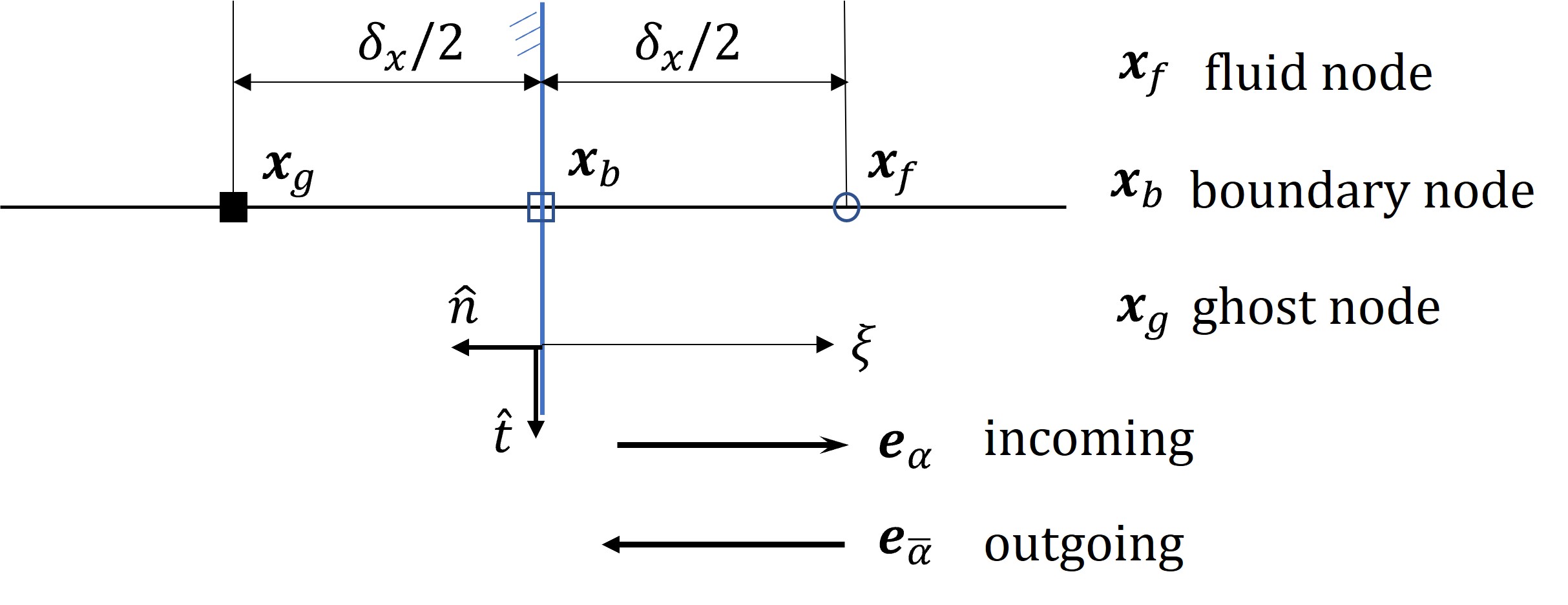}
\caption{Schematic of the arrangement of nodes and lattice links around conducting wall boundary for the link-based boundary condition scheme.}
\label{fig:fig_schematic_link_BC_representation} 
\end{figure}

We will now present a link-based boundary scheme to determine the unknown incoming distribution function from conducting walls for the LKS given in Sec.~\ref{sec:LKSmagneticfield} so that the boundary condition for the induced magnetic field as presented in Eq.~(\ref{eq:ShercliffBClocalcoordinate}) is enforced. A schematic of the arrangement of the nodes and lattice links around the boundary is shown in Fig.~\ref{fig:fig_schematic_link_BC_representation}.

Here, $\bm{x}_b$, $\bm{x}_f$, and $\bm{x}_g$ refer to the boundary node, nearest fluid node, and ghost node, respectively, and $\bm{e}_\alpha$ and $\bm{e}_{\bar{\alpha}}$ are the incoming and outgoing particle velocity directions,
where $\bm{e}_{\bar{\alpha}}=-\bm{e}_\alpha$. The outward unit normal and tangent to the wall at the boundary node are represented by $\hat{n}$ and $\hat{t}$, respectively, and the local wall normal coordinate pointing into the fluid is denoted by $\xi$. We assume that the wall is stationary in this section, and that the boundary node $\bm{x}_b$, where Eq.~(\ref{eq:ShercliffBClocalcoordinate}) is to be satisfied, is located half-way between the nearest fluid and ghost nodes, and our goal is to construct the unknown incoming distribution function $g_\alpha(\bm{x}_f,t+\delta_t)$ at the nearest fluid node. Let $B_i$ be the induced magnetic field component for which the Shercliff BC is applicable in the presence of an imposed external magnetic field.

First, using $\partial B_i/\partial \xi = \bm{e}_\alpha\cdot\bm{\nabla} B_i$, it follows that Eq.~(\ref{eq:ShercliffBClocalcoordinate}) can be rewritten in the following form that is more convenient for analysis:
\begin{equation}
-\bm{e}_\alpha\cdot\bm{\nabla} B_i+\frac{B_i}{c_wH} = 0\quad\quad \mbox{on}\;\; \bm{x}_b. \label{eq:ShercliffBClocalcoordinateLBM}
\end{equation}
Then, using the Chapman-Enskog (C-E) multiscale expansions~\cite{chapman1990mathematical}
\begin{equation}
g_{\alpha i}=g_{\alpha i}^{(0)}+\epsilon g_{\alpha i}^{(1)}+\cdots,\quad\quad \partial_t = \partial_{t_0}+\epsilon\partial_{t_1}+\cdots,\label{eq:CEexpansion}
\end{equation}
where $\epsilon=\delta_t$ is a small perturbation parameter, and applying the Taylor series expansion to the left side of Eq.~(\ref{eq:LKSformagneticfield}), we get the $O(\epsilon)$ evolution equation as
\begin{equation}
\partial_{t_0}g_{\alpha i}^{(0)}+e_{\alpha j}\partial_jg_{\alpha i}^{(0)}=-\frac{1}{\tau_m}g_{\alpha i}^{(1)},
\end{equation}
from which we obtain the non-equilibrium part of the distribution function $g_{\alpha i}^{(1)}$, which can be written as
\begin{equation}
g_{\alpha i}^{(1)}=-\tau_m\partial_{t_0}g_{\alpha i}^{(0)}-\tau_m e_{\alpha j}\partial_jg_{\alpha i}^{(0)}.\label{eq:noneqmpartofg}
\end{equation}
Substituting Eq.~(\ref{eq:noneqmpartofg}) in Eq.~(\ref{eq:CEexpansion}), we obtain an estimate of the distribution function $g_{\alpha i}$ in terms of its equilibrium part $g_{\alpha i}^{(0)}$ and its space and time derivatives as
\begin{equation}
g_{\alpha i}=g_{\alpha i}^{(0)}-\tau_m\delta_t\partial_{t_0}g_{\alpha i}^{(0)}-\tau_m\delta_te_{\alpha k}\partial_kg_{\alpha i}^{(0)}.\label{eq:gestimateintermsofgeq}
\end{equation}
Also, using $\bm{e}_{\bar{\alpha}}=-\bm{e}_{\alpha}$, the distribution function for the opposite direction to that given in Eq.~(\ref{eq:gestimateintermsofgeq}) follows as
\begin{equation}
g_{\bar{\alpha} i}=g_{\bar{\alpha} i}^{(0)}-\tau_m\delta_t\partial_{t_0}g_{\bar{\alpha} i}^{(0)}+\tau_m\delta_te_{\alpha k}\partial_kg_{\bar{\alpha} i}^{(0)}.\label{eq:goppestimateintermsofgeq}
\end{equation}
Now, the equilibrium distribution function given in Eq.~(\ref{eq:geqmagneticfield}) in view of Eq.~(\ref{eq:rank2antisymmtensoreq}) can be rewritten as a function of the magnetic field $B_i$ and the anti-symmetric tensor $\Lambda_{ji}^{(0)}$ as
\begin{equation}
g_{\alpha i}^{(0)}=W_\alpha\left\{B_i+\frac{e_{\alpha j}}{c_{sm}^2}\Lambda_{ji}^{(0)}\right\},\label{eq:geqmagneticfieldrank2antisymmtensor}
\end{equation}
and using $W_{\bar{\alpha}}=W_{\alpha}$ and $\bm{e}_{\bar{\alpha}}=-\bm{e}_\alpha$ again, the corresponding expression in the opposite direction reads as
\begin{equation}
g_{\bar{\alpha} i}^{(0)}=W_\alpha\left\{B_i-\frac{e_{\alpha j}}{c_{sm}^2}\Lambda_{ji}^{(0)}\right\}.\label{eq:geqoppmagneticfieldrank2antisymmtensor}
\end{equation}
From the last two equations, we can then write the sum and difference of the equilibria in the two opposite directions as
\begin{subequations}\label{eq:sumdiffgeq}
\begin{eqnarray}
 g_{\alpha i}^{(0)} + g_{\bar{\alpha} i}^{(0)} &=& 2W_{\alpha}B_i, \\
 g_{\alpha i}^{(0)} - g_{\bar{\alpha} i}^{(0)} &=& 2W_{\alpha}\frac{e_{\alpha j}}{c_{sm}^2}\Lambda_{ji}^{(0)}.
\end{eqnarray}
\end{subequations}
On the other hand, combining Eqs.~(\ref{eq:gestimateintermsofgeq}) and (\ref{eq:goppestimateintermsofgeq}), the sum and difference of the distribution function in the opposite directions follow as
\begin{subequations}\label{eq:sumdiffg}
\begin{eqnarray}
 g_{\alpha i} + g_{\bar{\alpha} i} &=& (g_{\alpha i}^{(0)} + g_{\bar{\alpha} i}^{(0)})-\tau_m\delta_t\left[\partial_{t_0}(g_{\alpha i}^{(0)} + g_{\bar{\alpha} i}^{(0)})+e_{\alpha_k}\partial_k(g_{\alpha i}^{(0)} - g_{\bar{\alpha} i}^{(0)})\right], \\
 g_{\alpha i} - g_{\bar{\alpha} i} &=& (g_{\alpha i}^{(0)} - g_{\bar{\alpha} i}^{(0)})-\tau_m\delta_t\left[\partial_{t_0}(g_{\alpha i}^{(0)} - g_{\bar{\alpha} i}^{(0)})+e_{\alpha_k}\partial_k(g_{\alpha i}^{(0)} + g_{\bar{\alpha} i}^{(0)})\right].
\end{eqnarray}
\end{subequations}
Substituting for $(g_{\alpha i}^{(0)} + g_{\bar{\alpha} i}^{(0)})$ and $(g_{\alpha i}^{(0)} - g_{\bar{\alpha} i}^{(0)})$ from Eq.~(\ref{eq:sumdiffgeq}) in Eq.~(\ref{eq:sumdiffg}), we finally obtain $g_{\alpha i} \pm g_{\bar{\alpha} i}$ in terms of the macroscopic field variables in their derivatives as
\begin{subequations}\label{sumdiffgmacrovars}
\begin{eqnarray}
 g_{\alpha i} + g_{\bar{\alpha} i} &=& 2W_{\alpha}B_i-2W_{\alpha}\tau_m\delta_t\left[\partial_{t_0}B_i+\frac{e_{\alpha_j}e_{\alpha_k}}{c_{sm}^2}\partial_k\Lambda_{ji}^{(0)}\right], \\
 g_{\alpha i} - g_{\bar{\alpha} i} &=& 2W_\alpha\frac{e_{\alpha_j}}{c_{sm}^2}\Lambda_{ji}^{(0)}-2W_\alpha\tau_m\delta_t\left[\frac{e_{\alpha_j}}{c_{sm}^2}\partial_{t_0}\Lambda_{ji}^{(0)}+e_{\alpha_k}\partial_kB_i\right].
\end{eqnarray}
\end{subequations}
Next, we evaluate Eq.~(\ref{sumdiffgmacrovars}) on the boundary node $\bm{x}_b$, where due to the no-slip BC $u_i(\bm{x}_b)=0$ and from the scaling $O(B_i)\sim O(u_i)$, we neglect terms of $O(u_i^2)$ or smaller. Thus, $\partial_{t_0}B_i(\bm{x}_b)\sim \partial_k\Lambda_{ji}^{(0)}\sim O(u_i^2)$ and $\Lambda_{ji}^{(0)}(\bm{x}_b)=0$ due to the no-slip BC. Hence, the expressions in Eq.~(\ref{sumdiffgmacrovars}) for $g_{\alpha i} \pm g_{\bar{\alpha} i} $ at $\bm{x}_b$ reduce to
\begin{subequations}\label{sumdiffgmacrovarsbnode}
\begin{eqnarray}
 g_{\alpha i}(\bm{x}_b) + g_{\bar{\alpha} i}(\bm{x}_b) &\approx& 2W_{\alpha}B_i(\bm{x}_b), \\
 g_{\alpha i}(\bm{x}_b) - g_{\bar{\alpha} i}(\bm{x}_b) &\approx&-2W_\alpha\tau_m\delta_te_{\alpha_k}\partial_kB_i(\bm{x}_b),
\end{eqnarray}
\end{subequations}
which expresses the fact the sum is related to the magnitude of the magnetic field, while the difference is related to its gradient, with both of them being evaluated at the boundary node.

Now, the key idea in constructing a link-based boundary condition implementation for the LKS is to express the boundary node estimates of the combinations of the distribution functions $g_{\alpha i}(\bm{x}_b) \pm g_{\bar{\alpha} i}(\bm{x}_b)$ given above in terms of the distribution functions at the nearest fluid node $\bm{x}_f$, i.e., $g_{\alpha i}(\bm{x}_f,t+\delta_t)$, which is unknown and the required incoming quantity from the wall, and $\tilde{g}_{\bar{\alpha} i}(\bm{x}_f,t)$, which is the outgoing and a known quantity. Here, $\tilde{g}_{\bar{\alpha} i}$ refers to the post-collision value of $g_{\bar{\alpha} i}$, which follows from Eq.~(\ref{eq:LKSformagneticfield}) as
\begin{equation}
\tilde{g}_{\bar{\alpha} i}(\bm{x}_f,t)=g_{\bar{\alpha} i}(\bm{x}_f,t)-\frac{1}{\tau_m}[g_{\bar{\alpha} i}(\bm{x}_f,t)-g_{\bar{\alpha} i}^{(0)}(\bm{x}_f,t)].
\end{equation}
In this regard, we apply the Taylor series in space to the above two quantities $g_{\alpha i}(\bm{x}_f,t+\delta_t)$ and $\tilde{g}_{\bar{\alpha} i}(\bm{x}_f,t)$ by assuming the boundary node $\bm{x}_b$ is located half-way from the nearest fluid node $\bm{x}_f$, which yield
\begin{subequations}\label{eq:Taylorseriesnearfluidnode}
\begin{eqnarray}
 g_{\alpha i}(\bm{x}_f,t+\delta_t) &\approx& g_{\alpha i}(\bm{x}_b,t+\delta_t)+ \frac{\delta_t}{2}e_{\alpha j}\partial_j g_{\alpha i}(\bm{x}_b,t+\delta_t), \\
 \tilde{g}_{\bar{\alpha} i}(\bm{x}_f,t) &\approx& \tilde{g}_{\bar{\alpha} i}(\bm{x}_b,t) - \frac{\delta_t}{2}e_{\alpha j}\partial_j  \tilde{g}_{\bar{\alpha} i}(\bm{x}_b,t),
\end{eqnarray}
\end{subequations}
where the last equation exploited the fact that $\bm{e}_{\bar{\alpha} j}=-\bm{e}_{\alpha j}$. Now, on $\bm{x}_b$, we take $g_{\alpha i}(\bm{x}_b,t+\delta_t)=g_{\alpha i}(\bm{x}_b,t)=g_{\alpha i}(\bm{x}_b)$ and $\tilde{g}_{\bar{\alpha} i}(\bm{x}_b,t)=g_{\bar{\alpha} i}(\bm{x}_b,t)=g_{\bar{\alpha} i}(\bm{x}_b)$. Moreover, in evaluating the spatial gradients $\partial_j g_{\alpha i}$ and $\partial_j \tilde{g}_{\bar{\alpha} i}$ on $\bm{x}_b$ in the above last two equations, we consider the leading order equilibrium contributions for the distribution functions, i.e., $\partial_j g_{\alpha i}(\bm{x}_b)\approx \partial_j g_{\alpha i}^{(0)}(\bm{x}_b)\approx W_\alpha \partial_j B_i (\bm{x}_b)$ and $\partial_j \tilde{g}_{\bar{\alpha} i}(\bm{x}_b)\approx \partial_j g_{\bar{\alpha} i}^{(0)}(\bm{x}_b)\approx W_\alpha \partial_j B_i(\bm{x}_b)$. Thus, we can simplify the Eqs.~(\ref{eq:Taylorseriesnearfluidnode}) to
\begin{subequations}\label{eq:Taylorseriesnearfluidnodesimplified}
\begin{eqnarray}
 g_{\alpha i}(\bm{x}_f,t+\delta_t) &\approx& g_{\alpha i}(\bm{x}_b)+ W_\alpha\frac{\delta_t}{2}e_{\alpha j}\partial_j B_i(\bm{x}_b), \\
 \tilde{g}_{\bar{\alpha} i}(\bm{x}_f,t) &\approx& \tilde{g}_{\bar{\alpha} i}(\bm{x}_b) - W_\alpha\frac{\delta_t}{2}e_{\alpha j}\partial_j  B_i(\bm{x}_b).
\end{eqnarray}
\end{subequations}
These last two equations in Eq.~(\ref{eq:Taylorseriesnearfluidnodesimplified}) can be rearranged further as
\begin{subequations}\label{eq:Taylorseriesnearfluidnodesimplifiedrearrg}
\begin{eqnarray}
 g_{\alpha i}(\bm{x}_b) &\approx& g_{\alpha i}(\bm{x}_f,t+\delta_t)- W_\alpha\frac{\delta_t}{2}e_{\alpha j}\partial_j B_i(\bm{x}_b), \\
 g_{\bar{\alpha} i}(\bm{x}_b) &\approx& \tilde{g}_{\bar{\alpha} i}(\bm{x}_f,t) + W_\alpha\frac{\delta_t}{2}e_{\alpha j}\partial_j  B_i(\bm{x}_b).
\end{eqnarray}
\end{subequations}
Now, combining these two equations in Eq.~(\ref{eq:Taylorseriesnearfluidnodesimplifiedrearrg}) as their sum and difference, we can then express them in the following equivalent forms:
\begin{subequations}\label{eq:sumdiffTaylorseriessimprearrg}
\begin{eqnarray}
 g_{\alpha i}(\bm{x}_b) + g_{\bar{\alpha} i}(\bm{x}_b) &\approx& g_{\alpha i}(\bm{x}_f,t+\delta_t)+\tilde{g}_{\bar{\alpha} i}(\bm{x}_f,t), \\
 g_{\alpha i}(\bm{x}_b) - g_{\bar{\alpha} i}(\bm{x}_b) &\approx& g_{\alpha i}(\bm{x}_f,t+\delta_t)-\tilde{g}_{\bar{\alpha} i}(\bm{x}_f,t)- W_\alpha\delta_te_{\alpha j}\partial_j  B_i(\bm{x}_b).
\end{eqnarray}
\end{subequations}
Then, using Eq.~(\ref{eq:sumdiffTaylorseriessimprearrg}) in Eq.~(\ref{sumdiffgmacrovarsbnode}), we eliminate the intermediate quantities $g_{\alpha i}(\bm{x}_b)\pm g_{\bar{\alpha} i}(\bm{x}_b)$ and relate the unknown incoming distribution function $g_{\alpha i}(\bm{x}_f,t+\delta_t)$ to the known outgoing distribution function $\tilde{g}_{\bar{\alpha} i}(\bm{x}_f,t)$ and the boundary node values of the magnetic field $B_i(\bm{x}_b)$ and its gradient $\partial_jB_i(\bm{x}_b)$. As results of these steps, we get
\begin{subequations}\label{eq:magneticfieldanditsgradientrelation}
\begin{eqnarray}
  \frac{1}{2}\left[g_{\alpha i}(\bm{x}_f,t+\delta_t)+\tilde{g}_{\bar{\alpha} i}(\bm{x}_f,t)\right] &=& W_\alpha B_i(\bm{x}_b), \label{eq:magneticfieldanditsgradientrelation1}\\
  -\frac{1}{2\left(\tau_m-\frac{1}{2}\right)\delta_t}\left[g_{\alpha i}(\bm{x}_f,t+\delta_t)-\tilde{g}_{\bar{\alpha} i}(\bm{x}_f,t)\right] &=& W_\alpha\delta_te_{\alpha j}\partial_j  B_i(\bm{x}_b).\label{eq:magneticfieldanditsgradientrelation2}
\end{eqnarray}
\end{subequations}
The relations in Eq.~(\ref{eq:magneticfieldanditsgradientrelation}) together need to satisfy the constraint given by the Shercliff boundary condition for thin conducting walls, which can be rewritten from Eq.~(\ref{eq:ShercliffBClocalcoordinateLBM}) by multiplying it with the weighting factor $W_\alpha$ as
\begin{equation}
W_\alpha B_i(\bm{x}_b)-c_w H W_\alpha e_{\alpha j}\partial_j B_i(\bm{x}_b) = 0.\label{eq:ShercliffBClocalcoordinateLBMmodified}
\end{equation}
Substituting the expressions in Eq.~(\ref{eq:magneticfieldanditsgradientrelation}) into Eq.~(\ref{eq:ShercliffBClocalcoordinateLBMmodified}), we get
\begin{equation}\label{eq:linkbasedBCscheme}
\left[1+\frac{c_wH}{\left(\tau_m-\frac{1}{2}\right)\delta_t}\right]g_{\alpha i}(\bm{x}_f,t+\delta_t)+\left[1-\frac{c_wH}{\left(\tau_m-\frac{1}{2}\right)\delta_t}\right]\tilde{g}_{\bar{\alpha} i}(\bm{x}_f,t)=0.
\end{equation}
Based on Eq.~(\ref{eq:linkbasedBCscheme}), we then introduce a weighting factor $\theta$ related to the wall conductance ratio $c_w$ and the relaxation time $\tau_m$ as
\begin{equation}\label{eq:weightingfactor}
\theta = \frac{c_w H}{\left(\tau_m-\frac{1}{2}\right)\delta_t}.
\end{equation}
Then, Eq.~(\ref{eq:linkbasedBCscheme}) can be rewritten as
\begin{equation}\label{eq:linkbasedBCschemefinal}
g_{\alpha i}(\bm{x}_f,t+\delta_t) = \frac{1}{1+\theta}[-\tilde{g}_{\bar{\alpha} i}(\bm{x}_f,t)]+\frac{\theta}{1+\theta}[+\tilde{g}_{\bar{\alpha} i}(\bm{x}_f,t)], \quad\quad 0 \le \theta \le \infty.
\end{equation}
Equation~(\ref{eq:linkbasedBCschemefinal}) represents the new link-based BC closure which reconstructs the incoming distribution function for the induced magnetic field component from thin conducting walls via a weighted combination of the anti-bounce back (first term) and bounce back (second term) of the outgoing post-collision distribution function. It follows that the above boundary scheme reduces to the following limiting cases:

(a) Insulated wall: $c_w=\theta = 0$, which corresponds to
\begin{equation*}
   g_{\alpha i}(\bm{x}_f,t+\delta_t)=-\tilde{g}_{\bar{\alpha} i}(\bm{x}_f,t)\quad \mbox{(Anti-bounce back scheme)},
\end{equation*}

(b) Perfectly conducting wall: $c_w=\theta \rightarrow \infty$, which corresponds to
\begin{equation*}
   g_{\alpha i}(\bm{x}_f,t+\delta_t)=+\tilde{g}_{\bar{\alpha} i}(\bm{x}_f,t)\quad \mbox{(Bounce back scheme)},
\end{equation*}
On the other hand, the implementation of the link-based no-slip velocity boundary condition via the standard half-way bounce back scheme~\cite{ladd1994numerical} implies that $f_{\alpha}(\bm{x}_f,t+\delta_t)=\tilde{f}_{\bar{\alpha}}(\bm{x}_f,t)$. A derivation of the extension of the above boundary formulation given for a stationary conducting wall to a moving conducting wall is presented in~\ref{sec:bcsmovingwallslinkbased}.

\subsection{Example implementation: Hartmann flow through a channel bounded by conducting walls}\label{subsec:Hartmannflowconductingwalls}
\begin{figure}[ht]
\centering
\includegraphics[width=0.5\linewidth]{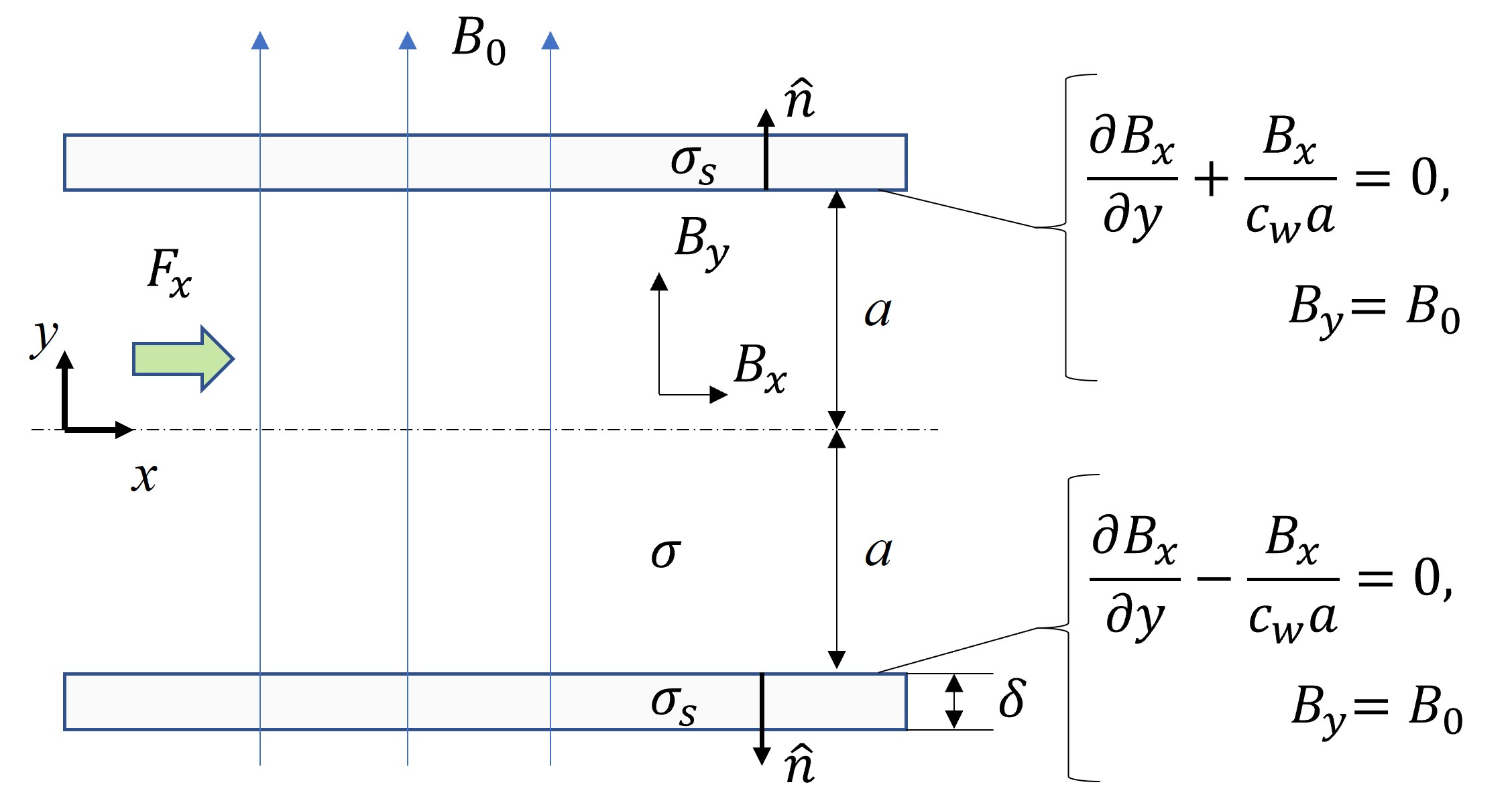}
\caption{Sketch of the Hartmann flow bounded by two parallel plates with finite electrical conductivities.}
\label{fig:fig_schematic_Hartmann_flow_conducting_BC} 
\end{figure}

We will now illustrate the implementation of the link-based boundary scheme for the body force-driven flow between two conducting plates in the presence of an external magnetic field. A schematic of this flow set up is shown in Fig.~\ref{fig:fig_schematic_Hartmann_flow_conducting_BC}, where an electrically conducting fluid with conductivity $\sigma$ bounded by plates, each of which has a thickness $\delta$ and electrical conductivity $\sigma_s$ and a separated by a distance $2a$. The fluid is in motion due to a body force $F_x$ applied in the $x$ direction in the presence of an imposed magnetic field $B_0$ normal to the bounding walls, i.e., in the $y$ direction.

The analytical solution for this generalized Hartmann flow problem with finite wall conductivity effects was given by Chang and Lundgren~\cite{chang1961duct}, which will be summarized later in Sec.~\ref{sec:resultsanddiscussion} when discussing the numerical results. For this case, as shown in Fig.~\ref{fig:fig_schematic_Hartmann_flow_conducting_BC}, at each boundary, the Shercliff thin wall boundary condition is applicable for the induced magnetic in the $x$ direction, i.e., $B_x$, while the magnetic field in the $y$ direction $B_y$ is based on the imposed magnetic field $B_0$. Here, the wall conductance ratio $c_w$ and the weighting factor $\theta$ can be written as
\begin{equation*}
c_w = \frac{\sigma_s\delta}{\sigma a},\quad\quad \theta = \frac{c_w a}{(\tau_m-\frac{1}{2})\delta_t}.
\end{equation*}
Referring to Fig.~\ref{fig:fig_D2Q9_D2Q5_lattice}(a) for the lattice directions, the link-based boundary closure relations for the incoming distribution functions corresponding to the components of the magnetic field $B_x$ and $B_y$ for the bottom and top walls follow from the above considerations read as
\begin{eqnarray*}
  y=-a: \quad g_{2x} &=& \frac{\theta - 1}{\theta + 1} \tilde{g}_{4x}, \\
              g_{2y} &=& -\tilde{g}_{4y} + 2W_2B_0,
\end{eqnarray*}
\begin{eqnarray*}
  y=+a: \quad g_{4x} &=& \frac{\theta - 1}{\theta + 1} \tilde{g}_{2x}, \\
              g_{4y} &=& -\tilde{g}_{4y} + 2W_4B_0.
\end{eqnarray*}
That is, the components of $g_{2x}$ and $g_{4x}$ for the bottom and top walls, respectively, follow from Eq.~(\ref{eq:linkbasedBCschemefinal}), while that for the wall normal components $g_{2y}$ and $g_{4y}$ that satisfy the Dirichlet boundary condition based on $B_0$ directly result from Eq.~(\ref{eq:magneticfieldanditsgradientrelation1}).

\section{On-Node Moment-Based Boundary Scheme for Implementation of Shercliff Boundary Condition for the Magnetic Field}\label{sec:momentbasedShercliffBC}
We will now present an alternative moment-based approach, which implements the boundary conditions directly by means of constraints on the moments of the distribution functions on the nodes rather than on some cut-link locations. Originally proposed by Noble~\cite{noble1995consistent} for the D2Q7 lattice and then applied for the D2Q9 lattice by Bennett~\cite{bennett2010lattice} for specifying the no-slip velocity boundary conditions, it was subsequently extended for the distribution functions for the magnetic fields for representing the specific case of the insulated wall MHD by Dellar~\cite{dellar2013moment}. Here, we will consider a more general situation involving MHD with finite wall conductivities, which require the specification of the mixed or Robin-type boundary condition.

Consider again the Hartmann flow case bounded by thin conducting walls discussed at the end of the previous section (see Fig.~\ref{fig:fig_schematic_Hartmann_flow_conducting_BC}). Our goal is to impose the following boundary exactly via the necessary moment constraints:
\begin{subequations}\label{eq:ShercliffBCforHartmannFlow}
\begin{eqnarray}
  -\frac{\partial B_x}{\partial y}+\frac{B_x}{c_w a} &=& 0, \quad B_y = B_0 \quad\quad \mbox{at}\;\;y = -a, \label{eq:ShercliffBCforHartmannFlowbottom}\\
  +\frac{\partial B_x}{\partial y}+\frac{B_x}{c_w a} &=& 0, \quad B_y = B_0 \quad\quad \mbox{at}\;\;y = +a. \label{eq:ShercliffBCforHartmannFlowtop}
\end{eqnarray}
\end{subequations}
Unlike the previous approach, here $\bm{x}_b$ is located on directly on the lattice node and not on a link, i.e., we consider the boundary to be `wet’. Decomposing the tensor component $\Lambda_{yx}$ by means of its equilibrium and non-equilibrium parts, i.e., $\Lambda_{yx}=\Lambda_{yx}^{(0)}+\delta_t\Lambda_{yx}^{(1)}$, and evaluating this at the wall boundary where $\Lambda_{yx}^{(0)}=(u_yB_x-B_yu_x)(\bm{x}_b)=0$ via the no-slip condition, we get $\Lambda_{yx}=\delta_t\Lambda_{yx}^{(1)}$. From Eq.~(\ref{eq:rank2antisymmtensornoneq}), which follows from a C-E analysis as given in the derivation in Sec.~\ref{sec:linkbasedShercliffBC}, we have $\Lambda_{yx}^{(1)}=-\tau_mc_{sm}^2\partial_yB_x$. Thus, we can express the boundary node evaluation of the tensor component $\Lambda_{yx}$ as
\begin{equation}
\Lambda_{yx}(\bm{x}_b)=-\tau_mc_{sm}^2\delta_t\frac{\partial B_x}{\partial y}(\bm{x}_b),
\end{equation}
which implies that the wall normal gradient of the induced magnetic field at the boundary can be written in terms of the first moment involving the velocity in the $y$ direction of the distribution function in the $x$ direction $g_{\alpha x}$ at $\bm{x}_b$ as follows:
\begin{equation}
\frac{\partial B_x}{\partial y}(\bm{x}_b)=-\frac{1}{\tau_mc_{sm}^2\delta_t}\Lambda_{yx}(\bm{x}_b)=-\frac{1}{\tau_mc_{sm}^2\delta_t}\sum_{\alpha=0}^{4}e_{\alpha y}g_{\alpha x}(\bm{x}_b),\label{eq:maggradientestimate}
\end{equation}
On the other hand, the zeroth moment of $g_{\alpha x}$ at the boundary node gives the magnetic field component $B_x$ as
\begin{equation}
B_x(\bm{x}_b)=\sum_{\alpha=0}^{4}g_{\alpha x}(\bm{x}_b).\label{eq:magestimate}
\end{equation}
Then, using Eqs.~(\ref{eq:maggradientestimate}) and (\ref{eq:magestimate}) in the Shercliff boundary condition for the bottom wall in Eq.~(\ref{eq:ShercliffBCforHartmannFlowbottom}), we get
\begin{equation*}
\frac{1}{\tau_mc_{sm}^2\delta_t}\sum_{\alpha=0}^{4}e_{\alpha y}g_{\alpha x}(\bm{x}_b)+\frac{1}{c_w a}\sum_{\alpha=0}^{4}g_{\alpha x}(\bm{x}_b) = 0,
\end{equation*}
which simplifies to
\begin{equation}
\frac{1}{\tau_mc_{sm}^2\delta_t}(g_{2 x}-g_{4 x})+\frac{1}{c_w a}(g_{0 x}+g_{1 x}+g_{2 x}+g_{3 x}+g_{4 x})=0 \quad\quad\mbox{at}\;\;y=-a.\label{eq:momentconstraintgx}
\end{equation}
On the other hand $B_y=B_0$ at $\bm{x}_b$ (see Eq.~(\ref{eq:ShercliffBCforHartmannFlowbottom})) implies
\begin{equation}
g_{0 y}+g_{1 y}+g_{2 y}+g_{3 y}+g_{4 y} = B_0\quad\quad\mbox{at}\;\;y=-a.\label{eq:momentconstraintgy}
\end{equation}
Then, solving for the unknowns $g_{2 x}$ and $g_{2 y}$ from Eqs.~(\ref{eq:momentconstraintgx}) and (\ref{eq:momentconstraintgy}), respectively, we finally get the moment-based closures for the distribution functions at the bottom wall as
\begin{subequations}
\begin{eqnarray}\label{eq:momentbasedBCslowerboundary}
  g_{2 x} &=& \left[\frac{1}{\tau_mc_{sm}^2\delta_t}+\frac{1}{c_w a}\right]^{-1}\left\{\left[\frac{1}{\tau_mc_{sm}^2\delta_t}-\frac{1}{c_w a}\right]g_{4 x}-\frac{1}{c_w a}(g_{0 x}+g_{1 x}+g_{3 x})\right\}\quad\quad\mbox{at}\;\;y=-a, \\
  g_{2 y} &=& B_0-(g_{0 y}+g_{1 y}+g_{3 y}+g_{4 y}).
\end{eqnarray}
\end{subequations}

Similarly, at $y=+a$, from Eq.~(\ref{eq:ShercliffBCforHartmannFlowtop}) and using Eqs.~(\ref{eq:maggradientestimate}) and (\ref{eq:magestimate}), we get
\begin{equation}
-\frac{1}{\tau_mc_{sm}^2\delta_t}(g_{2 x}-g_{4 x})+\frac{1}{c_w a}(g_{0 x}+g_{1 x}+g_{2 x}+g_{3 x}+g_{4 x})=0 \quad\quad\mbox{at}\;\;y=+a,\label{eq:momentconstraintgx1}
\end{equation}
and $B_y=B_0$ implies that
\begin{equation}
g_{0 y}+g_{1 y}+g_{2 y}+g_{3 y}+g_{4 y} = B_0\quad\quad\mbox{at}\;\;y=+a.\label{eq:momentconstraintgy1}
\end{equation}
Rearranging Eqs.~(\ref{eq:momentconstraintgx1}) and (\ref{eq:momentconstraintgy1}) finally yield the incoming distribution functions from the top wall for the moment-based formulation as
\begin{subequations}
\begin{eqnarray}\label{eq:momentbasedBCsupperboundary}
  g_{4 x} &=& \left[\frac{1}{\tau_mc_{sm}^2\delta_t}+\frac{1}{c_w a}\right]^{-1}\left\{\left[\frac{1}{\tau_mc_{sm}^2\delta_t}-\frac{1}{c_w a}\right]g_{2 x}-\frac{1}{c_w a}(g_{0 x}+g_{1 x}+g_{3 x})\right\}\quad\quad\mbox{at}\;\;y=+a, \\
  g_{4 y} &=& B_0-(g_{0 y}+g_{1 y}+g_{2 y}+g_{3 y}).
\end{eqnarray}
\end{subequations}
As in the previous section, we now analyze the limiting cases involving the wall conductance ratio $c_w$, and in what follows, and will express below just the $x$ component of the distribution function which are dependent in this quantity. The $y$ component expressions given above are based on the imposed magnetic field $B_0$ and are invariant under changes in $c_w$. For the lower boundary at $y=-a$, from Eq.~(\ref{eq:momentbasedBCslowerboundary}), we get
\begin{eqnarray*}
c_w=0:\quad g_{2x} &=& -(g_{0x}+g_{1x}+g_{3x}+g_{4x})\quad\quad\quad\quad \mbox{(Perfectly insulating wall)},\nonumber\\
c_w\rightarrow\infty: g_{2x} &=& g_{4x}\qquad\qquad\qquad\qquad\qquad\qquad\quad\;\; \mbox{(Perfectly conducting wall)}.\nonumber
\end{eqnarray*}
Similarly, degenerate cases at the upper boundary ($y=+a$) follow from Eq.~(\ref{eq:momentbasedBCsupperboundary}) as
\begin{eqnarray*}
c_w=0:\quad g_{4x} &=& -(g_{0x}+g_{1x}+g_{2x}+g_{3x})\quad\quad\quad\quad \mbox{(Perfectly insulating wall)},\nonumber\\
c_w\rightarrow\infty: g_{4x} &=& g_{2x}\qquad\qquad\qquad\qquad\qquad\qquad\quad\;\; \mbox{(Perfectly conducting wall)}.\nonumber
\end{eqnarray*}
We note here that the results given above for the insulated wall condition match with those presented in Ref.~\cite{dellar2013moment}. For completeness, the moment-based implementation of the velocity boundary conditions needed for the coupled MHD solution is briefly summarized in~\ref{sec:momentbasedBCvelocity}. Moreover,~\ref{sec:bcsmovingwallsmomentbased} discusses the extension of the above moment-based boundary scheme for MHD with moving conducting walls. It may be noted that the link-based boundary scheme as well as the moment-based boundary formulation
derived in the last two sections can be implemented to resolve the Hartmann layers as well as the Shercliff or side layers in MHD bounded by conducting walls, which arise in many practical problems especially under high magnetic field strengths, via using appropriate local grid refinement approaches.

\section{Results and Discussion}\label{sec:resultsanddiscussion}
\subsection{MHD bounded by conducting walls with stationary boundaries: Body force-driven flow}
We will first consider a case study involving the Hartmann flow through a channel bounded by thin conducting walls to perform a numerical investigations of the two new boundary closures for the LKS for representing the Shercliff BC that were discussed earlier. A schematic of this flow problem is shown in Fig.~\ref{fig:fig_schematic_Hartmann_flow_conducting_BC} and its description is given in Sec.~\ref{subsec:Hartmannflowconductingwalls}. For this liquid metal fluid of kinematic viscosity $\nu$, density $\rho$, electrical conductivity $\sigma$ flowing due a body force of magnitude $F_x$ in the presence of an external magnetic field $B_0$ and bounded by plates with conductance wall ratio $c_w$ and separated by a distance $2a$, the analytical solutions for the profiles of the velocity and the induced magnetic fields were derived by Chang and Lundgren~\cite{chang1961duct}, and are given by
\begin{eqnarray*}
  u^*(\xi) &=& \frac{u(y)}{U_{ref}}=\frac{1}{\mbox{Ha}}\left[\frac{c_w+1}{\mbox{Ha}\;c_w+\tanh(\mbox{Ha})}\right]\left[1-\frac{\cosh(\mbox{Ha}\xi)}{\cosh(\mbox{Ha})}\right],\\
  B_x^*(\xi) &=& \frac{B_x(y)}{B_{ref}}=-\frac{\xi}{\mbox{Ha}}+\frac{1}{\mbox{Ha}}\left[\frac{c_w+1}{\mbox{Ha}\;c_w+\tanh(\mbox{Ha})}\right]\frac{\sinh(\mbox{Ha}\xi)}{\cosh(\mbox{Ha})},
\end{eqnarray*}
where $\xi = y/a$ and $-1\le \xi \le 1$. See also Ref.~\cite{muller2013magnetofluiddynamics}. Here, the reference velocity $U_{ref}$, the reference scale for the magnetic field $B_{ref}$, as well as the Hartmann number $\mbox{Ha}$ are given by
\begin{equation*}
  U_{ref} = \frac{a^2}{\rho\nu}F_x,\quad\quad B_{ref}=a^2F_x\mu_m\left(\frac{\sigma}{\rho\nu}\right)^{1/2},\quad\quad \mbox{Ha}=B_0 a \left(\frac{\sigma}{\rho\nu}\right)^{1/2},
\end{equation*}
where the latter represents the ratio of the electromagnetic or the Lorentz force to the viscous force. We consider periodic boundary condition along the flow, i.e., $x$ direction, and the no-slip velocity boundary conditions as well as the conducting wall condition for the induced magnetic field in the $x$ direction and the external magnetic field in the $y$ direction at the bottom and top walls are imposed using the link-based and the moment-based boundary schemes discussed in the previous sections. For validating these two boundary schemes, we considered two different values of the Harmann number $\mbox{Ha}=5$ and $\mbox{Ha}=20$, and in each case, the wall conductance ratio $c_w$ is specified to have a range of values given by $c_w=\{0, 0.1, 0.5, 1.0, 10^{4}\}$. In this regard, the computational domain is resolved using $3\times 101$ grid nodes, and the relaxation times of the LB schemes used to compute the velocity field and magnetic field, respectively, are chosen as $\tau=0.57$ and $\tau_m=0.75$. In each of the chosen set of parameters, the simulations are run until steady state.

\begin{figure}[h!]
\centering
\advance\leftskip-1.7cm
    \subfloat[$\mbox{Ha}=5$] {
        \includegraphics[width=0.4\textwidth] {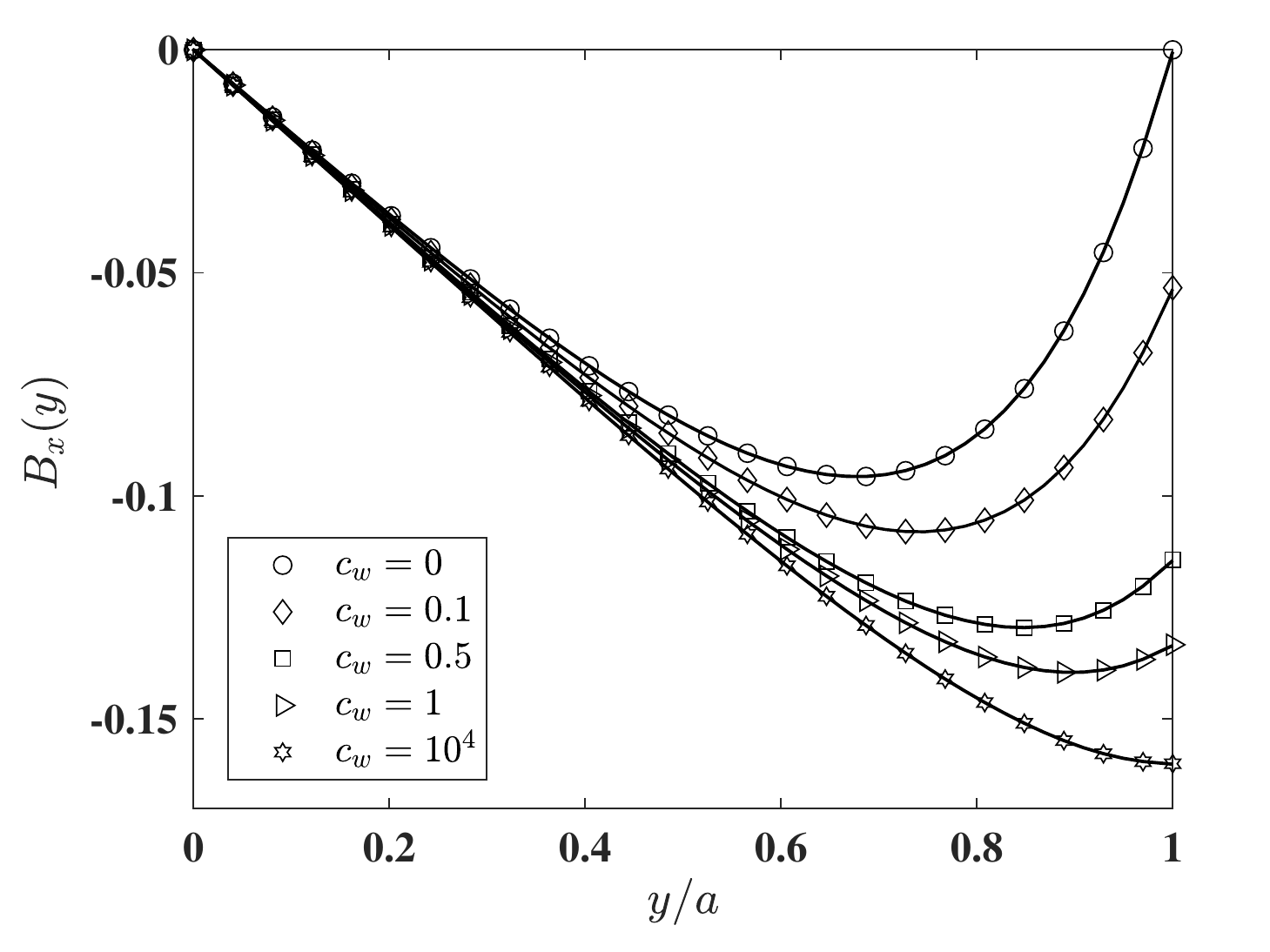}
        \label{fig:1a} } 
    \subfloat[$\mbox{Ha}=20$] {
        \includegraphics[width=0.4\textwidth] {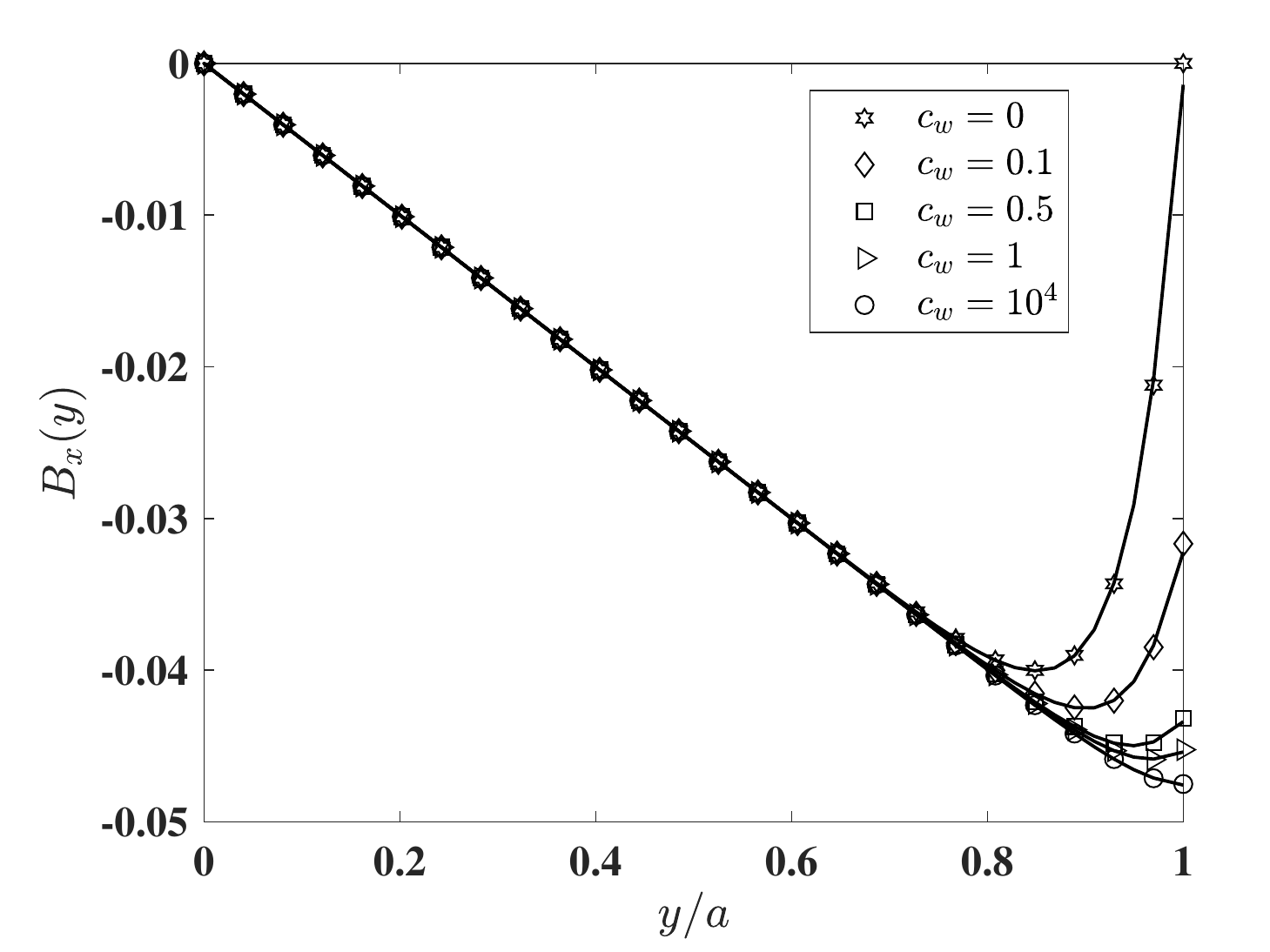}
        \label{fig:1b} } \\
                \advance\leftskip0cm
    \caption{The magnetic field profiles $B_x(y)$ across the top half of the channel for the Hartmann flow bounded by conducting walls for different values of the wall conductance ratio $c_w= 0, 0.1, 0.5, 1$, and $10^{4}$ at Hartmann numbers (a) $\mbox{Ha}=5$ and (b) $\mbox{Ha}=20$ computed via the link-based boundary scheme (lines) compared with the analytical solution of Chang and Lundren (1961)~\cite{chang1961duct} (symbols).}
    \label{fig:figBxHa5Ha20Linked}
\end{figure}

\begin{figure}[h!]
\centering
\advance\leftskip-1.7cm
    \subfloat[$\mbox{Ha}=5$] {
        \includegraphics[width=0.4\textwidth] {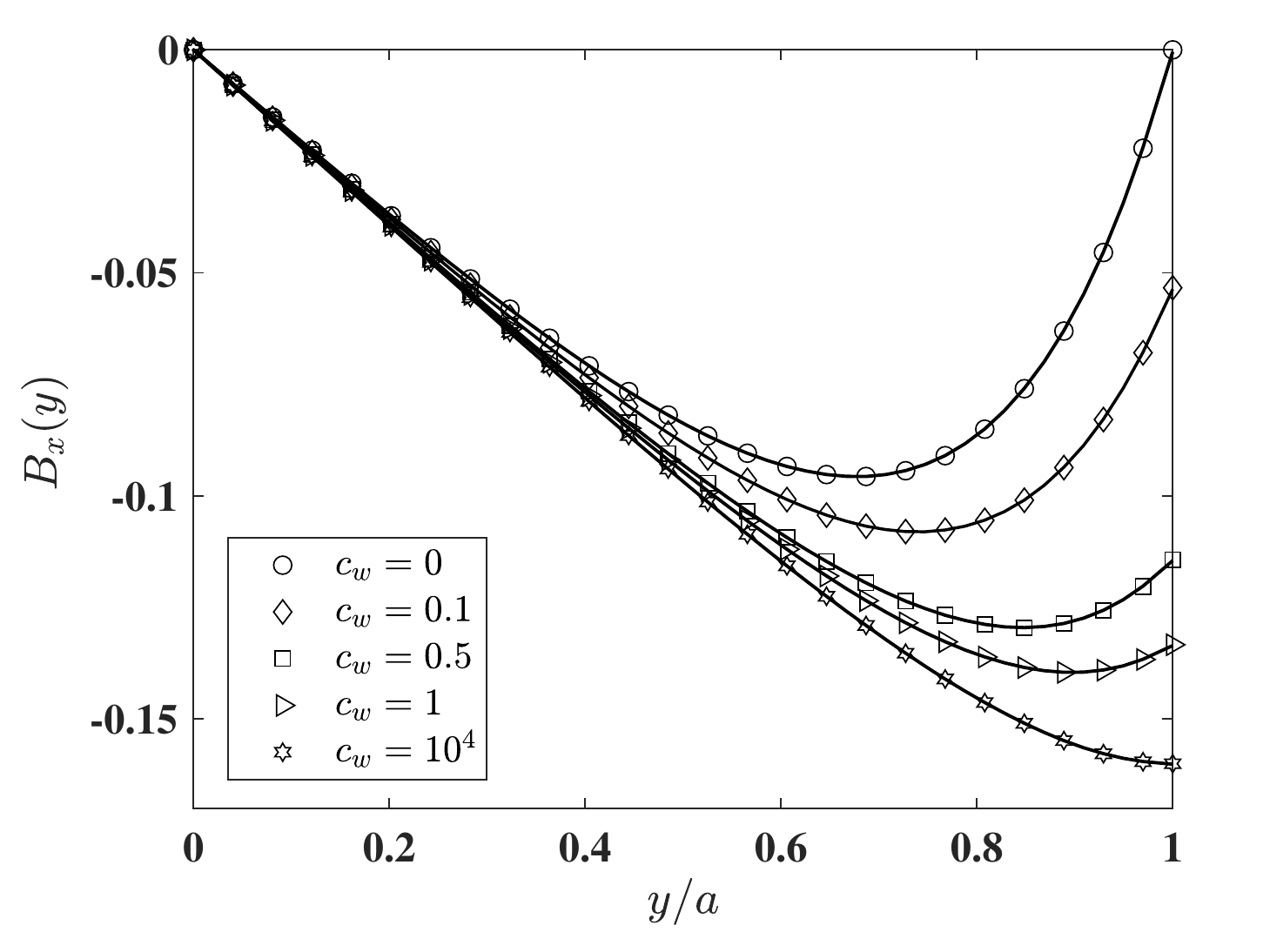}
        \label{fig:2a} } 
    \subfloat[$\mbox{Ha}=20$] {
        \includegraphics[width=0.4\textwidth] {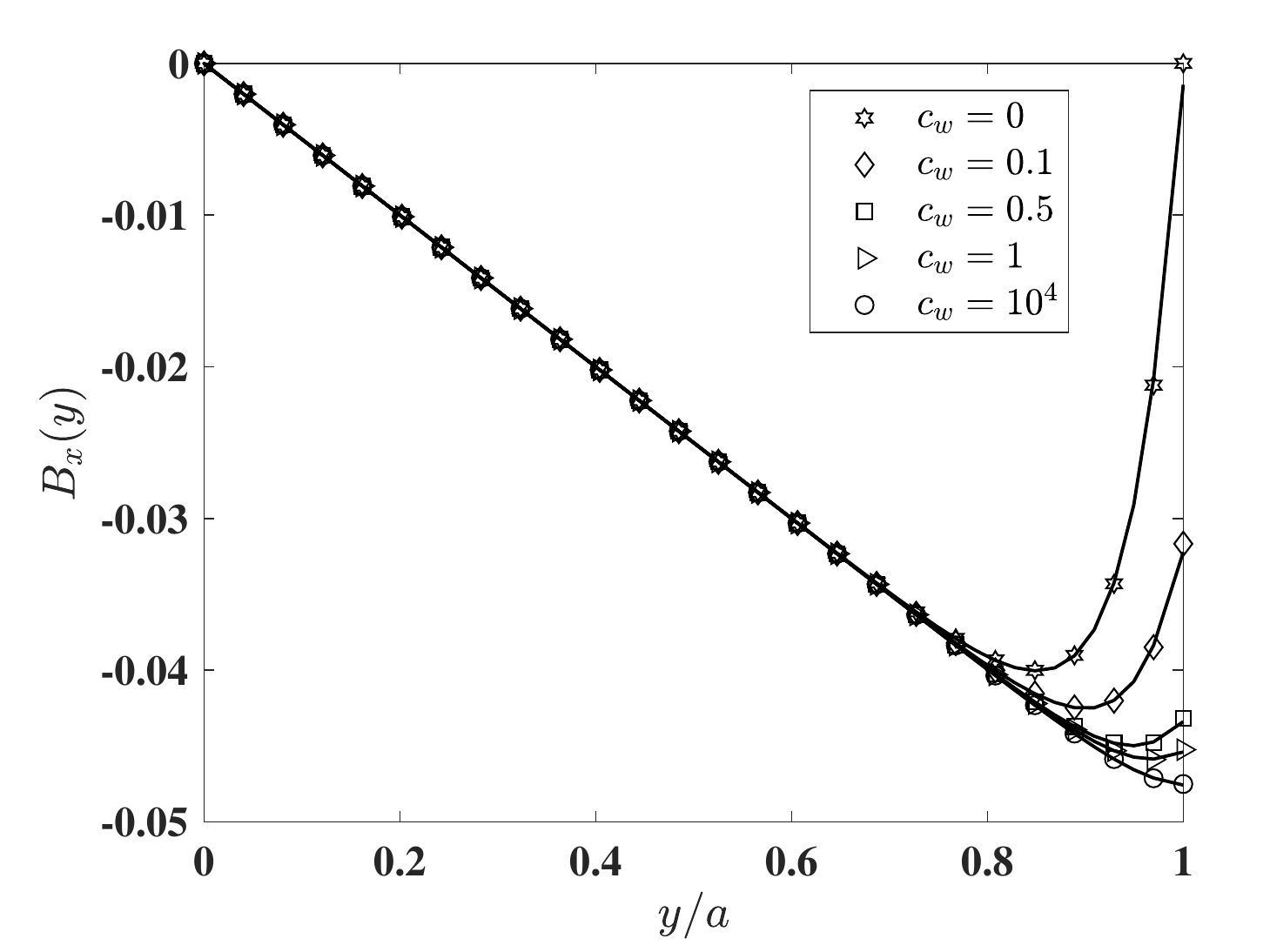}
        \label{fig:2b} } \\
                \advance\leftskip0cm
    \caption{The magnetic field profiles $B_x(y)$ across the top half of the channel for the Hartmann flow bounded by conducting walls for different values of the wall conductance ratio $c_w= 0, 0.1, 0.5, 1$, and $10^{4}$ at Hartmann numbers (a) $\mbox{Ha}=5$ and (b) $\mbox{Ha}=20$ computed via the moment-based boundary scheme (lines) compared with the analytical solution of Chang and Lundren (1961)~\cite{chang1961duct} (symbols).}
    \label{fig:figBxHa5Ha20Moment}
\end{figure}

Figures~\ref{fig:figBxHa5Ha20Linked} and~\ref{fig:figBxHa5Ha20Moment} present comparisons between the computed profiles of the magnetic field obtained using the link-based and moment-based boundary schemes, respectively, with the analytical solution at the two Hartmann numbers and for the range of the wall conductance ratio given above. For clarity, the profiles along only the top half of the channel are shown and it may be noted that in the bottom half of the channel, the magnetic field profiles are antisymmetric to that in its top half. When $c_w$ is zero, corresponding to the insulated wall case, as expected, the induced magnetic field is seen to zero. As $c_w$ increases, the finite wall conductivity effects become more prominent, with non-zero values of the induced magnetic field on the wall and accompanied by progressively decreasing magnitudes of its slope normal to the wall are observed. While both the boundary condition implementations are found to work for any values of the wall conductance ratio (i.e., $0\le c_w \le \infty$), it was noticed that when $c_w=10^4$, the wall becomes essentially perfectly conducting with zero wall-normal gradient for the induced magnetic field. For the smaller $\mbox{Ha}$ case, variations in the magnetic fields are seen to occur for the most part of the channel; by contrast, for the higher $\mbox{Ha}$ case, such variations are confined within the relatively thinner Hartmann layer. It is evident that there is an excellent agreement with the analytical solution for the structures of the magnetic field for both the boundary schemes for all the range of values considered for the electrical properties of the walls for both the Hartmann numbers.

\begin{figure}[h!]
\centering
\advance\leftskip-1.7cm
    \subfloat[$\mbox{Ha}=5$] {
        \includegraphics[width=0.4\textwidth] {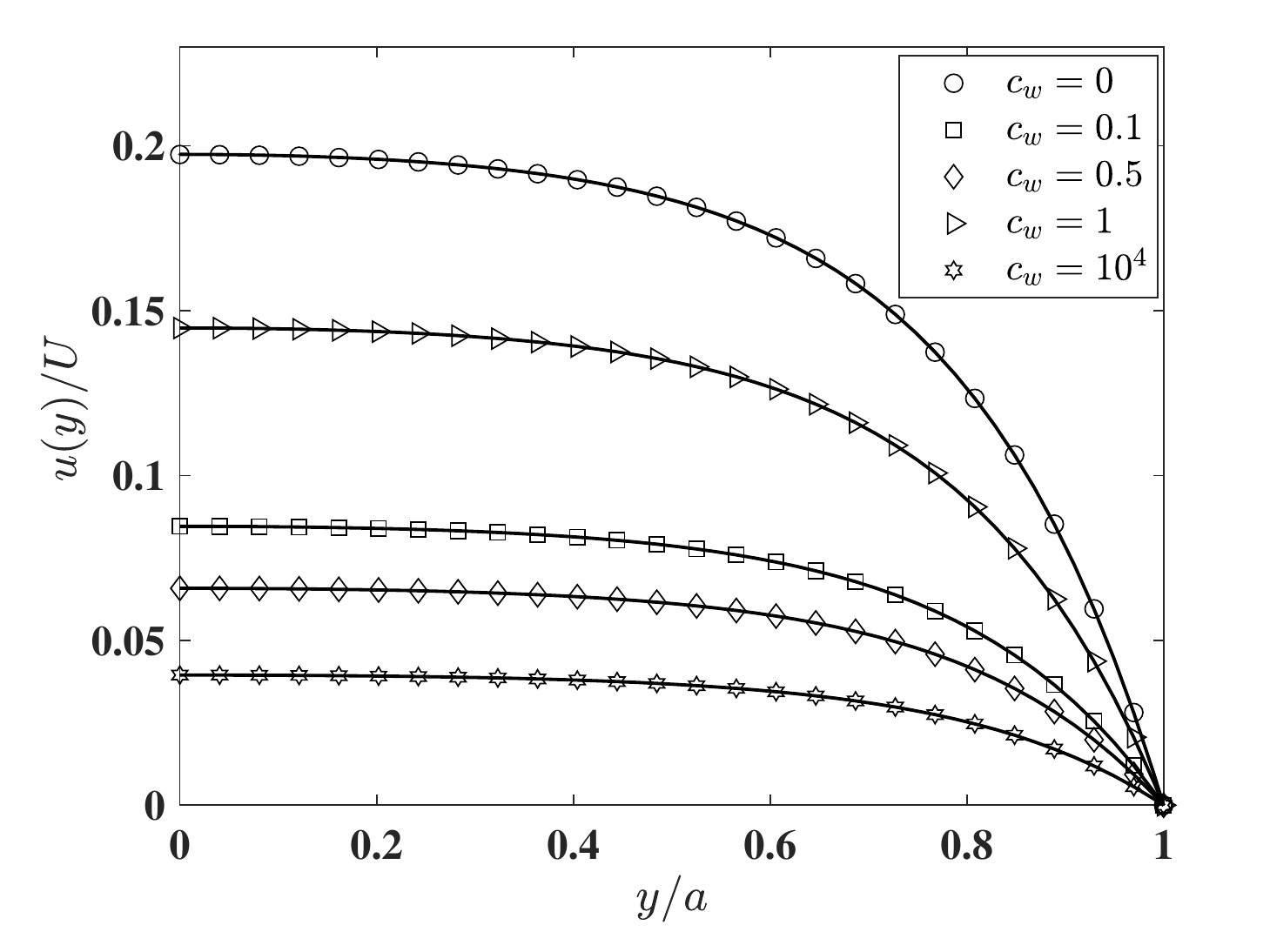}
        \label{fig:3a} } 
    \subfloat[$\mbox{Ha}=20$] {
        \includegraphics[width=0.4\textwidth] {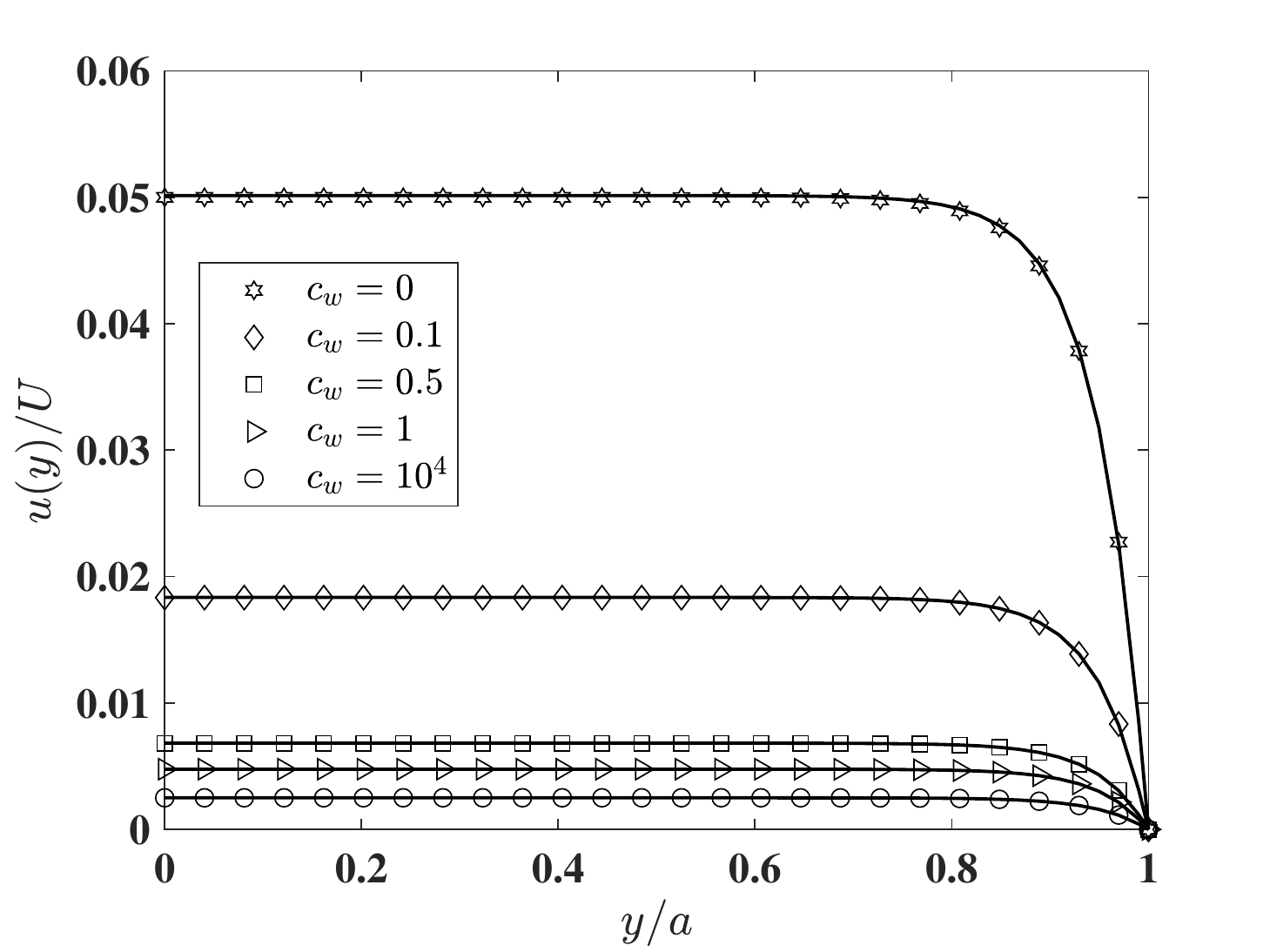}
        \label{fig:3b} } \\
                \advance\leftskip0cm
    \caption{The velocity field profiles $u(y)$ across the top half of the channel for the Hartmann flow bounded by conducting walls for different values of the wall conductance ratio $c_w= 0, 0.1, 0.5, 1$, and $10^{4}$ at Hartmann numbers (a) $\mbox{Ha}=5$ and (b) $\mbox{Ha}=20$ computed via the link-based boundary scheme (lines) compared with the analytical solution of Chang and Lundren (1961)~\cite{chang1961duct} (symbols).}
    \label{fig:figuxHa5Ha20Linked}
\end{figure}

\begin{figure}[h!]
\centering
\advance\leftskip-1.7cm
    \subfloat[$\mbox{Ha}=5$] {
        \includegraphics[width=0.4\textwidth] {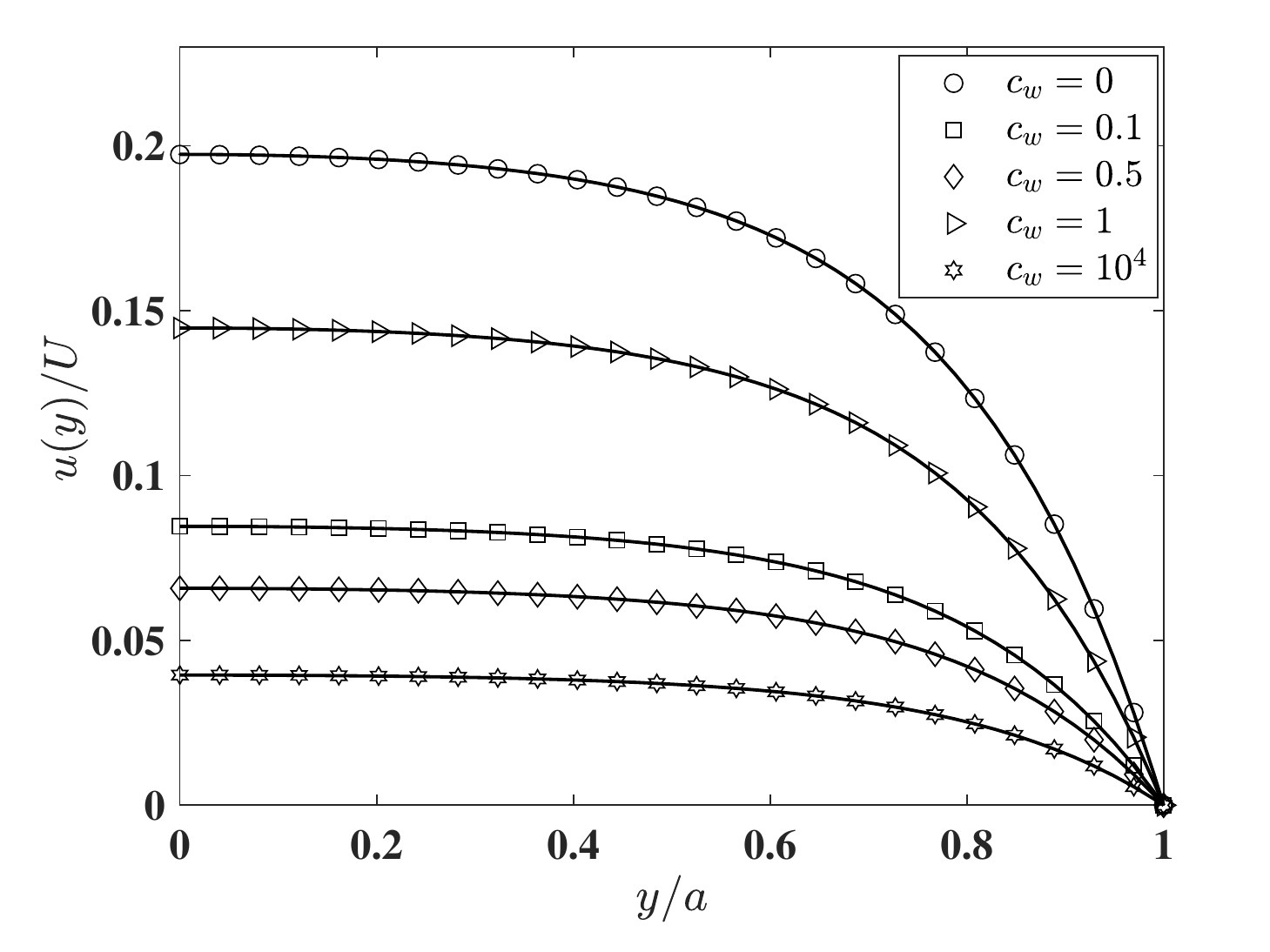}
        \label{fig:4a} } 
    \subfloat[$\mbox{Ha}=20$] {
        \includegraphics[width=0.4\textwidth] {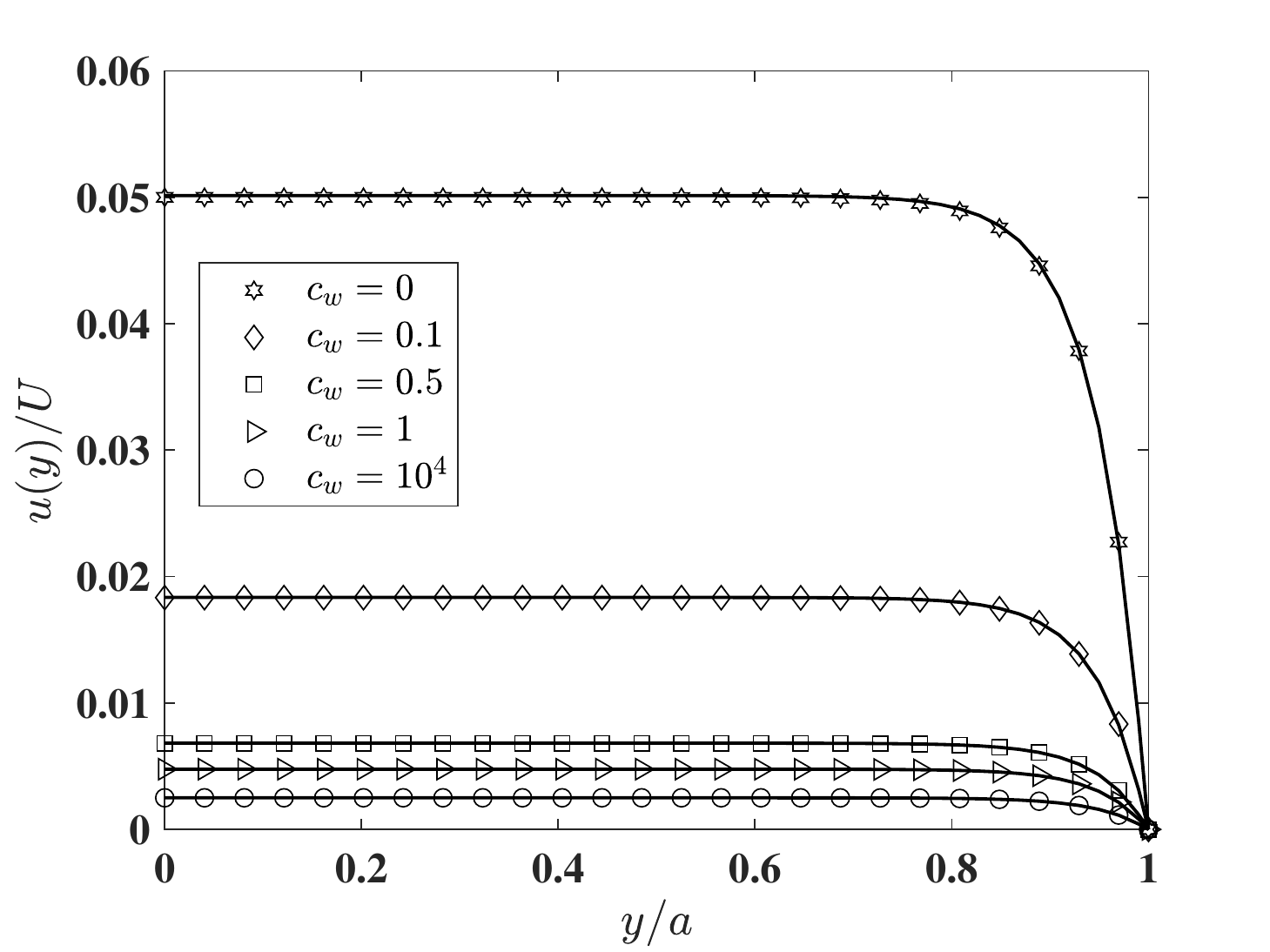}
        \label{fig:4b} } \\
                \advance\leftskip0cm
    \caption{The velocity field profiles $u(y)$ across the top half of the channel for the Hartmann flow bounded by conducting walls for different values of the wall conductance ratio $c_w= 0, 0.1, 0.5, 1$, and $10^{4}$ at Hartmann numbers (a) $\mbox{Ha}=5$ and (b) $\mbox{Ha}=20$ computed via the moment-based boundary scheme (lines) compared with the analytical solution of Chang and Lundren (1961)~\cite{chang1961duct} (symbols).}
    \label{fig:figuxHa5Ha20Moment}
\end{figure}

Moreover, in Figs.~\ref{fig:figuxHa5Ha20Linked} and~\ref{fig:figuxHa5Ha20Moment} the respective comparisons for the velocity field obtained using link-based and moment-based boundary implementations are shown. Since the Lorentz force couples such strong variations in the induced magnetic fields at different values of the wall conductance ratio, it is expected that the latter will also have significant effect on the velocity field across the channel. In general, the higher the wall conductance ratio, the smaller are the magnitudes of the velocity field for any given $\mbox{Ha}$. This is consistent with the fact that walls with finite conductivities ($c_w\ne 0$) supports a tangential flow of current within them, whose interaction with the external magnetic field generates larger Lorentz force that resists the fluid motion when compared to the insulated wall case ($c_w=0$), where no current flow exists within the wall. Moreover, as $\mbox{Ha}$ is increased, as the Hartmann layer becomes thinner, there is a significant flattening of the velocity profiles within the core of the flow. Again, the analytical predictions of these variations in the velocity field profiles at different $c_w$ and $\mbox{Ha}$ are found to be very well reproduced by both the link-based and moment-based boundary schemes constructed in the previous sections.


\begin{figure}[h!]
\centering
\advance\leftskip-1.7cm
    \subfloat[] {
        \includegraphics[width=0.4\textwidth] {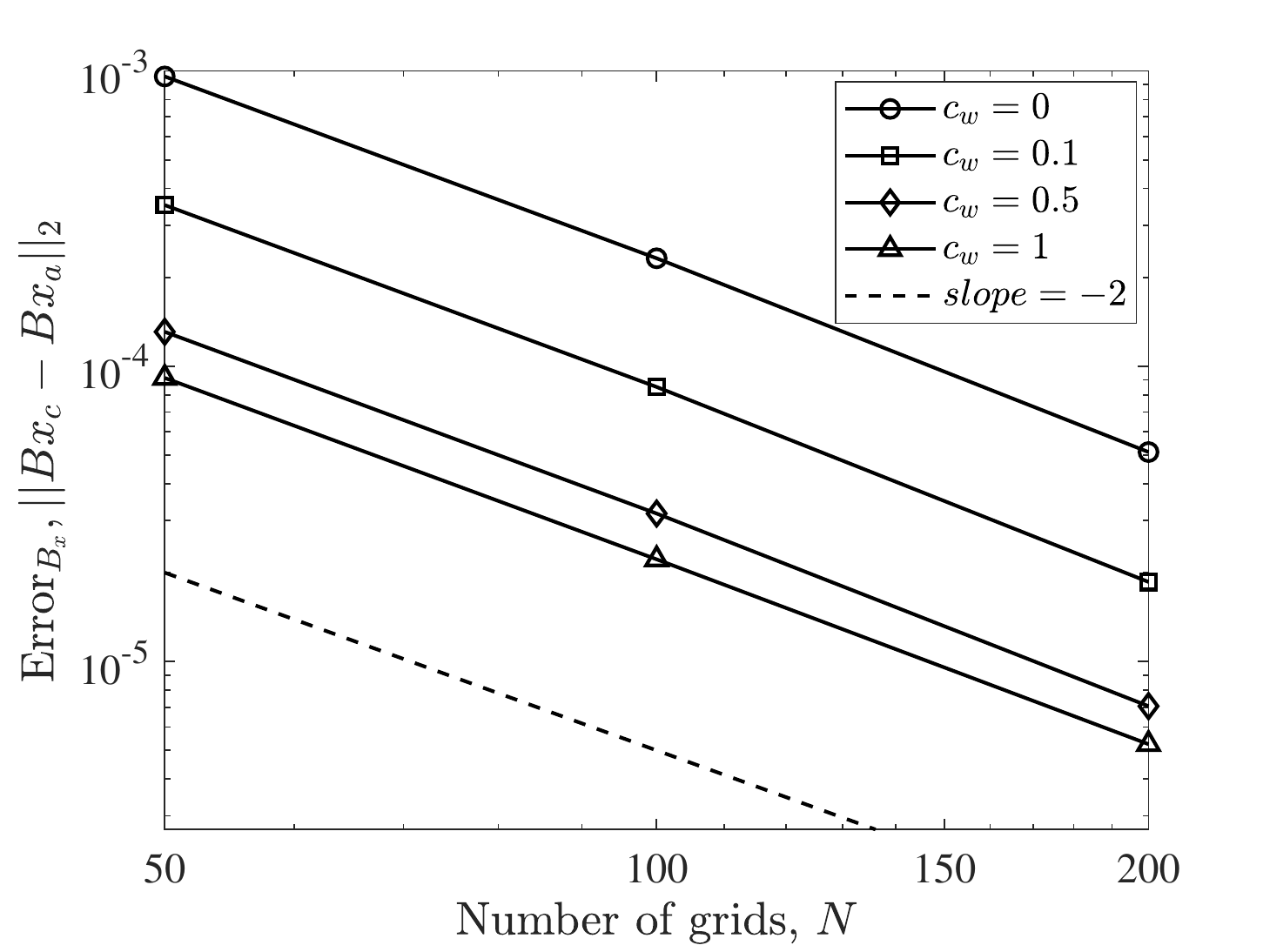}
        \label{fig:5a} } 
    \subfloat[] {
        \includegraphics[width=0.4\textwidth] {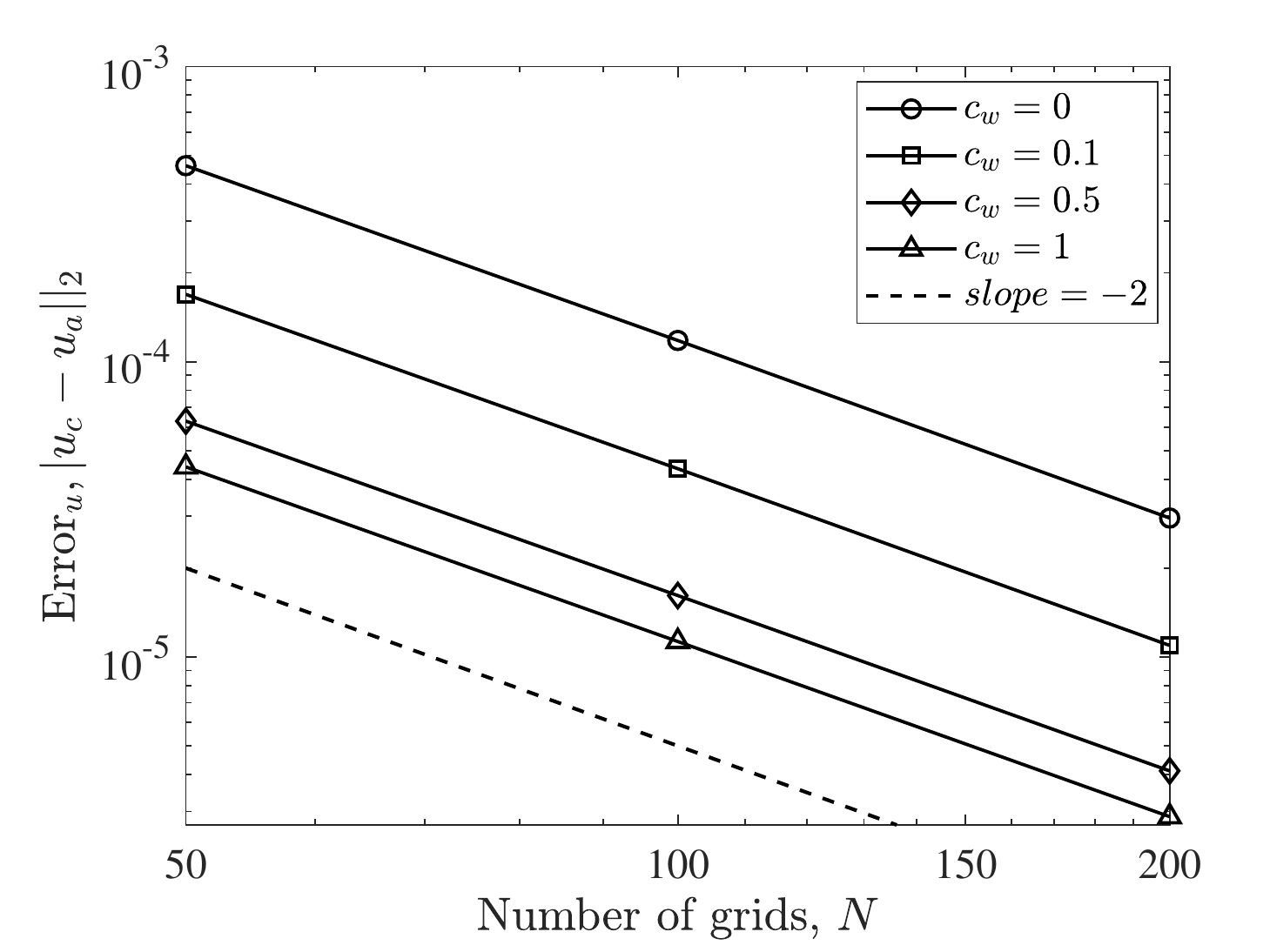}
        \label{fig:5b} } \\
                \advance\leftskip0cm
    \caption{Grid convergence of the (a) magnetic field $B_x$ and (b) velocity field $u$ using the LB method for MHD using the link-based boundary scheme as a function of the number of grid nodes in the wall normal direction $N$ at Hartmann number $\mbox{Ha}=20$ for different values of the wall ratio ($c_w= 0, 0.1, 0.5, 1$) with $\tau=0.9$ and $\tau_m=0.95$.}
    \label{fig:gridconvergenceLinkbased}
\end{figure}

\begin{figure}[h!]
\centering
\advance\leftskip-1.7cm
    \subfloat[] {
        \includegraphics[width=0.4\textwidth] {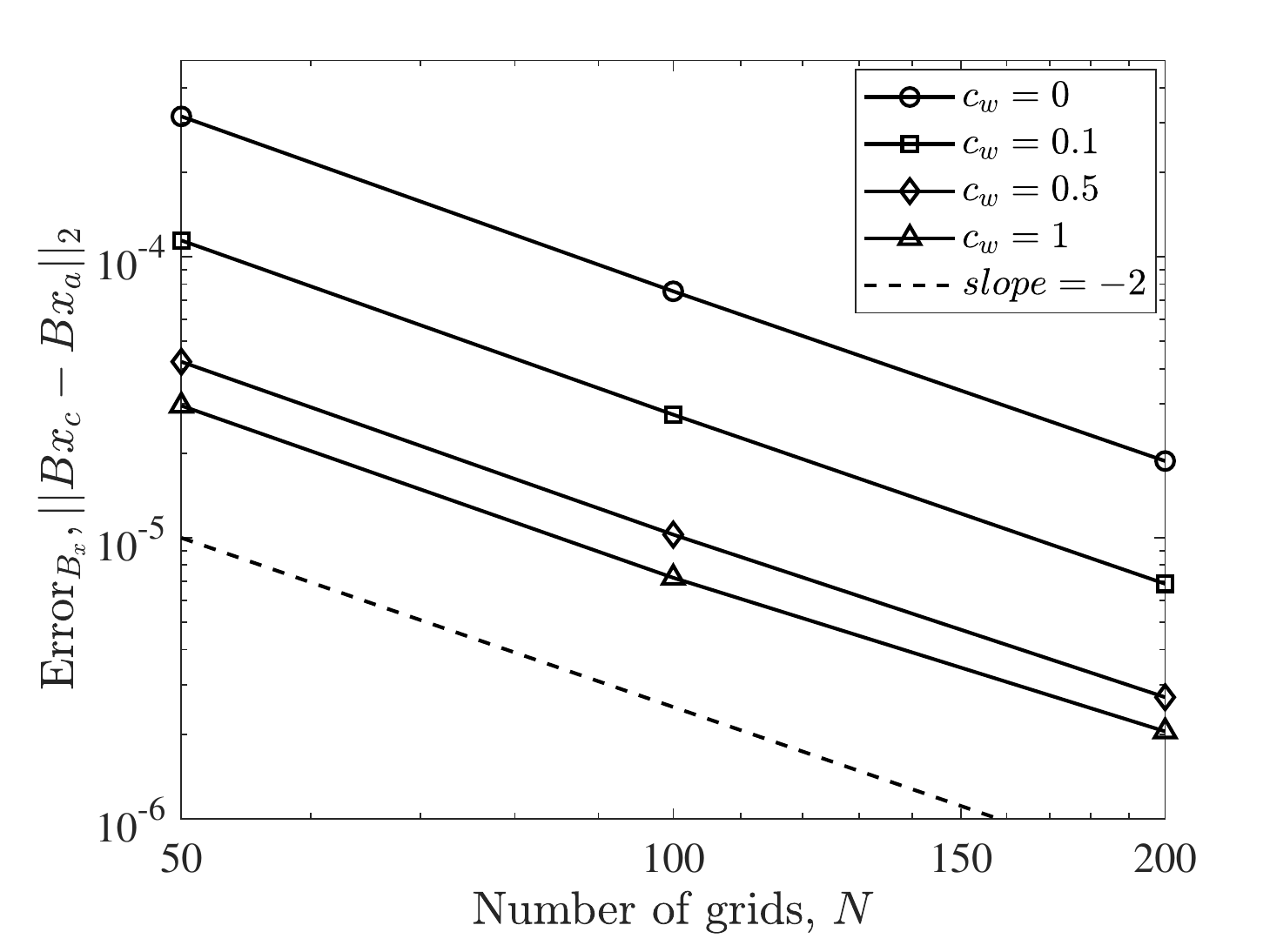}
        \label{fig:6a} } 
    \subfloat[] {
        \includegraphics[width=0.4\textwidth] {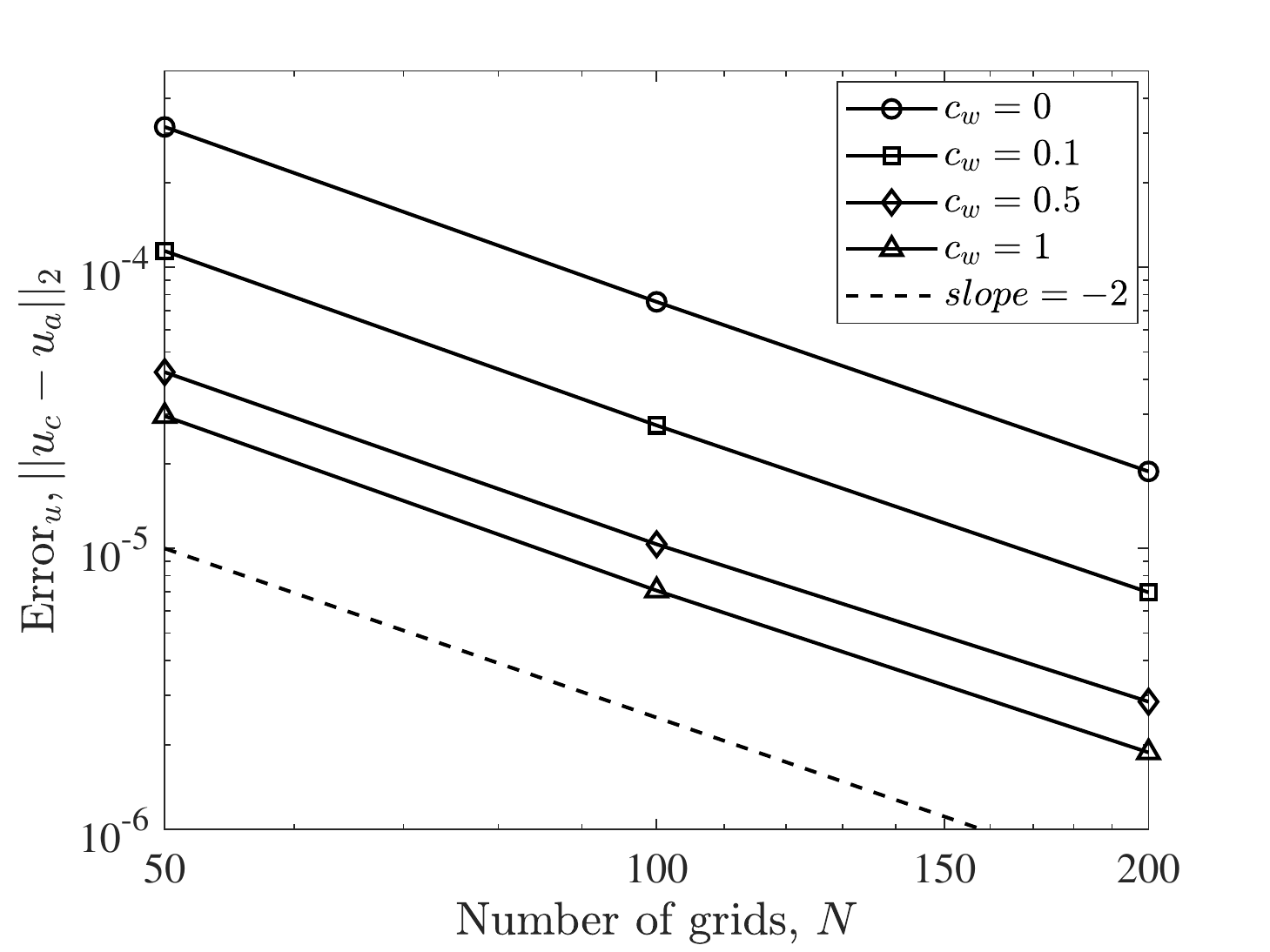}
        \label{fig:6b} } \\
                \advance\leftskip0cm
    \caption{Grid convergence of the (a) magnetic field $B_x$ and (b) velocity field $u$ using the LB method for MHD using the moment-based boundary scheme as a function of the number of grid nodes in the wall normal direction $N$ at Hartmann number $\mbox{Ha}=20$ for different values of the wall ratio ($c_w= 0, 0.1, 0.5, 1$) with $\tau=0.9$ and $\tau_m=0.95$.}
    \label{fig:gridconvergenceMomentbased}
\end{figure}

Next, let us estimate the order of accuracy under grid refinement for both the boundary implementation schemes. In this regard, we define the following error estimates between the computed and analytical solutions of the velocity and magnetic fields under the second norm:
\begin{eqnarray*}
  \mbox{Error}_u &=& ||u_c-u_a||_2=\left[\frac{\sum_i(u_{c,i} -u_{a,i})^2}{N}\right]^{1/2}, \\
  \mbox{Error}_{B_x} &=& ||B_{xc}-B_{xa}||_2=\left[\frac{\sum_i(B_{xc,i} -B_{xa,i})^2}{N}\right]^{1/2},
\end{eqnarray*}
where $u_c$ and $u_a$ represent the computed and analytical solutions of the velocity field, while $B_{xc}$ and $B_{xa}$ denote the computed and analytical results for the magnetic field. The subscript $i$ and $\sum_i$ denote the summation is carried out across the width of the channel resolved by $N$ number of grid nodes. For consistent convergence to the incompressible limit, we employ the usual diffusive scaling, where for fixed relaxation times $\tau$ and $\tau_m$, the scales of the velocity and magnetic fields are varied in proportion to the spatial discretization length scales given by $1/N$. In other words, the compressibility errors decrease at the same rate as the spatial truncation errors in this limit process to represent the incompressible MHD solutions. In particular, for each of the boundary scheme, we consider $\mbox{Ha}=20$ and fix the relaxation times $\tau=0.9$ and $\tau_m=0.95$, and vary the grid resolution by choosing $N=50, 100$ and $200$ for the following different values of the wall conductance ratio: $c_w= 0, 0.1, 0.5$, and $1.0$. The errors between the computed solutions obtained using the link-based boundary scheme and analytical solutions for the magnetic field and the velocity field at different values of the wall conductance ratios for various choices of the grid nodes are plotted in Figs.~\ref{fig:gridconvergenceLinkbased}(a) and~\ref{fig:gridconvergenceLinkbased}(b), respectively. Similar errors versus the number of grid nodes for the magnetic field and velocity field in the case of the moment-based boundary scheme are shown in Figs.~\ref{fig:gridconvergenceMomentbased}(a) and~\ref{fig:gridconvergenceMomentbased}(b), respectively. In both these figures, a reference slope of $-2$ is also displayed. Clearly, both the link-based and moment-based boundary schemes for LB methods for MHD show second order accuracy in computing solutions for the magnetic and velocity fields for the Hartmann flow with conducting walls. These are also consistent with the second order accurate results arising from an error analysis made for the special case involving the insulated walls in Ref.~\cite{dellar2013moment}.


\subsection{MHD bounded by conducting walls with a moving boundary: Shear-driven flow}
We will now present a case study that provides a validation of the extension of the boundary implementation schemes for moving walls given in~\ref{sec:bcsmovingwalls}, where~\ref{sec:bcsmovingwallslinkbased} focuses on the link-based formulation and~\ref{sec:bcsmovingwallsmomentbased} on the moment-based scheme. In this regard, we consider the Hartmann-Couette flow of an electrically conducting fluid with conductivity $\sigma$ enclosed between lower and upper plates having the wall conductance ratios of $c_{wl}$ and $c_{wu}$, respectively, with a spacing of $2a$ between them. The shear flow is set up by the motion of the top plate at a constant velocity $U_w$ in the $x$ direction, with the bottom plate being at rest, and it is subjected to an imposed magnetic field $B_0$ in the $y$ direction. For this general MHD flow configuration involving contrasts in the electrical properties of the two bounding plates, we have presented a new derivation of the analytical solution in~\ref{sec:analyticalsolutionMHDCouetteflow}, which will be utilized as a benchmark reference against which the numerical results obtained using our two boundary implementation approaches will be compared. The simulations are set up with a grid resolution of $3\times 101$ grid nodes, where the top plate velocity is chosen to be $U_w = 0.05$ and using the relaxation parameters $\tau=0.6$ and $\tau_m=0.65$ for the LB schemes for solution of the velocity field and magnetic field, respectively. Two different values of the Hartmann number $\mbox{Ha}=5$ and $20$, based on which the imposed magnetic field $B_0$ is determined, and different combinations of the wall conductance ratios $(c_{wu},c_{wl})$ for the bounding plates are considered, and the computations in each case are run until the results converge to a steady state.

\begin{figure}[h!]
\centering
\advance\leftskip-1.7cm
    \subfloat[$\mbox{Ha}=5$] {
        \includegraphics[width=0.4\textwidth] {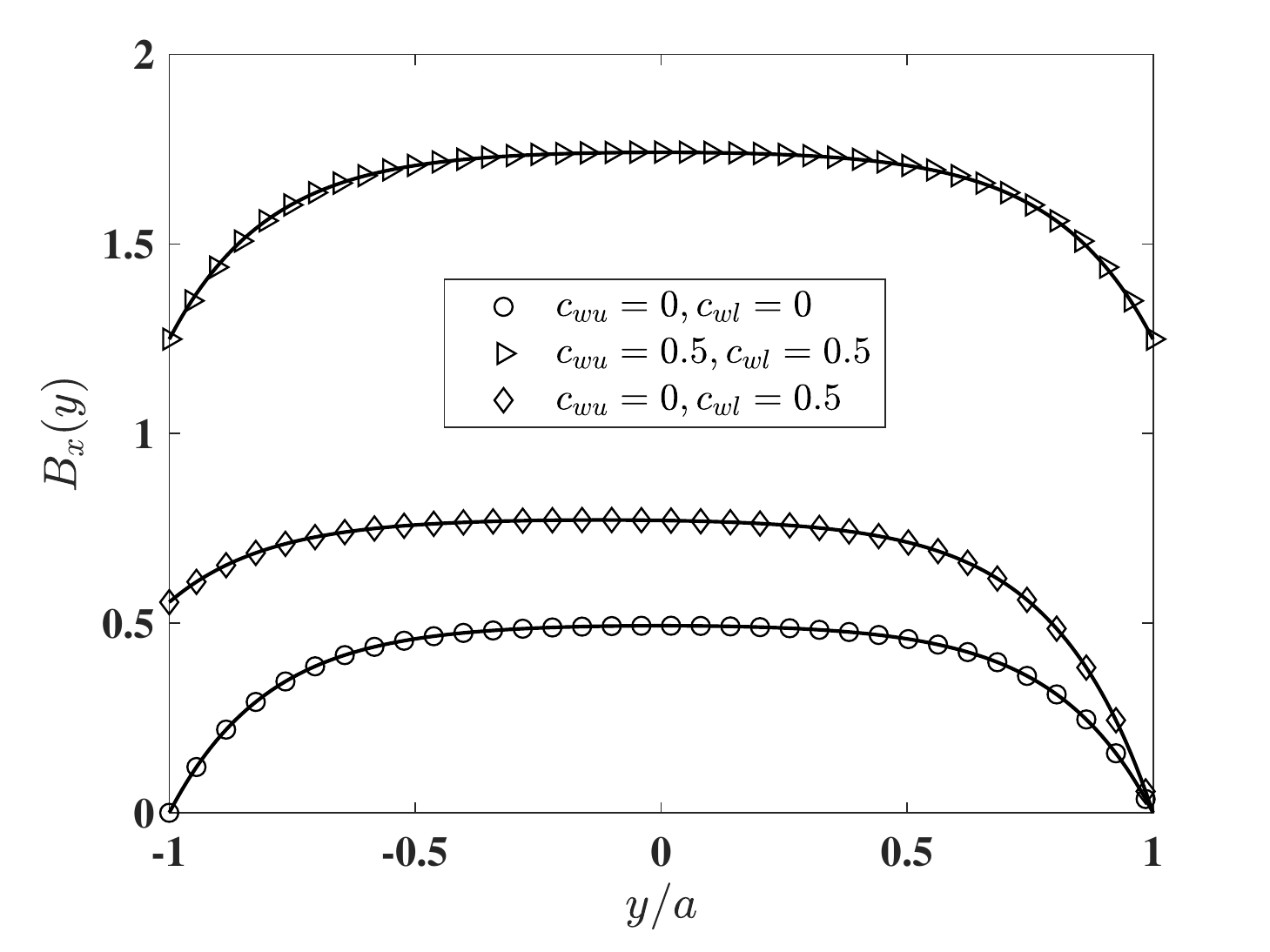}
        \label{fig:7a} } 
    \subfloat[$\mbox{Ha}=20$] {
        \includegraphics[width=0.4\textwidth] {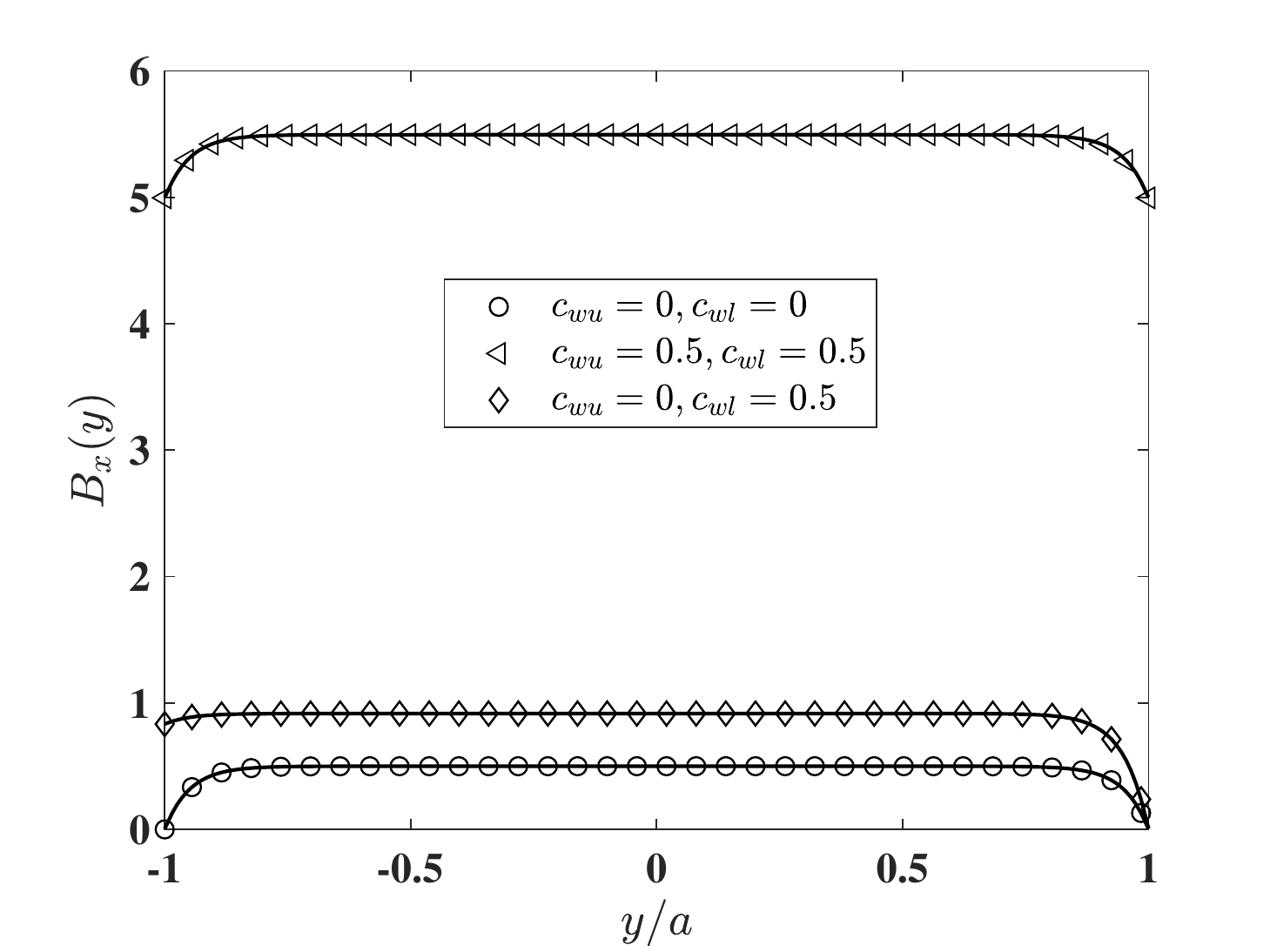}
        \label{fig:7b} } \\
                \advance\leftskip0cm
    \caption{The magnetic field profiles $B_x(y)$ across the channel for the shear-driven Hartmann-Couette flow bounded by upper (u) and lower (l) conducting walls for different pairs of values of the wall conductance ratios $(c_{wu}, c_{wl})=(0, 0)$, $(c_{wu}, c_{wl})=(0.5, 0.5)$, and $(c_{wu}, c_{wl})=(0, 0.5)$ at Hartmann numbers (a) $\mbox{Ha}=5$ and (b) $\mbox{Ha}=20$ computed via the link-based boundary scheme (lines) compared with the analytical solution (symbols) given in Eqs.~(\ref{eq:1DMHDmovingwallfinalsolution}) and~(\ref{eq:1DMHDmovingwallintconstants}) in~\ref{sec:analyticalsolutionMHDCouetteflow}.}
    \label{fig:figBxHa5Ha20LinkedMHDCouetteflow}
\end{figure}

Figure~\ref{fig:figBxHa5Ha20LinkedMHDCouetteflow} presents a comparison between the computed magnetic field profiles obtained using the link-based boundary scheme and the analytical solution for two values of the Hartmann number ($\mbox{Ha}=5$ and $\mbox{Ha}=20$) and for the following combinations of the values of the upper and lower wall conductance ratios: $(c_{wu}, c_{wl})=(0, 0)$, $(c_{wu}, c_{wl})=(0.5, 0.5)$, and $(c_{wu}, c_{wl})=(0, 0.5)$. Clearly, computed solutions are in excellent agreement with the analytical solutions for all the choices of wall conductance ratios and the Hartmann numbers. In general, the strengths of the induced magnetic fields are seen to increase for cases where one or both the walls are conducting when compared to the case where both the walls are electrically insulated. Moreover, as the Hartmann number is increased for a given pair of the conductance ratios for the top and bottom walls, the induced magnetic field becomes larger in magnitude, and a flattening effect of their profiles is observed and accompanied by the wall boundary layer becoming thinner.

\begin{figure}[h!]
\centering
\advance\leftskip-1.7cm
    \subfloat[$\mbox{Ha}=5$] {
        \includegraphics[width=0.4\textwidth] {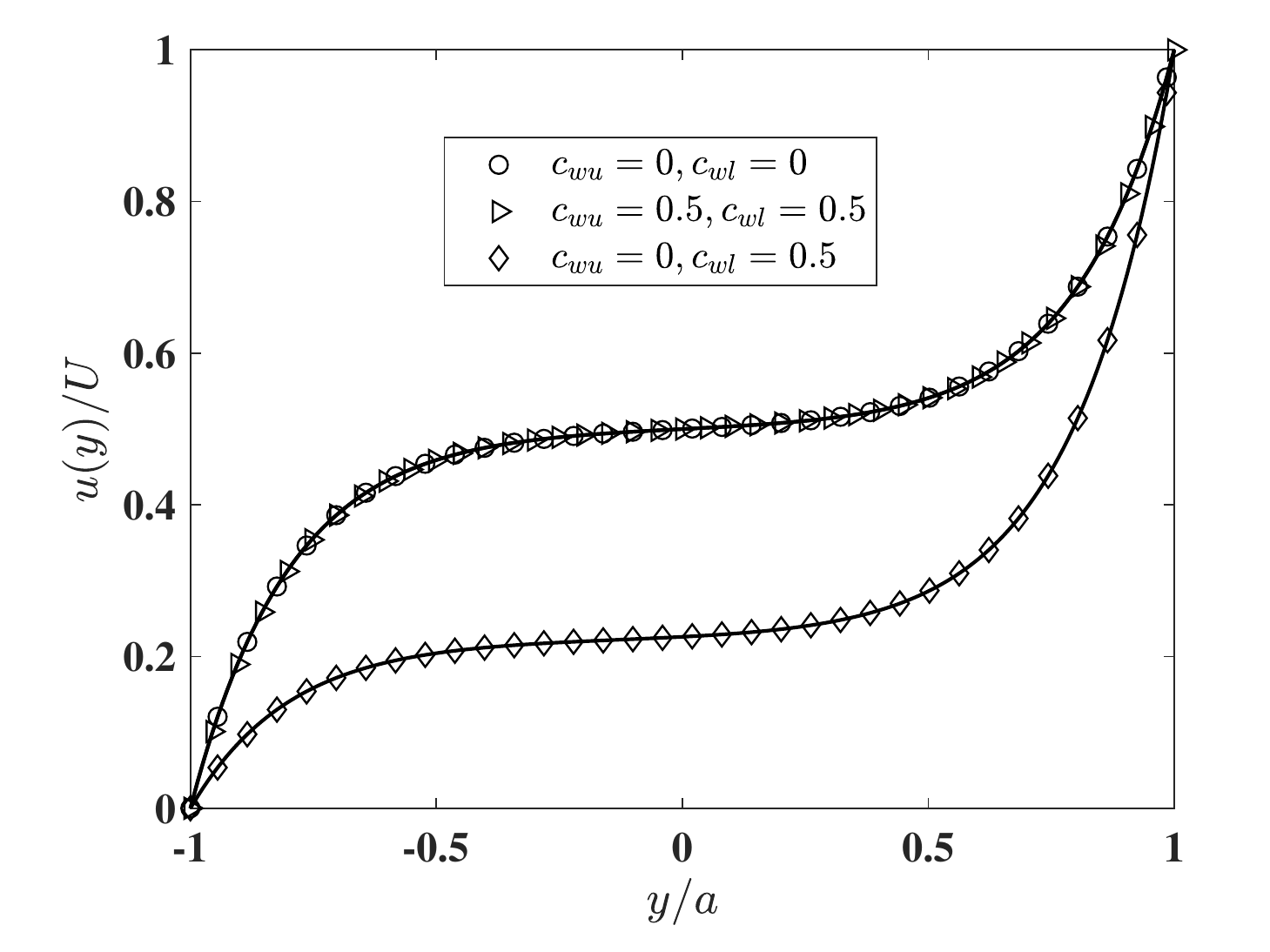}
        \label{fig:8a} } 
    \subfloat[$\mbox{Ha}=20$] {
        \includegraphics[width=0.4\textwidth] {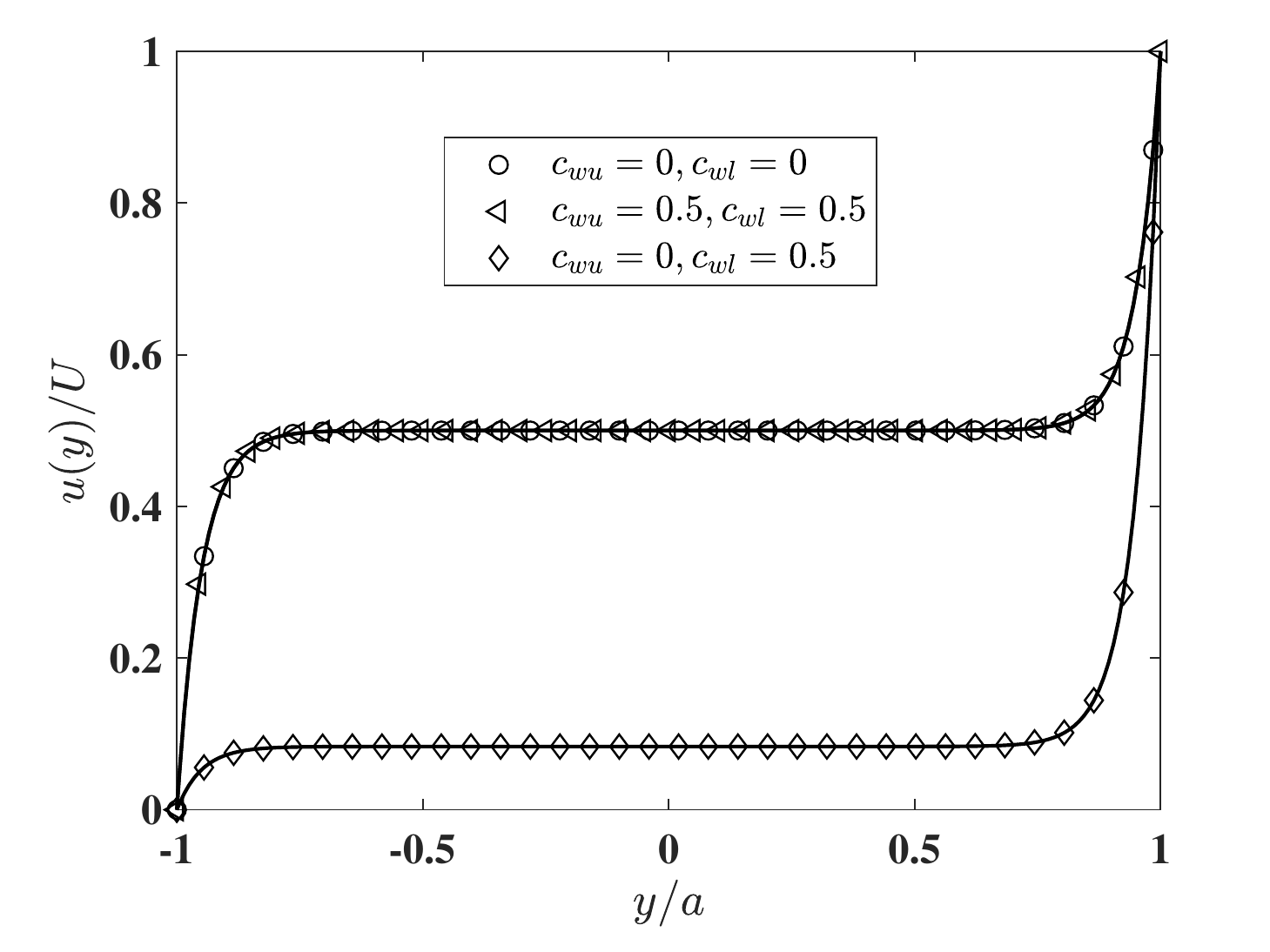}
        \label{fig:8b} } \\
                \advance\leftskip0cm
    \caption{The velocity field profiles $u(y)$ across the channel for the shear-driven Hartmann-Couette flow bounded by upper (u) and lower (l) conducting walls for different pairs of values of the wall conductance ratios $(c_{wu}, c_{wl})=(0, 0)$, $(c_{wu}, c_{wl})=(0.5, 0.5)$, and $(c_{wu}, c_{wl})=(0, 0.5)$ at Hartmann numbers (a) $\mbox{Ha}=5$ and (b) $\mbox{Ha}=20$ computed via the link-based boundary scheme (lines) compared with the analytical solution (symbols) given in Eqs.~(\ref{eq:1DMHDmovingwallfinalsolution}) and~(\ref{eq:1DMHDmovingwallintconstants}) in~\ref{sec:analyticalsolutionMHDCouetteflow}.}
    \label{fig:figuxHa5Ha20LinkedMHDCouetteflow}
\end{figure}

The comparisons made for the corresponding velocity profiles for the same choices of $(c_{wu}, c_{wl})$ and $\mbox{Ha}$ in Fig.~\ref{fig:figuxHa5Ha20LinkedMHDCouetteflow} show that the numerical solutions based on the link-based scheme for a moving wall are again in very good agreement with the reference exact solution derived in~\ref{sec:bcsmovingwalls}.

Notice that imposing an external magnetic field in a shear flow results in velocity profiles with a pair of inflection points away from each wall, which can be contrasted with the linear velocity profile arising in the absence of an imposed magnetic field. The layer formed between the wall and such an inflection point is seen to become narrower as $\mbox{Ha}$ increases. Moreover, unlike the case for the induced magnetic field profiles in Fig.~\ref{fig:figBxHa5Ha20LinkedMHDCouetteflow}, the velocity profiles shown in Fig.~\ref{fig:figuxHa5Ha20LinkedMHDCouetteflow} are invariant if the values of the conductance ratio of the top and bottom walls are the same, and variations in them are observed only when there is a contrast in the conductance ratios at the top and bottom walls. Thus, the velocity profiles for the cases $(c_{wu}, c_{wl})=(0, 0)$ and $(c_{wu}, c_{wl})=(0.5, 0.5)$ are identical, which differ from that arising from the choice $(c_{wu}, c_{wl})=(0, 0.5)$ that involves a contrast in the electrical properties of the bounding walls. This interesting finding could be useful in modulating the structure of shear-driven MHD flows based on manipulating the contrasts in the conductance ratios of the confining walls.

\begin{figure}[h!]
\centering
\advance\leftskip-1.7cm
    \subfloat[$\mbox{Ha}=5$] {
        \includegraphics[width=0.4\textwidth] {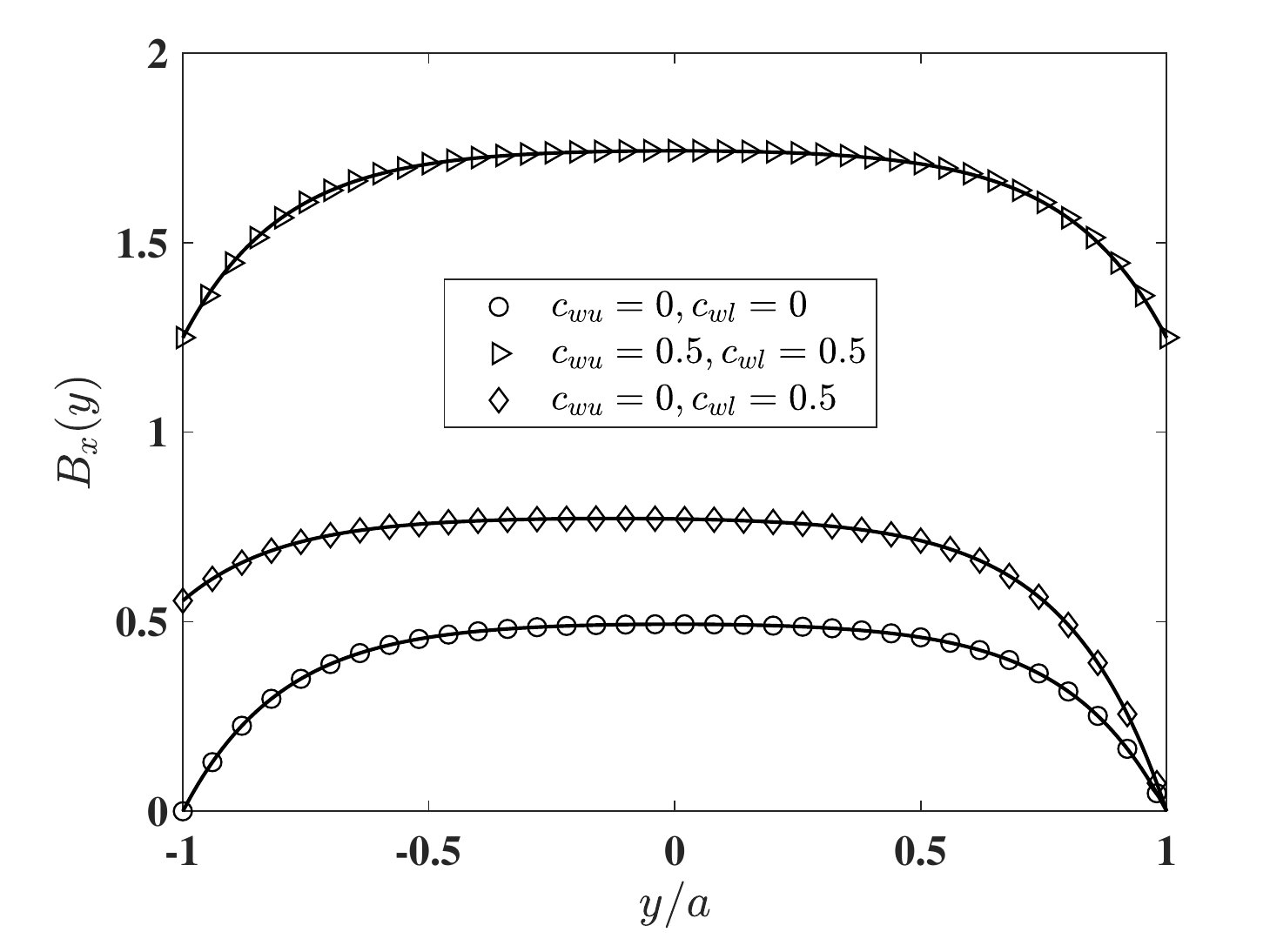}
        \label{fig:9a} } 
    \subfloat[$\mbox{Ha}=20$] {
        \includegraphics[width=0.4\textwidth] {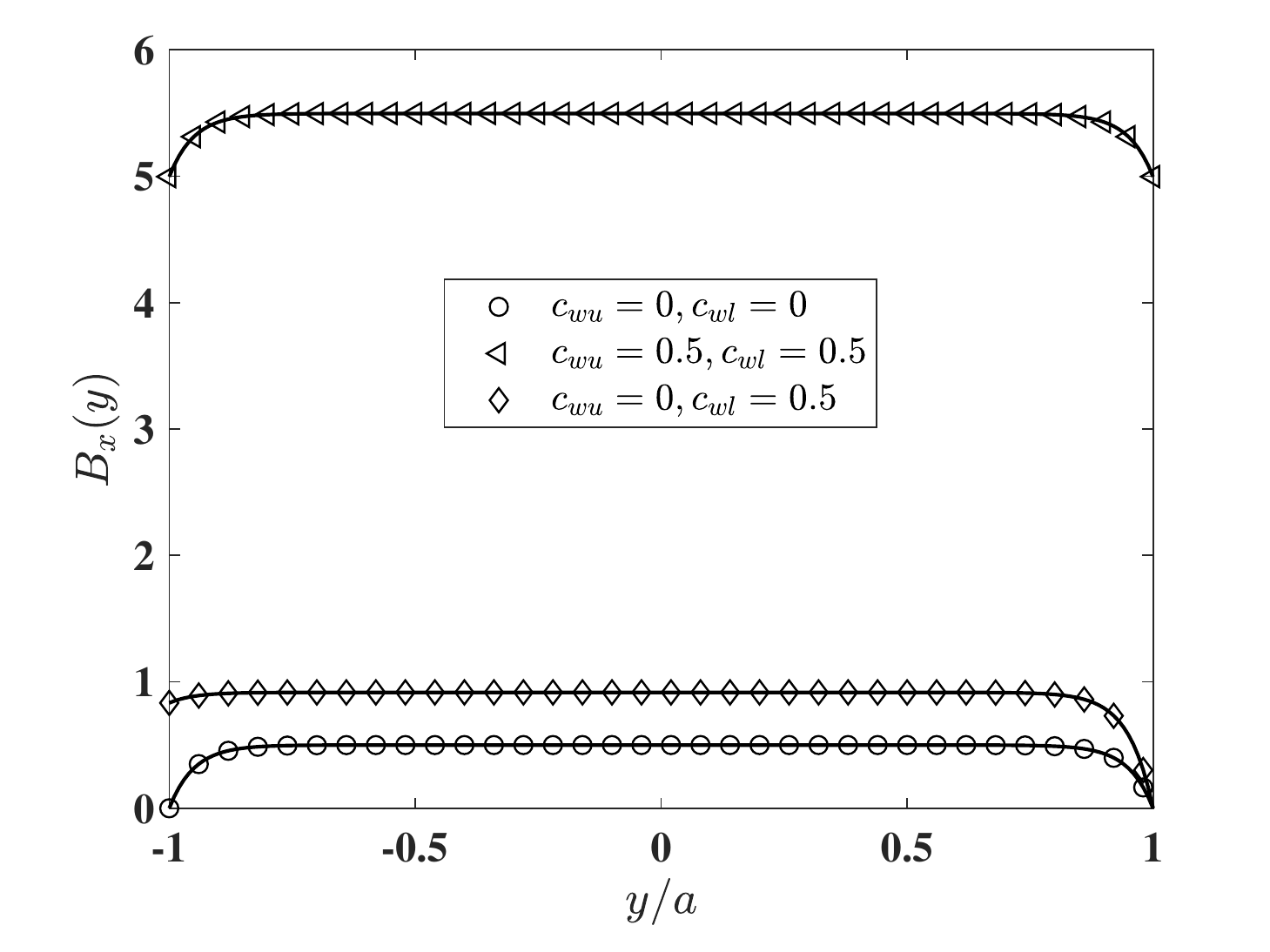}
        \label{fig:9b} } \\
                \advance\leftskip0cm
    \caption{The magnetic field profiles $B_x(y)$ across the channel for the shear-driven Hartmann-Couette flow bounded by upper (u) and lower (l) conducting walls for different pairs of values of the wall conductance ratios $(c_{wu}, c_{wl})=(0, 0)$, $(c_{wu}, c_{wl})=(0.5, 0.5)$, and $(c_{wu}, c_{wl})=(0, 0.5)$ at Hartmann numbers (a) $\mbox{Ha}=5$ and (b) $\mbox{Ha}=20$ computed via the moment-based boundary scheme (lines) compared with the analytical solution (symbols) given in Eqs.~(\ref{eq:1DMHDmovingwallfinalsolution}) and~(\ref{eq:1DMHDmovingwallintconstants}) in~\ref{sec:analyticalsolutionMHDCouetteflow}.}
    \label{fig:figBxHa5Ha20MomentMHDCouetteflow}
\end{figure}

\begin{figure}[h!]
\centering
\advance\leftskip-1.7cm
    \subfloat[$\mbox{Ha}=5$] {
        \includegraphics[width=0.4\textwidth] {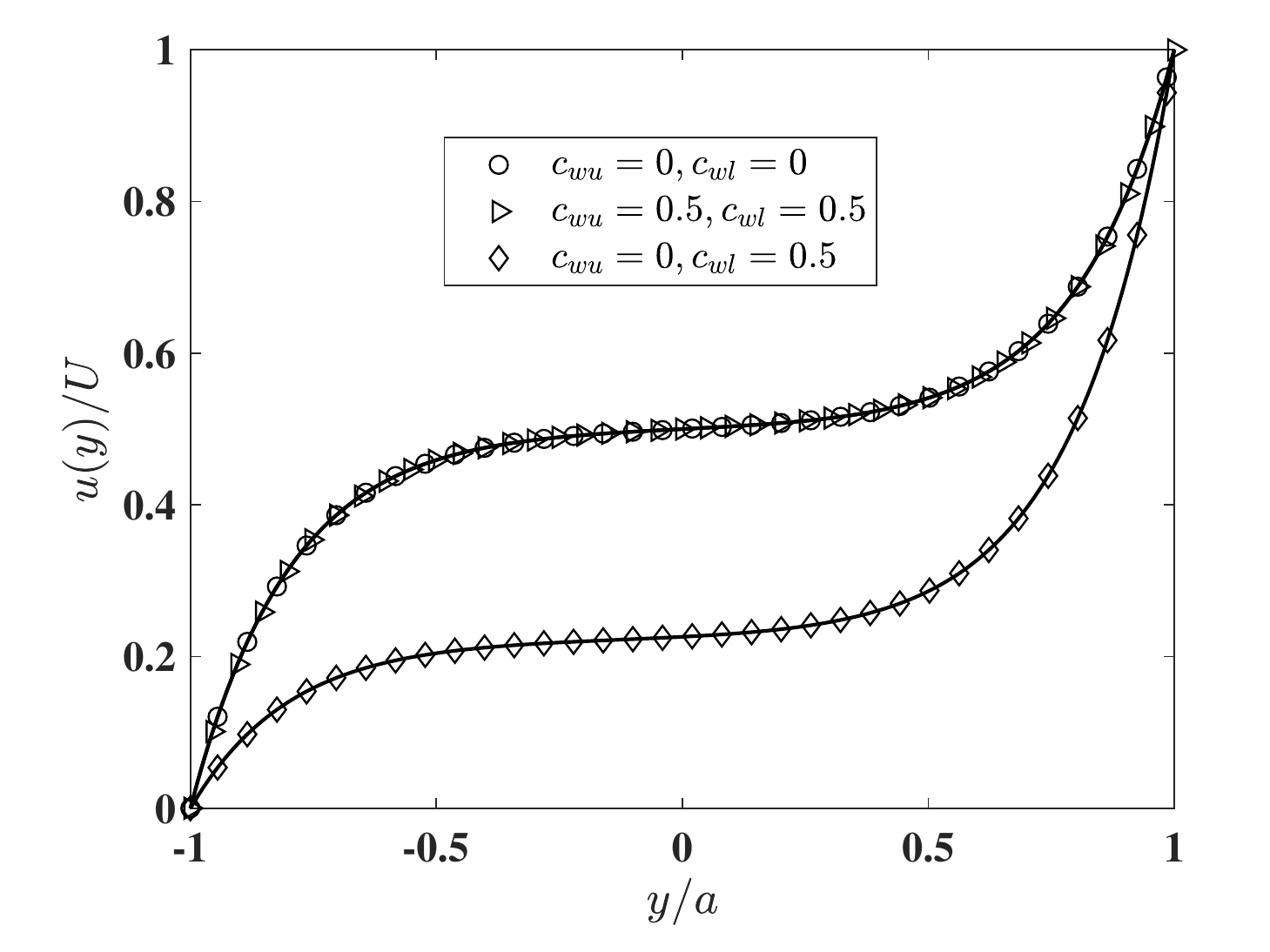}
        \label{fig:10a} } 
    \subfloat[$\mbox{Ha}=20$] {
        \includegraphics[width=0.4\textwidth] {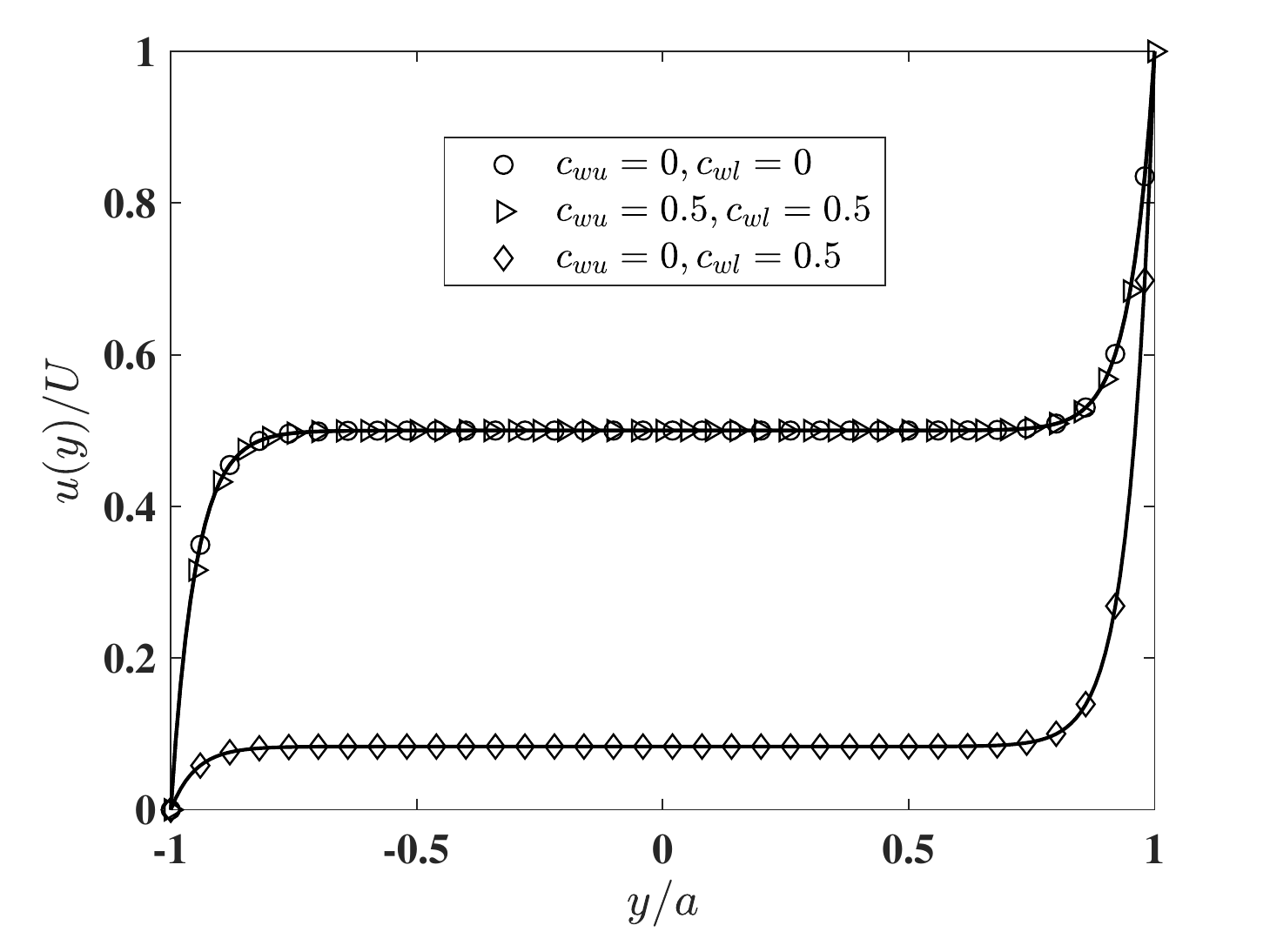}
        \label{fig:10b} } \\
                \advance\leftskip0cm
    \caption{The velocity field profiles $u(y)$ across the channel for the shear-driven Hartmann-Couette flow bounded by upper (u) and lower (l) conducting walls for different pairs of values of the wall conductance ratios $(c_{wu}, c_{wl})=(0, 0)$, $(c_{wu}, c_{wl})=(0.5, 0.5)$, and $(c_{wu}, c_{wl})=(0, 0.5)$ at Hartmann numbers (a) $\mbox{Ha}=5$ and (b) $\mbox{Ha}=20$ computed via the moment-based boundary scheme (lines) compared with the analytical solution (symbols) given in Eqs.~(\ref{eq:1DMHDmovingwallfinalsolution}) and~(\ref{eq:1DMHDmovingwallintconstants}) in~\ref{sec:analyticalsolutionMHDCouetteflow}.}
\label{fig:figuxHa5Ha20MomentMHDCouetteflow}
\end{figure}

Simulations performed using the same set of parameters as mentioned above using the moment-based scheme for moving walls given in~\ref{sec:bcsmovingwallsmomentbased} demonstrate that the resulting induced magnetic field and the velocity field compare very well with the analytical solutions as shown in Figs.~\ref{fig:figBxHa5Ha20MomentMHDCouetteflow} and~\ref{fig:figuxHa5Ha20MomentMHDCouetteflow}.

\begin{figure}[h!]
\centering
\advance\leftskip-1.7cm
    \subfloat[] {
        \includegraphics[width=0.4\textwidth] {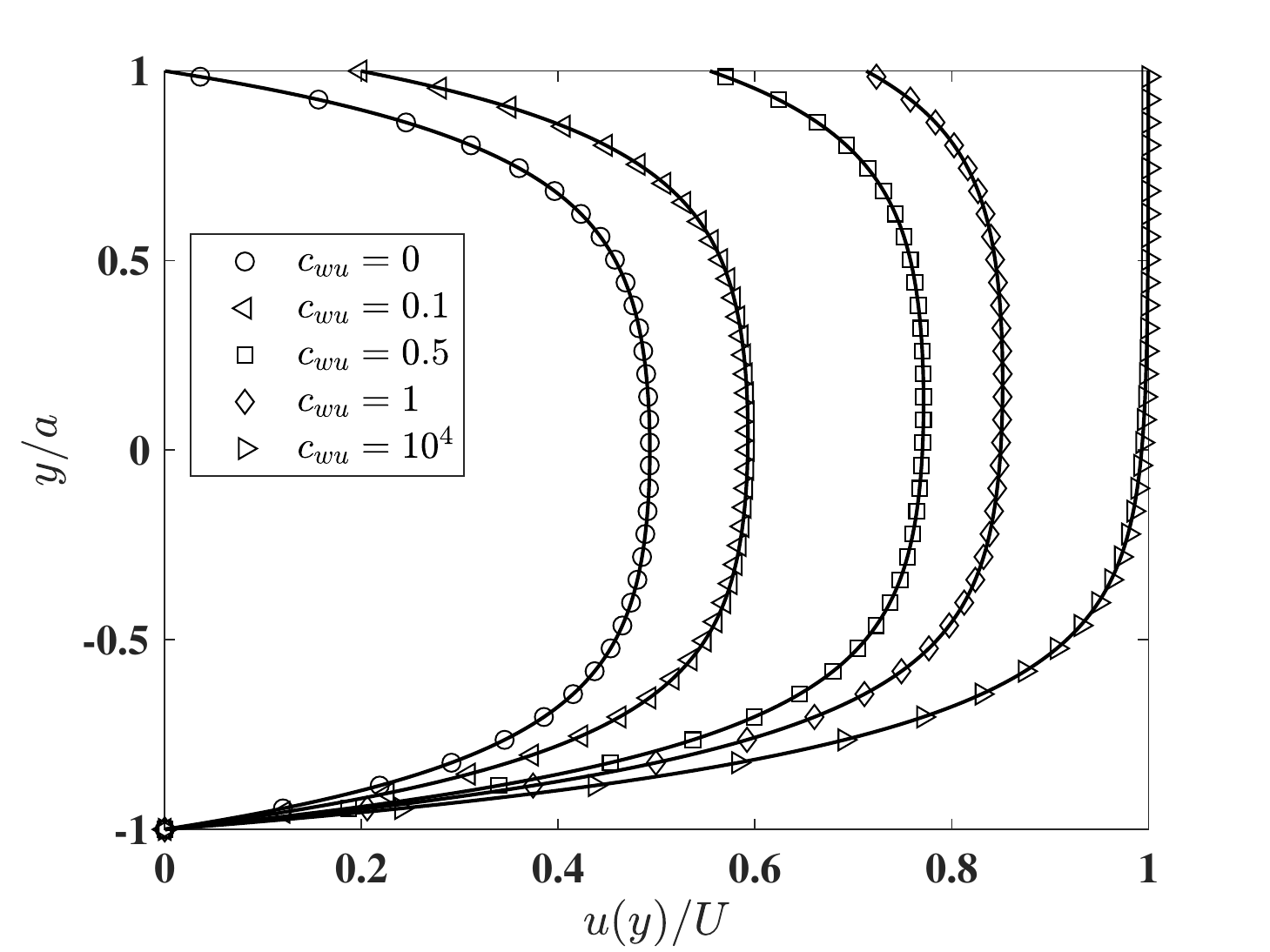}
        \label{fig:11a} } 
    \subfloat[] {
        \includegraphics[width=0.4\textwidth] {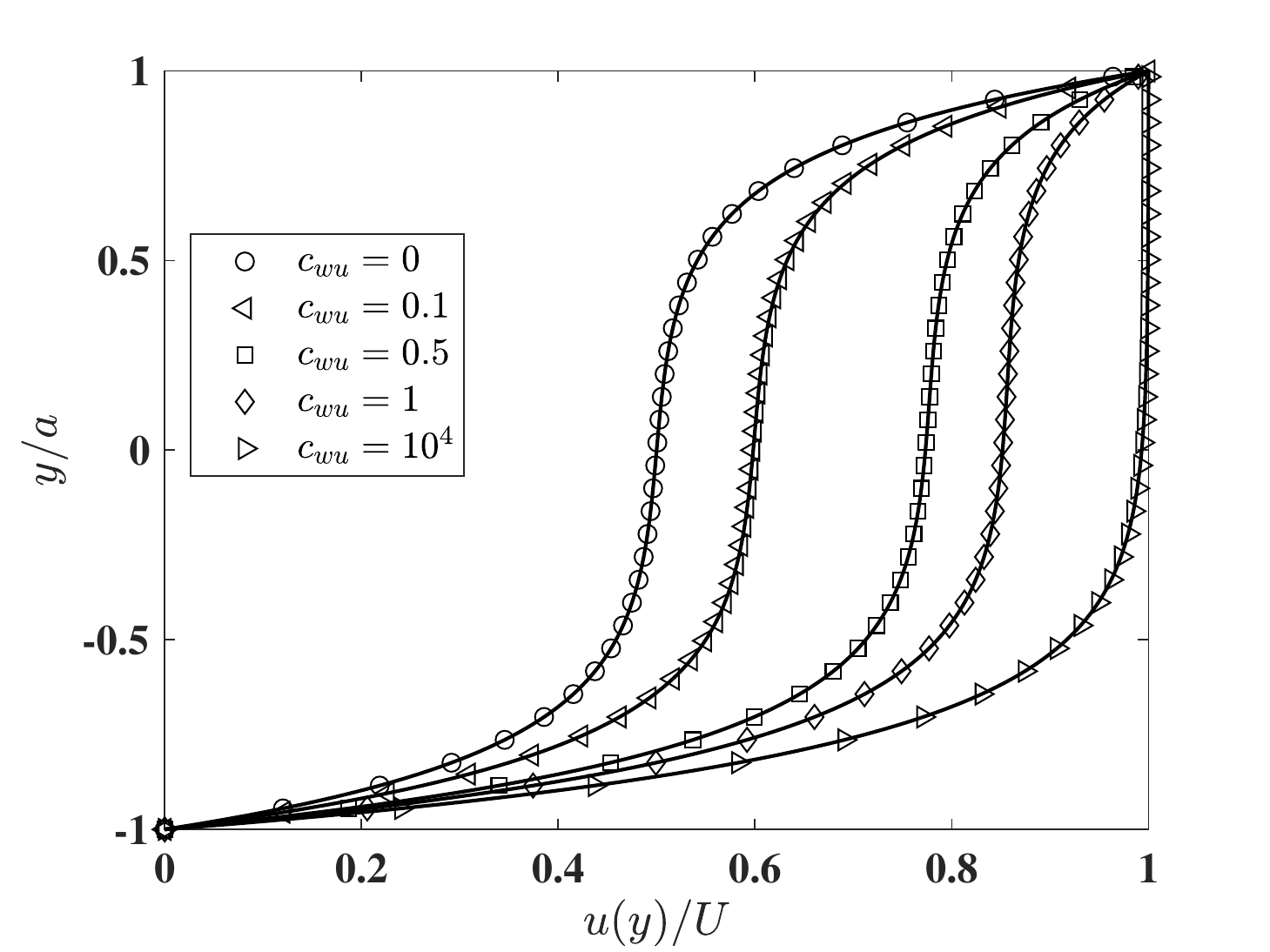}
        \label{fig:11b} } \\
                \advance\leftskip0cm
    \caption{The (a) magnetic field $B_x(y)$ and (b) velocity field $u(y)$ across the channel for the shear-driven Hartmann-Couette flow bounded by upper (u) and lower (l) conducting walls at a fixed $\mbox{Ha}=5$ and lower wall conductance ratio $c_{wl}=0.0$ (insulated) with varying values of the upper wall conductance ratio $c_{wu}=0, 0.1, 0.5, 1.0$ and $10^4$ computed via the moment-based boundary scheme (lines) compared with the analytical solution (symbols) given in Eqs.~(\ref{eq:1DMHDmovingwallfinalsolution}) and~(\ref{eq:1DMHDmovingwallintconstants}) in~\ref{sec:analyticalsolutionMHDCouetteflow}.}
    \label{fig:figBxuxHa5varyingcwuMomentMHDCouetteflow}
\end{figure}

\begin{figure}[h!]
\centering
\advance\leftskip-1.7cm
    \subfloat[] {
        \includegraphics[width=0.4\textwidth] {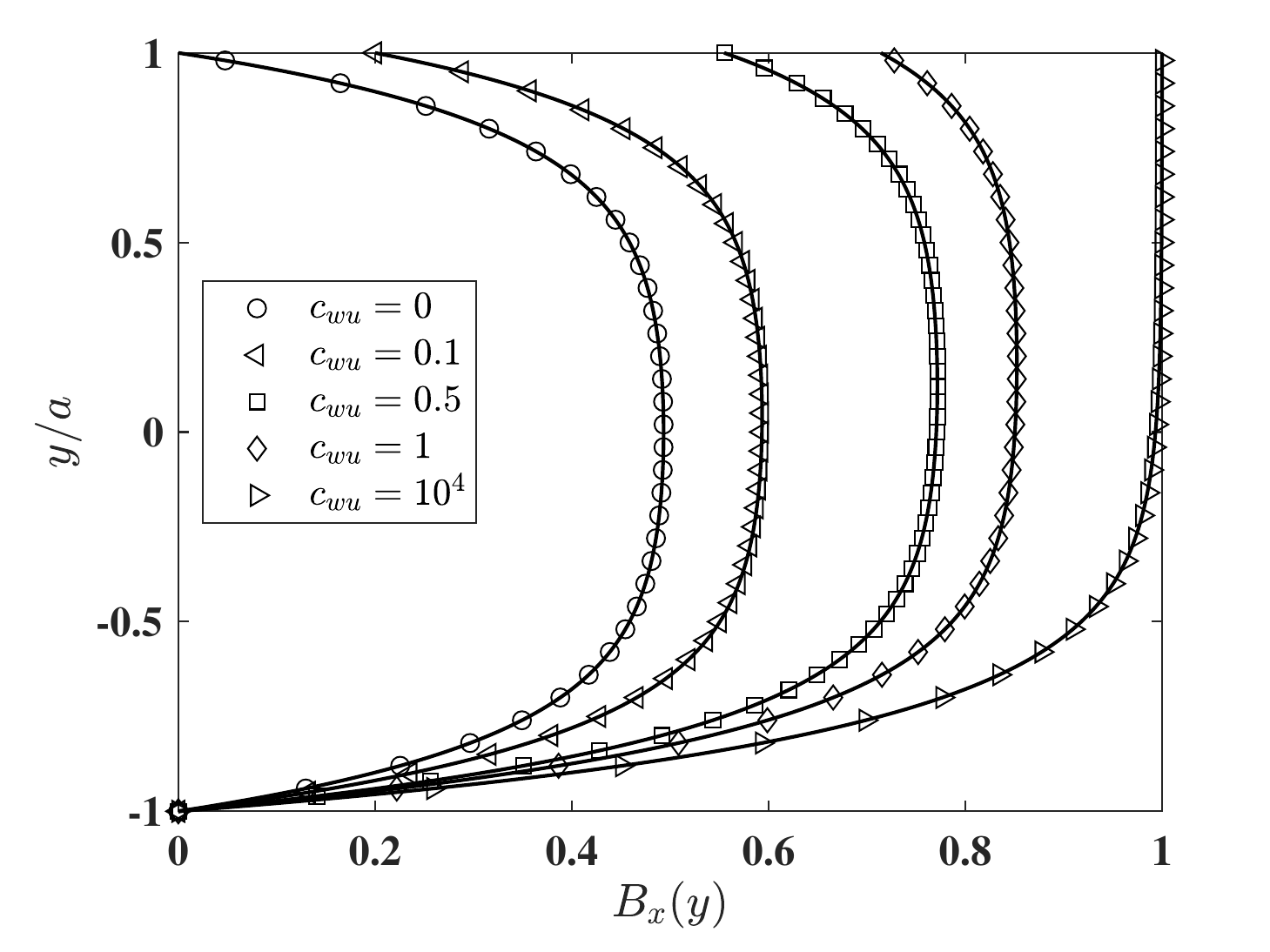}
        \label{fig:11a} } 
    \subfloat[] {
        \includegraphics[width=0.4\textwidth] {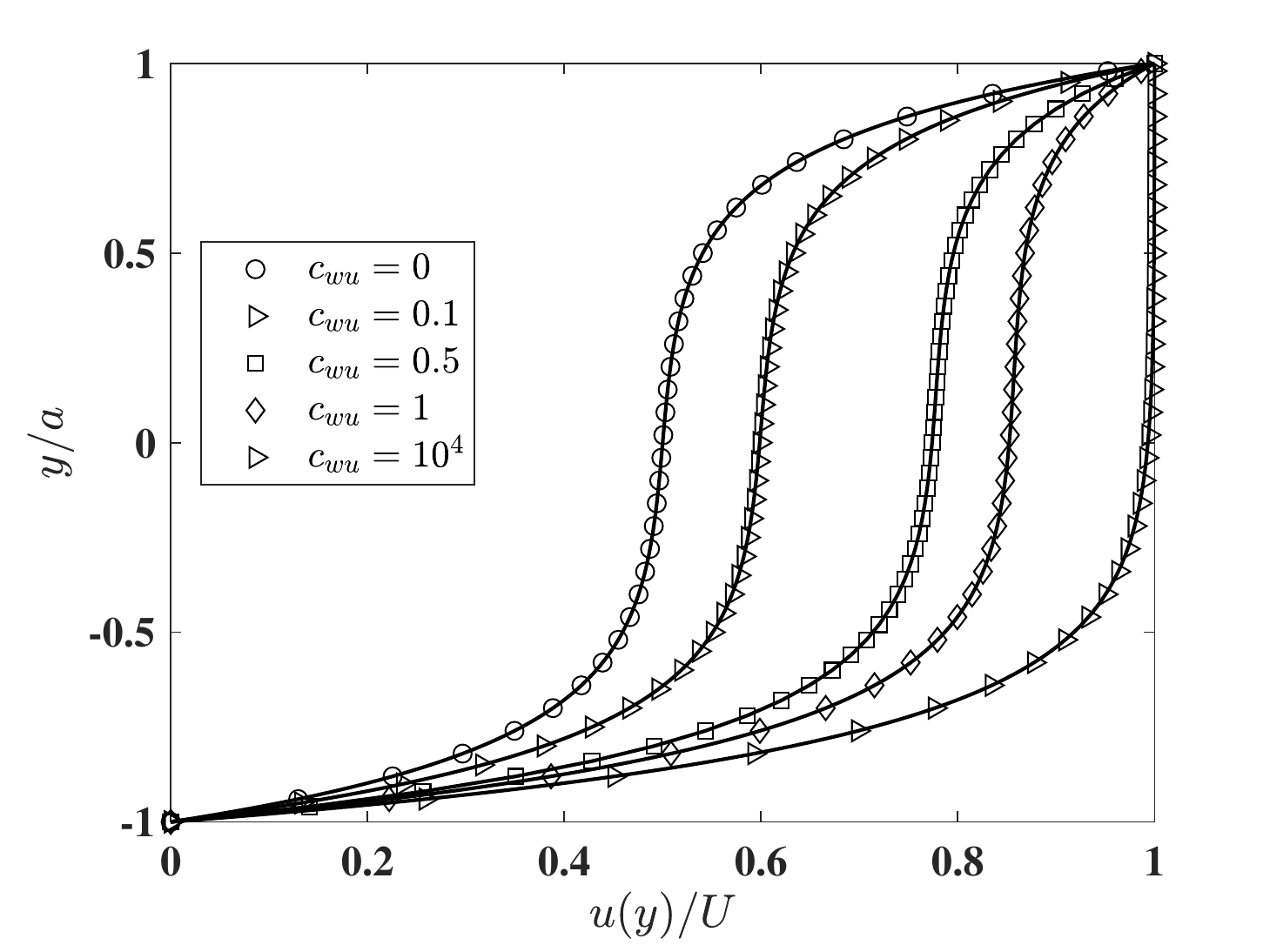}
        \label{fig:11b} } \\
                \advance\leftskip0cm
    \caption{The (a) magnetic field $B_x(y)$ and (b) velocity field $u(y)$ across the channel for the shear-driven Hartmann-Couette flow bounded by upper (u) and lower (l) conducting walls at a fixed $\mbox{Ha}=5$ and lower wall conductance ratio $c_{wl}=0.0$ (insulated) with varying values of the upper wall conductance ratio $c_{wu}=0, 0.1, 0.5, 1.0$ and $10^4$ computed via the link-based boundary scheme (lines) compared with the analytical solution (symbols) given in Eqs.~(\ref{eq:1DMHDmovingwallfinalsolution}) and~(\ref{eq:1DMHDmovingwallintconstants}) in~\ref{sec:analyticalsolutionMHDCouetteflow}.}
    \label{fig:figBxuxHa5varyingcwulinkMHDCouetteflow}
\end{figure}

Figure~\ref{fig:figBxuxHa5varyingcwuMomentMHDCouetteflow} demonstrates the ability of the moment-based boundary scheme in accurately representing the analytical predictions for a case with fixing $\mbox{Ha}=5$ and maintaining one of the walls insulated (i.e., $c_{wl}=0$), and varying the wall conductance ratio of the other wall across a wide range of values (i.e., $c_{wu}=0, 0.1, 0.5, 1.0$ and $10^4$). Notice that there are dramatic variations in the structures of both the induced magnetic fields and velocity fields as the top wall's electrical property varies from being insulated ($c_{wu}=0$) to its close to being a perfect conductor ($c_{wu}=10^4$). In particular, there is a qualitative change in the velocity profile as $c_{wu}$ is increased with the inflection points shifting towards the moving top wall and ultimately vanishing when it is perfectly conducting where a monotonic variation is observed. For completeness, shown in Fig.~\ref{fig:figBxuxHa5varyingcwulinkMHDCouetteflow} are comparisons between the results obtained using the link-based boundary scheme and the analytical solution for the induced magnetic field and velocity profiles for this MHD shear flow problem using the same conditions given below, which confirms excellent agreement between them. Clearly, the complex interactions among the magnetic and velocity fields, as well as the electrical properties of the walls in such MHD flows are well reproduced by our new boundary schemes for LB approaches.


\section{Summary, Conclusions and Outlook}\label{sec:summaryandconclusions}
The electrical properties of the bounding solid wall surfaces in MHD play an important role in determining the structure of the induced magnetic field and the velocity field. The finite electrical conductivities of the walls can be effectively represented using the Shercliff thin wall boundary condition, which is parameterized by a wall conductance ratio. In this paper, we have presented two new boundary implementation schemes for such a conducting wall condition for the LB algorithm for the magnetic field induction. These include a link-based approach and an on-node moment-based method which provide the necessary closure relations for the incoming distribution functions for both stationary and moving conducting walls. Numerical simulations involving body force-driven as well as shear-driven MHD flows validate the accuracy of both the boundary schemes for a wide range of the wall conductance ratio. For the latter shear-driven MHD flow case, we have derived new analytical solutions under a more general situation involving different conductance ratios for different bounding wall surfaces, which exhibit interesting flow physics dependent on the contrasts in such properties. The numerical predications based on the LB boundary schemes are also shown to be in good agreement with exact solutions for such more general cases. In addition, a second order grid convergence of the induced magnetic field as well as the velocity field using both the proposed boundary formulations is demonstrated.

We note here that the moment-based boundary scheme preserves the necessary constraints the moments at the boundary, which coincides with the lattice nodes, exactly, but are limited to straight boundaries. On the other hand, the present link-based boundary formulation is applicable for curved boundaries provided the boundary location is considered half-way between the lattice nodes. Both the boundary schemes are readily applicable in three-dimensions and with other more general collision models.

Several extensions of this work are envisaged, which are subjects for future investigations. These include extending the link-based approach for curved boundaries with general cut-link distances via suitable interpolations for performing LB simulations of MHD with conducting walls in complex geometries. Moreover, it may be beneficial to formulate link-based boundary schemes involving multiple relaxation times for the magnetic field~\cite{dellar2009moment}, or to develop a new two-relaxation time based LB scheme~\cite{ginzburg2008two} for the magnetic induction equation for simulations of wall-bounded MHD flows. Also, adapting the LB methods based on central moments~\cite{geier2006cascaded,premnath2011three} that have been extended for inhomogeneous and anisotropic flows~\cite{yahia2021central,yahia2021three,yahia2021phdthesis}, could enable efficient and robust simulations of MHD flows associated with boundary layers. In addition, performing MHD simulations by solving the magnetic induction equations within both the liquid and the solid directly, with generally different electrical conductivities in each domain, and then imposing the appropriate conjugate continuity conditions for the electric, magnetic and velocity fields at their boundary locations may be useful to consider in certain situations. With significant increase in focus on fusion energy related research currently and in the near future, and complemented by other applications, there is a major potential for LB approaches for simulating and studying wall-confined MHD flow physics with general conducting walls under a wide range of conditions.

\section*{Acknowledgements}
The second author would like to acknowledge the support of the US National Science Foundation (NSF) for research under Grant CBET-1705630 and for support of the development of a computer cluster infrastructure through the project ``CC* Compute: Accelerating Science and Education by Campus and Grid Computing'' under Award 2019089.

\appendix
\section{Moment-based Velocity Boundary Condition Formulation}\label{sec:momentbasedBCvelocity}
Defining the moment of order $(m+n)$ as $\Pi_{mn}=\sum_{\alpha=0}^{8}f_\alpha e_{\alpha x}^me_{\alpha y}^n$, the standard D2Q9 lattice supports the following nine independent moments $\Pi_{00}$, $\Pi_{10}$, $\Pi_{01}$, $\Pi_{20}$, $\Pi_{02}$, $\Pi_{11}$, $\Pi_{21}$, $\Pi_{12}$, and $\Pi_{22}$, out of which the first six influence the hydrodynamics directly and the remaining three are the kinetic or ghost moments. Let's now first consider the bottom wall at $y=-a$, for which the unknown distribution functions are $f_2$, $f_5$, and $f_6$ (see Fig.~\ref{fig:fig_D2Q9_D2Q5_lattice}(b)), requiring the specification of three independent moment relations. In this regard, we present the groupings of the moments based on linearly independent combinations of the unknowns as follows:

\begin{table}[h]
\small
\centering
\captionsetup{justification=centering}
\begin{tabular}{c | c}
\hline
 Combinations of unknown distributions &  Moments \\
\hline\hline
 $f_2+f_5+f_6$ & $\Pi_{00},\boxed{\Pi_{01}}, \Pi_{02}$\\
  $f_5-f_6$ & $\boxed{\Pi_{10}},\Pi_{11}, \Pi_{12}$\\
  $f_5+f_6$ & $\boxed{\Pi_{20}},\Pi_{21}, \Pi_{22}$\\
\end{tabular}
\label{tab:1}
\end{table}

We choose one moment from each category or row to establish a linearly independent system of equations for the unknowns. When appropriate, we choose only the lower order, i.e., the hydrodynamic moments~\cite{bennett2010lattice}. In this case, we constrain the wall normal component of momentum, and wall parallel components of the momentum and momentum flux based on the conditions at the stationary boundary, which are boxed in the above listings. Thus, invoking the no-slip boundary conditions on $\bm{x}_b$, i.e., $\bm{u}=(u_x,u_y)=\bm{0}$, and taking into account $\Pi_{20}^{(1)}(\bm{x}_b)=0$, since via a C-E analyais it is related to $\partial_x u_x$ which vanishes on the boundary, we specify
\begin{equation}
\Pi_{10}(\bm{x}_b)=0,\quad\quad \Pi_{01}(\bm{x}_b)=0, \quad\quad \Pi_{20}(\bm{x}_b)=\Pi_{20}^{(0)}(\bm{x}_b)=c_s^2\rho(\bm{x}_b).\label{eq:hydmomentconstraints}
\end{equation}
Here, we need an estimate for the boundary density $\rho(\bm{x}_b)$. Writing $\rho(\bm{x}_b)=\Pi_{00}(\bm{x}_b)=\sum_{K}f_\alpha(\bm{x}_b)+\sum_{U}f_\alpha(\bm{x}_b)$, where $K=\{0,1,3,4,7,8\}$ and $U=\{2,5,6\}$ are the sets of the known and unknown directions, respectively. From $\Pi_{01}(\bm{x}_b)=0$, it follows that $\sum_{U}f_\alpha(\bm{x}_b)=\sum_{K_1}f_\alpha(\bm{x}_b)$, where $K_1=\{4,7,8\}$ is a subset of the known directions. Based on these last two equations, it follows that $\rho=\sum_{K}f_\alpha(\bm{x}_b)+2\sum_{K_1}f_\alpha(\bm{x}_b)$. So, we can exactly estimate the boundary density as
\begin{equation}
\rho(\bm{x}_b)=f_0(\bm{x}_b)+f_1(\bm{x}_b)+f_3(\bm{x}_b)+2\left[f_4(\bm{x}_b)+f_7(\bm{x}_b)+f_8(\bm{x}_b)\right].
\end{equation}
Then, writing the left sides of each of the three sub-equations in Eq.~(\ref{eq:hydmomentconstraints}) based on their definitions in terms of the distribution functions given at the beginning of this section and evaluated at the boundary, we get
\begin{subequations}\label{eq:hydmomentconstraints1}
\begin{eqnarray}
  \Pi_{10}(\bm{x}_b) &=& f_1(\bm{x}_b)-f_3(\bm{x}_b)+\boxed{f_5(\bm{x}_b)}-\boxed{f_6(\bm{x}_b)}-f_7(\bm{x}_b)+f_8(\bm{x}_b)=0, \\
  \Pi_{01}(\bm{x}_b) &=& \boxed{f_2(\bm{x}_b)}-f_4(\bm{x}_b)+\boxed{f_5(\bm{x}_b)}+\boxed{f_6(\bm{x}_b)}-f_7(\bm{x}_b)-f_8(\bm{x}_b)=0, \\
  \Pi_{20}(\bm{x}_b) &=& f_1(\bm{x}_b)+f_3(\bm{x}_b)+\boxed{f_5(\bm{x}_b)}+\boxed{f_6(\bm{x}_b)}+f_7(\bm{x}_b)+f_8(\bm{x}_b)=c_s^2\rho(\bm{x}_b),
\end{eqnarray}
\end{subequations}
where the variables within the boxes are the unknown quantities to be determined, while the rest are known. Solving the above three moment-closure equations given in Eq.~(\ref{eq:hydmomentconstraints1}) for the unknown or incoming distribution functions shown within the boxes, we finally get
\begin{eqnarray*}
  f_2(\bm{x}_b) &=& -c_s^2\rho(\bm{x}_b)+ f_1(\bm{x}_b)+f_3(\bm{x}_b)+f_4(\bm{x}_b)+2\left[f_7(\bm{x}_b)+f_8(\bm{x}_b)\right], \\
  f_5(\bm{x}_b) &=&  \frac{1}{2}c_s^2\rho(\bm{x}_b)-\left[f_1(\bm{x}_b)+f_8(\bm{x}_b)\right], \\
  f_6(\bm{x}_b) &=&  \frac{1}{2}c_s^2\rho(\bm{x}_b)-\left[f_3(\bm{x}_b)+f_7(\bm{x}_b)\right].
\end{eqnarray*}

Similar approach can be used for obtaining the necessary relations for the incoming distribution functions at the top wall at $y=+a$, which yield
\begin{equation*}
\rho(\bm{x}_b)=f_0(\bm{x}_b)+f_1(\bm{x}_b)+f_3(\bm{x}_b)+2\left[f_2(\bm{x}_b)+f_5(\bm{x}_b)+f_6(\bm{x}_b)\right],
\end{equation*}
and
\begin{eqnarray*}
  f_4(\bm{x}_b) &=& -c_s^2\rho(\bm{x}_b)+ f_1(\bm{x}_b)+f_2(\bm{x}_b)+f_3(\bm{x}_b)+2\left[f_5(\bm{x}_b)+f_6(\bm{x}_b)\right], \\
  f_7(\bm{x}_b) &=&  \frac{1}{2}c_s^2\rho(\bm{x}_b)-\left[f_3(\bm{x}_b)+f_6(\bm{x}_b)\right], \\
  f_8(\bm{x}_b) &=&  \frac{1}{2}c_s^2\rho(\bm{x}_b)-\left[f_1(\bm{x}_b)+f_5(\bm{x}_b)\right].
\end{eqnarray*}

\section{Boundary Schemes for the Lattice Kinetic Equation for the Magnetic Field for Moving Walls with Finite Electrical Conductivities}\label{sec:bcsmovingwalls}

\subsection{Link-based implementation}\label{sec:bcsmovingwallslinkbased}
Let us consider that the boundary is moving at a velocity $U_w$ parallel to the tangential direction $\hat{t}$ of the wall (see Fig.~\ref{fig:fig_schematic_Shercliff_BC}). Then, the boundary conditions on the fluid velocity components normal and tangential directions are $u_n(\bm{x}_b)=0$ and $u_t(\bm{x}_b)=U_w$, respectively. In the following, for convenience, we take $i$ and $j$ directions to correspond to the tangential and normal directions, respectively, i.e., $i=\hat{t}$ and $j=\hat{n}$. The imposed magnetic field $B_0$ is along the normal direction, i.e., $B_n=B_0$. Then, the equilibrium part of the anti-symmetric tensor $\Lambda_{ji}$ on moving boundary is given by $\Lambda_{ji}^{(0)}(\bm{x}_b)=u_n(\bm{x}_b)B_t(\bm{x}_b)-B_n(\bm{x}_b)u_t(\bm{x}_b)=-B_n(\bm{x}_b)u_t(\bm{x}_b)$. As a result, for a moving wall, from the above it follows that the equilibrium part of the $\Lambda_{ji}$ has a non-zero contribution given by
\begin{equation}
\Lambda_{ji}^{(0)}(\bm{x}_b)=-B_0U_w.
\end{equation}
Using this expression in the two equations in Eq.~(\ref{sumdiffgmacrovars}) and simplifying under the standard assumptions made in Sec.~\ref{sec:linkbasedShercliffBC}, we get
\begin{subequations}\label{eq:sumdiffgmovingwall}
\begin{eqnarray}
 g_{\alpha i}(\bm{x}_b) + g_{\bar{\alpha} i}(\bm{x}_b) &\approx& 2W_{\alpha}B_i(\bm{x}_b), \\
 g_{\alpha i}(\bm{x}_b) - g_{\bar{\alpha} i}(\bm{x}_b) &\approx&-2W_\alpha\frac{e_{\alpha n}}{c_{sm}^2}B_0U_w-2W_\alpha\tau_m\delta_te_{\alpha_k}\partial_kB_i(\bm{x}_b).
\end{eqnarray}
\end{subequations}
Then, following the steps given in Sec.~\ref{sec:linkbasedShercliffBC} and using the above updated equations for the moving walls, the two equations in Eq.~(\ref{eq:magneticfieldanditsgradientrelation}) modify to
\begin{subequations}\label{eq:magfieldanditsgradientgmovingwall}
\begin{eqnarray}
  \frac{1}{2}\left[g_{\alpha i}(\bm{x}_f,t+\delta_t)+\tilde{g}_{\bar{\alpha} i}(\bm{x}_f,t)\right] &=& W_\alpha B_i(\bm{x}_b), \\
  -\frac{1}{2\left(\tau_m-\frac{1}{2}\right)\delta_t}\left[g_{\alpha i}(\bm{x}_f,t+\delta_t)-\tilde{g}_{\bar{\alpha} i}(\bm{x}_f,t)+2W_\alpha\frac{e_{\alpha n}}{c_{sm}^2}B_0U_w\right] &=& W_\alpha\delta_te_{\alpha j}\partial_j  B_i(\bm{x}_b).
\end{eqnarray}
\end{subequations}
When these two equations in the above Eq.~(\ref{eq:magfieldanditsgradientgmovingwall}) are substituted in the Shercliff boundary condition (see Eq.~(\ref{eq:ShercliffBClocalcoordinateLBMmodified}) given earlier) and taking into account the weighting factor $\theta$ defined in Eq.~(\ref{eq:weightingfactor}), we obtain the constraint relation between the unknown incoming distribution function $g_{\alpha i}(\bm{x}_f,t+\delta_t)$ and the known outgoing post-collision distribution function $\tilde{g}_{\bar{\alpha} i}(\bm{x}_f,t)$ augmented for the effect of the moving wall as
\begin{equation}\label{eq:constraintrelationmovingwall}
(1+\theta)g_{\alpha i}(\bm{x}_f,t+\delta_t)+(1-\theta)\tilde{g}_{\bar{\alpha} i}(\bm{x}_f,t)+2\theta W_\alpha\frac{e_{\alpha n}}{c_{sm}^2}B_0U_w=0.
\end{equation}
This equation (Eq.~(\ref{eq:constraintrelationmovingwall})) can then be rewritten as follows providing the required link-based closure for the distribution function for the magnetic field involving a moving wall:
\begin{equation}\label{eq:movingwalllinkbasedclosure}
g_{\alpha i}(\bm{x}_f,t+\delta_t)=\left(\frac{\theta-1}{\theta+1}\right)\tilde{g}_{\bar{\alpha} i}(\bm{x}_f,t)-\underline{\left(\frac{2\theta}{\theta+1}\right)  W_\alpha\frac{e_{\alpha n}}{c_{sm}^2}B_0U_w},
\end{equation}
where the underlined term expresses the effect of the shearing motion of the wall on the magnetic distribution function. The limiting or degenerate cases of this result can then be summarized as follows:

(a) Insulated wall: $c_w=\theta = 0$, which corresponds to
\begin{equation*}
   g_{\alpha i}(\bm{x}_f,t+\delta_t)=-\tilde{g}_{\bar{\alpha} i}(\bm{x}_f,t),
\end{equation*}

(b) Perfectly conducting wall: $c_w\;\mbox{or}\;\theta \rightarrow \infty$, which corresponds to
\begin{equation*}
   g_{\alpha i}(\bm{x}_f,t+\delta_t)=+\tilde{g}_{\bar{\alpha} i}(\bm{x}_f,t)-2W_\alpha\frac{e_{\alpha n}}{c_{sm}^2}B_0U_w,
\end{equation*}
Clearly, when the wall is insulated, its motion does not influence the incoming magnetic distribution function. On the other hand, the largest effect of the wall motion is introduced when the wall is perfectly conducting. Let's now illustrate the above result given in Eq.~(\ref{eq:movingwalllinkbasedclosure}) for the Hartmann-Couette flow between two parallel plates separated by a length $2a$ and subjected to an imposed magnetic field $B_0$ normal to the walls, where the top plate at $y=a$ is moving at a velocity $U_w$, while the bottom on at $y=-a$ is at rest. Focusing on the top wall, and using $\alpha=4$, $\bar{\alpha}=2$, $e_{\alpha n}=e_{4 y}=-1$ and $W_{\alpha}=1/6$, we get
\begin{eqnarray*}
  y=+a: \quad g_{4x} &=& \frac{\theta - 1}{\theta + 1} \tilde{g}_{2x}+\frac{\theta}{3(\theta+1)}\frac{B_0U_w}{c_{sm}^2}, \\
              g_{4y} &=& -\tilde{g}_{4y} + 2W_4B_0,
\end{eqnarray*}
which replaces the corresponding results given at the end of Sec.~\ref{sec:linkbasedShercliffBC}, while the closure for the stationary bottom wall remains the same as before.

It may be noted that the incoming distribution functions for the velocity field can be expressed via the standard half-way bounce back scheme augmented with a momentum correction for the moving wall~\cite{ladd1994numerical} given by $f_{\alpha}(\bm{x}_f,t+\delta_t)=\tilde{f}_{\bar{\alpha}}(\bm{x}_f,t)+2w_\alpha e_{\alpha t}U_w/c_{s}^2$.

\subsection{Moment-based implementation}\label{sec:bcsmovingwallsmomentbased}
As in the previous formulation given above, consider the Hartmann-Couette flow bounded by conducting walls, where the top plate at $y=a$ is moving at a velocity $U_w$. Then, the boundary conditions on the components of the velocity and magnetic fields at the moving boundary is given by
\begin{equation}
  u_x=U_w,\quad u_y=0, \quad +\frac{\partial B_x}{\partial y}+\frac{B_x}{c_w a} = 0, \quad B_y = B_0 \quad\quad \mbox{at}\;\;y = +a.
\end{equation}
As before (see Sec.~\ref{sec:momentbasedShercliffBC}), we first write the decomposition of the tensor component $\Lambda_{yx}$ as $\Lambda_{yx}=\Lambda_{yx}^{(0)}+\delta \Lambda_{yx}^{(1)}$. Here, unlike in the previous case for the moving boundary at $\bm{x}_b$, its equilibrium part becomes
\begin{equation}
\Lambda_{yx}^{(0)}(\bm{x}_b)=u_y(\bm{x}_b)B_x(\bm{x}_b)-B_y(\bm{x}_b)u_x(\bm{x}_b)=-B_y(\bm{x}_b)u_x(\bm{x}_b)=-B_0U_w.
\end{equation}
For the non-equilibrium part, we have $\Lambda_{yx}^{(1)}(\bm{x}_b)=-\tau_mc_{sm}^2\partial_yB_x(\bm{x}_b)$. Thus, it follow that
\begin{equation}
\Lambda_{yx}(\bm{x}_b)=-B_0U_w-\delta_t\tau_mc_{sm}^2\frac{\partial B_x}{\partial y}(\bm{x}_b),
\end{equation}
or
\begin{equation}\label{eq:magfieldgradientformovingwall}
\frac{\partial B_x}{\partial y}(\bm{x}_b)=-\frac{1}{\tau_mc_{sm}^2\delta_t}\left[\Lambda_{yx}(\bm{x}_b)+B_0U_w\right].
\end{equation}

Next, writing the definitions of the relevant moments of $g_{\alpha x}$ for $B_x$ and $\Lambda_{yx}$ (see Sec.~\ref{sec:momentbasedShercliffBC}) along with the above Eq.~(\ref{eq:magfieldgradientformovingwall}) in the Shercliff boundary condition $\partial B_x/\partial y+B_x/(c_w a)=0$ at $\bm{x}_b$, we get
\begin{equation}
-\frac{1}{\tau_mc_{sm}^2\delta_t}\left[(g_{2x}-g_{4x})+B_0U_w\right]+\frac{1}{c_w a}(g_{0x}+g_{1x}+g_{2x}+g_{3x}+g_{4x})=0.
\end{equation}
Then, solving for the unknown incoming distribution function $g_{4x}$, we finally obtain the required moment-based closure at the boundary as
\begin{equation}\label{eq:momentbasedclosureformovingwall}
  g_{4 x} = \left[\frac{1}{\tau_mc_{sm}^2\delta_t}+\frac{1}{c_w a}\right]^{-1}\left\{\left[\frac{1}{\tau_mc_{sm}^2\delta_t}-\frac{1}{c_w a}\right]g_{2 x}-\frac{1}{c_w a}(g_{0 x}+g_{1 x}+g_{3 x})+\underline{\frac{B_0U_w}{\tau_m c_{sm}^2\delta_t}}\right\},
\end{equation}
where the underlined term represents the effect of the moving wall. It may be noted that the expression of the unknown distribution function $g_{4y}$ based on maintaining the constraint $\sum_{\alpha=0}^4g_{\alpha y}=B_0$ remains the same as that given in Sec.~\ref{sec:momentbasedShercliffBC}. The limiting special cases of Eq.~(\ref{eq:momentbasedclosureformovingwall}) are then given as follows:
\begin{eqnarray*}
c_w=0:\quad g_{4x} &=& -(g_{0x}+g_{1x}+g_{2x}+g_{3x})\quad\quad\quad\quad \mbox{(Perfectly insulating wall)},\nonumber\\
c_w\rightarrow\infty: g_{4x} &=& g_{2x}+B_0U_w\qquad\qquad\qquad\qquad\quad\;\; \mbox{(Perfectly conducting wall)}.\nonumber
\end{eqnarray*}

On the other hand, the incoming distribution functions for the velocity field for representing the moving boundary with a velocity $U_w$ at $y=a$ can be obtained by imposing the following hydrodynamic moment constraints, amending those given in Eq.~(\ref{eq:hydmomentconstraints}) in~\ref{sec:momentbasedBCvelocity}:
\begin{equation*}
\Pi_{10}(\bm{x}_b)=0,\quad\quad \Pi_{01}(\bm{x}_b)=\rho(\bm{x}_b) U_w, \quad\quad \Pi_{20}(\bm{x}_b)=\Pi_{20}^{(0)}(\bm{x}_b)=\rho(\bm{x}_b)(c_s^2+U_w^2).
\end{equation*}
Then, following the approach presented in~\ref{sec:momentbasedBCvelocity}, we obtain the required closure relations that read as
\begin{eqnarray*}
  f_4(\bm{x}_b) &=& -\rho(\bm{x}_b)(c_s^2+U_w^2)+ f_1(\bm{x}_b)+f_2(\bm{x}_b)+f_3(\bm{x}_b)+2\left[f_5(\bm{x}_b)+f_6(\bm{x}_b)\right], \\
  f_7(\bm{x}_b) &=&  \frac{1}{2}\rho(\bm{x}_b)(c_s^2+U_w^2-U_w)-\left[f_3(\bm{x}_b)+f_6(\bm{x}_b)\right], \\
  f_8(\bm{x}_b) &=&  \frac{1}{2}\rho(\bm{x}_b)(c_s^2+U_w^2+U_w)-\left[f_1(\bm{x}_b)+f_5(\bm{x}_b)\right],
\end{eqnarray*}
where $\rho(\bm{x}_b)$ is computed from the same expression as that given at the end of~\ref{sec:momentbasedBCvelocity}.

\section{Derivation of the Analytical Solution for Hartmann-Couette Flow Bounded by Plates with Different Wall Conductance Ratios} \label{sec:analyticalsolutionMHDCouetteflow}
We will now present a systematic derivation of the analytical solution for the magnetic and velocity fields in a Hartmann-Couette flow bounded by plates with contrasts in the wall conductance ratios between them. It should be noted that exact solutions involving different wall conductance ratios on different bounding wall surfaces for the coupled MHD system are generally rare. The main objective here is to develop such an analytical solution and use it as a benchmark case for performing a validation of the new boundary schemes involving MHD shear flows under more general conditions on the electrical properties of the walls in Sec.~\ref{sec:resultsanddiscussion}. Nevertheless, we will point out later in this section that the latter leads to an interesting way to modulate the resulting structures of magnetic and velocity fields in a shear flow of liquid metals and the associated peculiar flow physics.

We consider an electrically conducting fluid with conductivity $\sigma$ and bounded by the lower and upper plates, with the wall conductance ratios of $c_{wl}$ and $c_{wu}$, respectively, and separated by a distance of $2a$. The fluid motion is set up due to shear generated by the steady motion of the top plate at a velocity $U_w$ in the $x$ direction under an external magnetic field $B_0$ applied normal to the plates in the $y$ direction, while the bottom plate is taken to be stationary. The origin of the coordinate system is placed midway between the two plates as in the other canonical problem shown in Fig.~\ref{fig:fig_schematic_Hartmann_flow_conducting_BC}. Then, under the assumption of one-dimensional, steady, laminar Stokes flow, the governing MHD equations given in Eqs.~(\ref{eq:NSE-MHD2}) and (\ref{eq:NSE-MHD3}), respectively, written for the respective $x$ component simplify to
\begin{equation}
\rho\nu\frac{d^2u_x}{dy^2}+\frac{B_0}{\mu_m}\frac{dB_x}{dy}=0.\label{eq:1DMHDsystem1}
\end{equation}
and
\begin{equation}
\frac{1}{\sigma\mu_m}\frac{d^2B_x}{dy^2}+B_0\frac{du_x}{dy}=0.\label{eq:1DMHDsystem2}
\end{equation}
The above two equations are subject to the following no-slip velocity boundary conditions for the stationary bottom wall and moving top wall at a velocity $U_w$, as well as the Shercliff conducting wall conditions for the $x$ component of the magnetic field with $c_{wl}$ and $c_{wu}$ for the wall conductance ratios at the lower and upper walls, respectively. These can be written as
\begin{subequations}\label{eq:1DMHDsystemBCs}
\begin{eqnarray}
u_x &=& 0,\quad \quad\quad -\frac{\partial B_x}{\partial y}+\frac{B_x}{c_{wl} a} = 0, \quad\quad \mbox{at}\;\;y = -a, \label{eq:1DMHDsystemBCs1}\\
u_x &=& U_w, \quad\quad +\frac{\partial B_x}{\partial y}+\frac{B_x}{c_{wu} a} = 0, \quad\quad \mbox{at}\;\;y = +a\label{eq:1DMHDsystemBCs2}.
\end{eqnarray}
\end{subequations}
Then, for convenience, defining the following dimensionless variables natural for this flow problem
\begin{equation}
\xi=\frac{y}{a},\qquad\qquad \mbox{Ha}=B_0 a \left(\frac{\sigma}{\rho\nu}\right)^{1/2}, \qquad\qquad u_x^*=\frac{u_x}{U_w}, \qquad\qquad B_x^*=\frac{B_x}{U_w(\rho\nu\sigma)^{1/2}\mu_m},
\end{equation}
Equations~(\ref{eq:1DMHDsystem1}) and (\ref{eq:1DMHDsystem2}) can be rewritten in their more compact non-dimensional forms as follows:
\begin{subequations}\label{eq:1DMHDsystemnondim}
\begin{eqnarray}
\frac{d^2u_x^*}{d\xi^2}+\mbox{Ha}\frac{dB_x^*}{d\xi}&=&0,\label{eq:1DMHDsystemnondim1}\\
\frac{d^2B_x^*}{d\xi^2}+\mbox{Ha}\frac{du_x^*}{d\xi}&=&0.\label{eq:1DMHDsystemnondim2}
\end{eqnarray}
\end{subequations}
and the boundary conditions given in Eq.~(\ref{eq:1DMHDsystemBCs}) become
\begin{subequations}\label{eq:1DMHDBCs}
\begin{eqnarray}
u_x^* &=& 0, \quad\quad -\frac{\partial B_x^*}{\partial \xi}+\frac{B_x^*}{c_{wl}} = 0, \quad\quad \mbox{at}\;\;\xi = -1, \label{eq:1DMHDBCs1}\\
u_x^* &=& 1, \quad\quad +\frac{\partial B_x^*}{\partial \xi}+\frac{B_x^*}{c_{wu}} = 0, \quad\quad \mbox{at}\;\;\xi = +1. \label{eq:1DMHDBCs2}
\end{eqnarray}
\end{subequations}

In order to solve the above set of coupled differential equations for $u_x^*(\xi)$ and $B_x^*(\xi)$, by exploiting their special structures, we introduce the following transformation variables (sometimes referred to as the Elsasser variables~\cite{elsasser1950hydromagnetic}):
\begin{equation}\label{eq:Elsassertransformation}
V_1^*=u_x^*+B_x^*,\quad\quad V_2^*=u_x^*-B_x^*.
\end{equation}
As a result, combining Eqs.~(\ref{eq:1DMHDsystemnondim1}) and (\ref{eq:1DMHDsystemnondim2}) via taking their sum and difference and using Eq.~(\ref{eq:Elsassertransformation}), we get the following intermediate de-coupled system:
\begin{eqnarray}
\frac{d^2V_1^*}{d\xi^2}+\mbox{Ha}\frac{dV_1^*}{d\xi}&=&0,\\
\frac{d^2V_2^*}{d\xi^2}-\mbox{Ha}\frac{dV_2^*}{d\xi}&=&0.
\end{eqnarray}
These two equations can be readily solved, which can be written as
\begin{equation}\label{eq:1DMHDsolutionmovingwallintrvar}
V_1^*(\xi)=m+n\exp(-\mbox{Ha}\xi), \quad\quad V_2^*(\xi)=m'+n'\exp(+\mbox{Ha}\xi),
\end{equation}
where $m$, $m'$, $n$ and $n'$ are arbitrary integration constants. We can then recover the solutions in terms of the original variables $u_x^*(\xi)$ and $B_x^*(\xi)$ via using $u_x^*=(V_1^*+V_2^*)/2$ and $B_x^*=(V_1^*-V_2^*)/2$, which yield
\begin{subequations}\label{eq:1DMHDsolutionmovingwallexpfunc}
\begin{eqnarray}
  u_x^*(\xi) &=& (p+p')+q\exp(-\mbox{Ha}\xi)+q'\exp(+\mbox{Ha}\xi), \label{eq:1DMHDsolutionmovingwallexpfunc1}\\
  B_x^*(\xi) &=& (p-p')+q\exp(-\mbox{Ha}\xi)-q'\exp(+\mbox{Ha}\xi), \label{eq:1DMHDsolutionmovingwallexpfunc2}
\end{eqnarray}
\end{subequations}
where the redefined integration constants are given by $p=m/2$, $p'=m'/2$, $q=n/2$ and $q'=n'/2$. Rewriting the last two equations (Eqs.~(\ref{eq:1DMHDsolutionmovingwallexpfunc1}) and (\ref{eq:1DMHDsolutionmovingwallexpfunc2})) further in terms of the hyperbolic functions for the bounded flow problem under consideration, and using $r=q+q'$ and $r'=q-q'$ for convenience, we finally get the general solutions for the velocity and magnetic fields as
\begin{subequations}\label{eq:1DMHDsolutionmovingwallhyprbolfunc}
\begin{eqnarray}
   u_x^*(\xi) &=& (p+p')+r\cosh(\mbox{Ha}\xi)-r'\sinh(\mbox{Ha}\xi), \label{eq:1DMHDsolutionmovingwallhyprbolfunc1}\\
  B_x^*(\xi) &=& (p-p')+r'\cosh(\mbox{Ha}\xi)-r\sinh(\mbox{Ha}\xi). \label{eq:1DMHDsolutionmovingwallhyprbolfunc2}
\end{eqnarray}
\end{subequations}
The four constants $p$, $p'$, $r$ and $r'$ can now be obtained from the four boundary conditions given above. Now, using the velocity field given above in the no-slip velocity boundary conditions at the bottom and top plates given in Eq.~(\ref{eq:1DMHDBCs}), it follows that
\begin{equation}
r=\frac{1/2-(p+p')}{\cosh(\mbox{Ha})},\quad\quad r'=-\frac{1}{2\sinh(\mbox{Ha})}.\label{eq:intermediateconstants}
\end{equation}
Then, using the magnetic field given in Eq.~(\ref{eq:1DMHDsolutionmovingwallhyprbolfunc2}) and substituting the Shercliff boundary conditions at the bottom and top plates, respectively (see Eq.~(\ref{eq:1DMHDBCs})), and using the expressions for the $r$ and $r'$ given above in Eq.~(\ref{eq:intermediateconstants}) and simplifying, we get the following coupled set of two equations for $p$ and $p'$ as
\begin{subequations}\label{eq:constraintsystemtforconstants}
\begin{eqnarray}
  -\lambda_1(p+p')+(p-p') &=& \gamma_1, \\
  -\lambda_2(p+p')+(p-p') &=& \gamma_2,
\end{eqnarray}
\end{subequations}
where
\begin{subequations}\label{eq:coefficientsexpressions}
\begin{eqnarray}
  \lambda_1 &=& c_{wl}\mbox{Ha}+\tanh(\mbox{Ha}),\quad\quad \gamma_1=\frac{1}{2}\left[\frac{1}{\tanh(\mbox{Ha})}-\tanh(\mbox{Ha})\right], \\
  \lambda_2 &=& c_{wu}\mbox{Ha}+\tanh(\mbox{Ha}),\quad\quad \gamma_2=\frac{1}{2}\left[\frac{1}{\tanh(\mbox{Ha})}+\tanh(\mbox{Ha})\right]+c_{wu}\mbox{Ha}.
\end{eqnarray}
\end{subequations}
Then, solving for $(p+p')$ and $(p-p')$ from the two equations given in Eq.~(\ref{eq:constraintsystemtforconstants}), we obtain
\begin{equation}\label{eq:pprimesumdiffexpression}
(p+p')=\frac{\gamma_2-\gamma_1}{\lambda_1+\lambda_2}, \quad\quad (p-p')=\frac{\lambda_1\gamma_2+\lambda_2\gamma_1}{\lambda_1+\lambda_2}.
\end{equation}

Using these results for $(p+p')$ and $(p-p')$ (see Eq.~(\ref{eq:pprimesumdiffexpression})) in Eqs.~(\ref{eq:1DMHDsolutionmovingwallhyprbolfunc1}) and (\ref{eq:1DMHDsolutionmovingwallhyprbolfunc2}), and along with the expressions for $r$ and $r'$ (see Eq.~(\ref{eq:intermediateconstants})) and simplifying and rearranging, we finally obtain the analytical solutions for the velocity and magnetic fields in the Hartmann-Couette flow bounded by plates with contrasts in electrical conductivities, which can be expressed as follows:
\begin{subequations}\label{eq:1DMHDmovingwallfinalsolution}
\begin{eqnarray}
  u_x^*(\xi) &=& K_1+\left(\frac{1}{2}-K_1\right)\frac{\cosh(\mbox{Ha}\xi)}{\cosh(\mbox{Ha})}+\frac{1}{2}\frac{\sinh(\mbox{Ha}\xi)}{\sinh(\mbox{Ha})}, \label{eq:1DMHDmovingwallfinalsolution1} \\
  B_x^*(\xi) &=& K_2-\left(\frac{1}{2}-K_1\right)\frac{\sinh(\mbox{Ha}\xi)}{\cosh(\mbox{Ha})}-\frac{1}{2}\frac{\cosh(\mbox{Ha}\xi)}{\sinh(\mbox{Ha})}, \label{eq:1DMHDmovingwallfinalsolution2}
\end{eqnarray}
\end{subequations}
where
\begin{subequations}\label{eq:1DMHDmovingwallintconstants}
\begin{eqnarray}
  K_1 &=& \frac{\tanh(\mbox{Ha})+c_{wu}\mbox{Ha}}{2\tanh(\mbox{Ha})+(1+\phi)c_{wu}\mbox{Ha}}, \label{eq:1DMHDmovingwallintconstants1}\\
  K_2 &=& \frac{1+((1+\phi)/2)c_{wu}\mbox{Ha}\left[\tanh(\mbox{Ha})+1/\tanh(\mbox{Ha})\right]+(c_{wu}\mbox{Ha})^2\phi}{2\tanh(\mbox{Ha})+(1+\phi)c_{wu}\mbox{Ha}}, \label{eq:1DMHDmovingwallintconstants2}\\
  \phi &=& \frac{c_{wl}}{c_{wu}}. \label{eq:1DMHDmovingwallintconstants3}
\end{eqnarray}
\end{subequations}
Here, the value of the factor $\phi$ represents the contrast between the wall conductance ratio at the bottom and top plates and thus the coefficients $K_1$ and $K_2$ parameterize the influence of electrical properties of the bounding walls. Equations~(\ref{eq:1DMHDmovingwallfinalsolution1}) and (\ref{eq:1DMHDmovingwallfinalsolution2}) represent new analytical solutions for the shear driven MHD flow between two plates with different electrical conductivities and can be used for validating computational techniques for MHD flows involving moving walls. Note that for the special case with the same conductance ratios for both the walls ($c_{wl}=c_{wu}=c_w$), i.e., $\phi=1$, the coefficient $K_1=1/2$ and the above analytical solutions simplify to
\begin{subequations}\label{eq:1DMHDmovingwallfinalsolutionsplcase}
\begin{eqnarray}
  u_x^*(\xi) &=& \frac{1}{2}\left[1+\frac{\sinh(\mbox{Ha}\xi)}{\sinh(\mbox{Ha})}\right], \label{eq:1DMHDmovingwallfinalsolutionsplcase1}\\
  B_x^*(\xi) &=& \frac{1+c_{w}\mbox{Ha}\left[\tanh(\mbox{Ha})+1/\tanh(\mbox{Ha})\right]+(c_{w}\mbox{Ha})^2}{2\left[\tanh(\mbox{Ha})+c_{w}\mbox{Ha}\right]}-\frac{1}{2}\frac{\cosh(\mbox{Ha}\xi)}{\sinh(\mbox{Ha})}. \label{eq:1DMHDmovingwallfinalsolutionsplcase2}
\end{eqnarray}
\end{subequations}
This leads to the important conclusion that while the induced magnetic field $B_x^*$ is always affected by the value of the wall conductance ratios, whether it is the same or different for the two walls, by contrast, the velocity field $u_x^*$ is sensitive to the electrical properties of the walls only if there is a contrast between the values of the wall conductance ratio at the bottom and top plates. This finding provides a potential mechanism to control the flow and transport in MHD shear flows by adjusting the electrical properties of the bounding plates. Also, as a further simplification, considering both the walls to be insulated, i.e., $c_w=0$, when we have $K_2=1/(2\tanh(\mbox{Ha}))$, the structure of the velocity and magnetic fields reduce further to
\begin{subequations}\label{eq:1DMHDmovingwallfinalsolutionsplinsulcase}
\begin{eqnarray}
  u_x^*(\xi) &=& \frac{1}{2}\left[1+\frac{\sinh(\mbox{Ha}\xi)}{\sinh(\mbox{Ha})}\right], \label{eq:1DMHDmovingwallfinalsolutionsplinsulcase1}\\
  B_x^*(\xi) &=& \frac{1}{2\tanh(\mbox{Ha})}\left[1-\frac{\cosh(\mbox{Ha}\xi)}{\cosh(\mbox{Ha})}\right]. \label{eq:1DMHDmovingwallfinalsolutionsplinsulcase2}
\end{eqnarray}
\end{subequations}

%

\end{document}